\documentclass[aps,prx,floatfix,nofootinbib,notitlepage,twocolumn,superscriptaddress]{revtex4-1}
\usepackage{booktabs}
\usepackage{lipsum}
\usepackage{graphicx}
\usepackage{amsmath,amssymb,amsthm}
\usepackage{braket}
\usepackage{mathtools}
\usepackage{hyperref}
\usepackage{color}
\usepackage{float}
\usepackage{hyperref}
\usepackage[normalem]{ulem}
\usepackage{colonequals}
\usepackage{siunitx}
\usepackage{enumitem}
\usepackage{bbm}
\usepackage[dvipsnames]{xcolor}
\usepackage{tabularx}
\usepackage{mathrsfs}
\usepackage{tikz-feynman}
\tikzfeynmanset{compat=1.1.0}
\usepackage{silence}
\usepackage{CJKutf8}

\makeatletter 
    
\renewcommand\onecolumngrid{
\do@columngrid{one}{\@ne}%
\def\set@footnotewidth{\onecolumngrid}
\def\footnoterule{\kern-6pt\hrule width 1.5in\kern6pt}%
}

\renewcommand\twocolumngrid{
        \def\footnoterule{
        \dimen@\skip\footins\divide\dimen@\thr@@
        \kern-\dimen@\hrule width.5in\kern\dimen@}
        \do@columngrid{mlt}{\tw@}
}%

\makeatother    

\newcommand{\ov}{\overline}

\newcommand{\GS}{\text{GS}}

\renewcommand{\v}[1]{\boldsymbol{#1}}
\DeclareMathOperator{\Tr}{Tr}
\DeclareMathOperator{\tr}{tr}
\DeclareMathOperator{\diag}{diag}
\let\Re\relax

\DeclareMathOperator{\Re}{Re}

\newcommand{\T}{\mathcal{T}}

\newcommand{\bz}{\mathrm{BZ}}
\newcommand{\uc}{\mathrm{UC}}

\newcommand{\BM}{\mathrm{BM}}
\newcommand{\SU}{\mathrm{SU}}

\newcommand{\U}{{\rm U}}

\newcommand{\br}{{\v r}}

\newcommand{\bR}{{\v R}}
\newcommand{\bk}{{\v k}}

\newcommand{\bq}{{\v q}}

\newcommand{\bG}{{\v G}}

\newcommand{\bA}{{\v A}}
\newcommand{\bQ}{{\v Q}}
\newcommand{\bsigma}{{\v \sigma}}
\newcommand{\bnabla}{{\v \nabla}}

\renewcommand{\H}{\mathcal{H}}

\DeclarePairedDelimiter\abs{\lvert}{\rvert}%
\DeclarePairedDelimiter\norm{\lVert}{\rVert}%
\makeatletter
\let\oldabs\abs
\def\abs{\@ifstar{\oldabs}{\oldabs*}}
\let\oldnorm\norm
\def\norm{\@ifstar{\oldnorm}{\oldnorm*}}
\makeatother
\begin{document}

\title{
Nonlocal Moments and Mott Semimetal in the Chern Bands of Twisted Bilayer Graphene
}

\begin{abstract}
    
    Twisted bilayer graphene (TBG) has elements in common with two paradigmatic examples of strongly correlated physics: quantum Hall physics and Hubbard physics. On one hand, TBG hosts flat topological Landau-level-like bands which exhibits quantum anomalous Hall effects under the right conditions. On the other hand, these bands are characterized by concentrated charge density and show signs of extensive entropy usually attributed to local moments. 
    The combination of these features leads to a question: can decoupled moments emerge in an isolated topological band, despite the lack of exponentially localized Wannier orbitals? In this work, we answer the question affirmatively by proposing a minimal model for the flat topological bands in TBG that combines topology and charge concentration at the AA sites, leading to analytic wavefunctions that closely approximate those of the Bistritzer-MacDonald model with realistic parameters. Importantly, charge concentration also leads to Berry curvature concentration at $\Gamma$, giving rise to a small parameter $s$ that makes the model analytically tractable. We show that, rather surprisingly, the model hosts nearly decoupled flavor moments without invoking any extra degrees of freedom. These moments are non-local due to topology-enforced power-law tails, yet have parametrically small overlap.  We develop a systematic diagrammatic expansion in which the self energy can be computed exactly to leading order in $s^2$ in the fluctuating moment regime and predict momentum-resolved spectral functions for future experiments to verify. Our key discovery is a charge neutrality state we refer to as the "Mott semimetal," characterized by high flavor entropy and a Mott gap that exists throughout most of the Brillouin zone but closes at the $\Gamma$ point. At $\Gamma$, the spectral function contains a \emph{single} Dirac cone per spin per valley and responds to perturbations in an exotic manner that is distinct from any other theoretical picture of TBG. Away from neutrality, the Mott semimetal gaps out in a spectrally imbalanced manner, with one Mott band having zero quasiparticle residue at the $\Gamma$ point. The model accurately reproduces results from finite temperature thermodynamic measurements, leads to new experimental predictions, and resolves the problem of the emergence of Hubbard physics in  isolated topological bands.

\end{abstract}

\author{Patrick J. Ledwith}
\affiliation{Department of Physics, Harvard University, Cambridge, MA 02138, USA}
\author{Junkai Dong (\begin{CJK*}{UTF8}{bsmi}董焌\end{CJK*}\begin{CJK*}{UTF8}{gbsn}锴\end{CJK*})}
\affiliation{Department of Physics, Harvard University, Cambridge, MA 02138, USA}
\affiliation{Kavli Institute for Theoretical Physics, University of California, Santa Barbara, California 93106, USA}
\author{Ashvin Vishwanath}
\affiliation{Department of Physics, Harvard University, Cambridge, MA 02138, USA}
\author{Eslam Khalaf}
\affiliation{Department of Physics, Harvard University, Cambridge, MA 02138, USA}

\maketitle

\section{Introduction}
The realization of flat bands in moir\'e systems has transformed the study of strong correlation physics and raised several fundamental conceptual questions about the role of band topology in interacting systems. While mapping to the Landau level problem -- the prototype for interacting topological bands -- has provided crucial insights in understanding the correlated phases in these platforms \cite{Po_2018,Tarnopolsky2019origin, BultinckHidden,LedwithFCI, ledwith2021strong, khalaf2020soft, LiuLiuDai, LiuDai, YahuiChernBands, YahuiFerromagnetism, LandauLevelWang}, this mapping does not conceptually account for unique qualitative ingredients of topological bands with no counterpart in Landau levels. These include the non-uniform charge distribution within the unit cell and the corresponding non-uniform quantum geometry in momentum space (non-uniform Berry curvature and quantum metric) \cite{LedwithFCI}.

A key question in the field of moir\'e materials is concerned with the nature of correlated insulating states observed in their topological flat bands and the relevance of Mott-Hubbard physics to describing them.
In a strongly correlated system with trivial bands where the Hubbard model is an appropriate description, there is a separation of scales between the Hubbard scale $U$, the largest energy scale in the problem where charge fluctuations are frozen, and other energy scales, such as the exchange and superexchange scales, under which spin moments are ordered. This leads to a broad intermediate temperature regime where charge is frozen but spins have not ordered yet, corresponding to a Mott insulator that does not break any symmetry and is characterized by a large entropy, $O(k_B)$ per lattice site. In contrast, in a topological band, any complete basis of real space orbitals has power-law tails whose asymptotics are universal (depending only on topology) \cite{Thouless1984wannier, Li2024constraints}, leading to a direct exchange contribution that naively seems to be of the same order as the interaction scale. This precludes any local moment description in a topological band and suggests that a fluctuating decoupled moment (Mott) phase is impossible.

This issue is particularly pertinent to explaining experimental observations in twisted bilayer graphene (TBG)~\cite{Cao_2019,Yankowitz_2019,Lu_2019,Stepanov_2020,Cao_2021,Liu_2021,ghoshThermopowerProbesEmergent2025,merinoEvidenceHeavyFermion2024,batlle-porroCryoNearFieldPhotovoltageMicroscopy2024,calugaruThermoelectricEffectIts2024}. On the one hand, there is ample evidence of band topology based on both theoretical consideration \cite{Po_2018,ahn2019failure, Po2019faithful, Tarnopolsky2019origin, ledwith2021strong, TBGII, KangStrongCoupling,koshinoMaximallyLocalizedWannier2018} as well as experimental observations: these include field-induced correlated Chern insulators~\cite{CorrelatedChernYoung, CorrelatedChernYazdani, CorrelatedChernNadjPerge, CorrelatedChernEvaAndrei, Pierce2021,VafekCorrelatedChern}, fractional Chern insulators at small magnetic fields~\cite{Xie2021fractional}, and quantum anomalous Hall effect in hBN-aligned samples~\cite{YoungQAH, DavidGGQAH}. On the other hand, STM shows that the electron densities of the flat bands are concentrated at AA moiré lattice sites~\cite{liObservationVanHove2010a,tilakFlatBandCarrier2021,kerelskyMaximizedElectronInteractions2019,xieSpectroscopicSignaturesManybody2019a}. Theory predicts that most of the Brillouin zone (BZ), except the topological $\Gamma$ point region, describes Bloch waves of AA-localized states \cite{CarrWannier,namLatticeRelaxationEnergy2017,guineaElectrostaticEffectsBand2018,koshinoMaximallyLocalizedWannier2018,carrExactContinuumModel2019,rademakerChargeSmootheningBand2019,hauleMottsemiconductingStateMagic2019b,goodwinHartreeTheoryCalculations2020}. Entropy measurements \cite{IlaniEntropy,saitoIsospinPomeranchukEffect2021} indicate a large entropy $O(k_B)$ per moir\'e unit cell which persists over a wide temperature range from $\sim$ $4$K to almost $100$K, suggesting the existence of a regime of decoupled fluctuating moments. 

Explaining these seemingly contradictory observations has motivated two promising sets of approaches. Both go beyond the isolated band limit by incorporating remote bands. First, a few studies have employed multiband models which combine the flat bands with a few remote bands such that the whole set of bands is topologically trivial~\cite{ZhoiWannier,Po2019faithful,KaxirasTightBinding,TBGII}. DMFT calculations on the multi-band model observed signatures of Hubbard physics and decoupled moment formation~\cite{hauleMottsemiconductingStateMagic2019b,Bascones2019, Bascones2020, Bascones2023}. The second is the topological heavy fermion (THF) model \cite{song2022TBGTHF,calugaru2023TBGTHF2,yu2023TTGTHF,HuRKKY,herzogarbeitman2024topologicalheavyfermionprinciple, TBGKondo, THFKondo, TBGKondoSDS, SongLian, lau2024topologicalmixedvalencemodel, ValentiDMFT} which decomposes the low energy degrees of freedom of the non-interacting Bistritzer-MacDonald (BM) model into localized $f$ electrons carrying the local moments and highly dispersing $c$ electrons carrying the topology. These $c$ and $f$ electrons are obtained by mixing the flat bands and part of the remote bands near $\Gamma$. In the limit where the interaction scale $U$ exceeds the hybridization scale between the $c$ and $f$ electrons, which determines the gap between the flat bands and the remote bands, the $f$ electron charge freezes and the model describes a set of local moments coupled to itinerant $c$ electrons. In the opposite limit, isolated flat bands of the BM model which can be thought of as a mixture of $c$ and $f$ electrons are recovered. Importantly, band topology forbids local moments in the latter limit. 

While the incorporation of remote bands represents a possible resolution to the conflicting requirements of band topology and Hubbard physics, it does not immediately explain \emph{decoupled moments} due to the presence of RKKY interaction between $f$ moments, induced by coupling to the $c$ electrons \cite{HuRKKY}. Furthermore, the hierarchy of scales where the local moments are weakly
hybridized (when the interaction scale far exceeds the
gap to the remote bands) is incompatible with the observed gaps at at full filling $\nu = \pm 4$ \cite{HuRKKY}. More importantly, approaches which incorporate remote bands do not address the fundamental question of whether the formation of decoupled moments is forbidden in an isolated topological band.  In fact, Monte Carlo simulations~\cite{hofmannFermionicMonteCarlo2022} of the projected interacting BM model at charge neutrality already indicates that aspects of Mott physics are realized without remote bands. Moreover, there are several conceptual advantages to working in the projected limit where remote bands are integrated out: the Hilbert space dimension is fixed without the need to introduce any cutoff, some constraints related to sum rules can be made very sharp \cite{MaoDiamagneticResponse,mao2023upperboundssuperconductingexcitonic,mendezvalderrama2023theorylowenergyopticalsumrule,Verma_2021}, and the physics can be related to LL physics by employing recently developed concepts of ideal bands~\cite{PhysRevB.90.165139,Parameswaran_2013,Jackson:2015aa,Martin_PositionMomentumDuality,Tarnopolsky2019origin,LedwithFCI,kahlerband1,kahlerband2,kahlerband3,JieWang_exactlldescription,Jie_hierarchyidealband,Eslam_highC_idealband,JW_origin_22,junkaidonghighC22,crepel_chiral_2023,valentin23ideal,andrews2023stability} and vortexability \cite{LedwithVishwanathParker22,okuma2024constructingvortexfunctionsbasis,fujimoto2024highervortexabilityzerofield}.

Here we work in the projected limit of fully isolated topological flat band and show, surprisingly, that it is possible to explain the physics of decoupled moments as well as  the experimental observation of large entropy. We further establish that this can be achieved using minimal ingredients, which are justified on phenomenological grounds and are present in the BM model. In particular, we show that the combination of symmetry, topology, and charge concentration at the AA site in TBG leads naturally to an analytic ansatz for the flat band wavefunctions that (i) reproduces BM flat band wavefunctions to remarkable accuracy and (ii) enables strong-coupling analytic tractability by treating the area of the unit cell where charge is concentrated as a small parameter. Despite the absence of local moments in the model, we can construct {\em non-local} moments whose localization length is infinite due to topology-enforced power law tails. However, these moments have most of their weight concentrated within the first unit cell~\cite{zang2022realspacerepresentationtopological} leading to a parametrically small exchange scale. 
These lead to a symmetric fluctuating moment regime at intermediate temperatures, whose entropy can be rigorously lower-bounded, and shows remarkable agreement with experiment~\cite{IlaniEntropy}. Our findings show how we can reconcile decoupled moment features seen in thermodynamic quantities with topological features seen in transport in \emph{fully isolated topological bands without incorporating any extra degrees of freedom}.

Furthermore, we use the small parameter of the model as the basis for a systematic diagrammatic expansion which, within the fluctuating moment regime, allows us to compute the self-energy to leading order exactly. To the best of our knowledge this is the first \emph{analytic} calculation of the electron spectral function in TBG in the fluctuating moment regime. Our most remarkable finding is a class of unusual Mott-like phases which describe fluctuating non-local moments. At charge neutrality, the Mott phase is a semimetal in the absence of particle-hole breaking perturbations, with a large Mott gap everywhere except in the vicinity of the $\Gamma$ point where the gap closes. The spectral function in the vicinity of $\Gamma$ describes a single Dirac cone per spin per valley and responds to perturbations in an exotic manner that is qualitatively incompatible with any non-interacting or mean-field description. A state with these characteristics has not been described before, to the best of our knowledge. Away from neutrality, the fluctuating moments give rise to a spectrally imbalanced state where the quasiparticle weight smoothly shifts between the lower and upper Mott bands, and vanishes on one side at $\Gamma$. 

The Mott spectra we obtain, and their response to perturbations, are sharply distinct from any other theoretical picture. These momentum-resolved spectra are directly observable with techniques such as the quantum twisting microscope \cite{QTM} (thus far only applied away from the magic angle). However, even without momentum resolved spectroscopy our predictions can be tested. For example, we show that the spectral function in the presence of heterostrain, at $\abs{\nu} = 1$ specifically, has a spectral gap that drastically increases with temperature. This is consistent with the experimentally observed Pomeranchuk effect in transport \cite{saitoIsospinPomeranchukEffect2021} and could be directly measured using STM or Dirac cone spectroscopy in trilayer~\cite{shenDiracSpectroscopyStrongly2023,bocarsly2024imagingcoulombinteractionsmigrating}.

\section{Summary of Results}

\begin{figure*}
    \centering
    \includegraphics{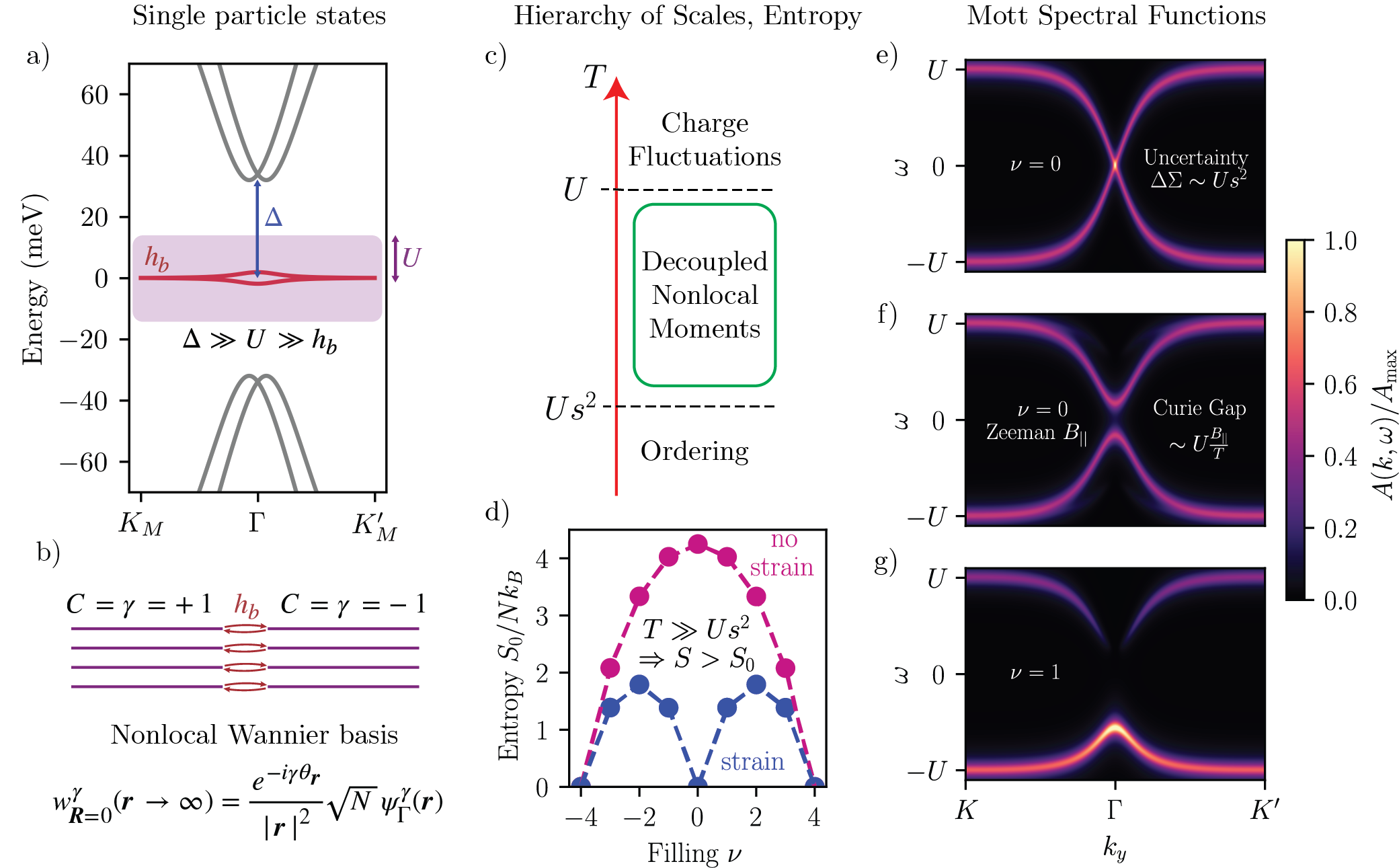}
    \caption{Summary of results. 
    Panel \textbf{{(a)}} plots the single particle BM band structure, with $w_0/w_1 = 0.75$ and $\theta = 1.06^\circ$, and highlights the relevant states associated with the separation of scales $\Delta \gg U \gg h_b$ that we use in this manuscript. Here $\Delta$ is the gap to the remote bands, $U$ is the Coulomb scale, and $h_b$ is the bare dispersion. We therefore project out remote bands and focus on strong interactions within the flat band subspace. 
    In panel \textbf{{(b)}}, we depict the Chern basis for the flat bands. The small non-interacting dispersion acts as a tunneling between Chern sectors, each of which has an approximate $\U(4)$ symmetry. Symmetric, exponentially localized, Wannier orbitals cannot be constructed due to the topology. We will instead use the ``nonlocal" Wannier states obtained by Wannierizing the bands in the Chern basis. Such states have power law asymptotics whose coefficient cannot be suppressed and infinite localization length $\sqrt{\langle r^2 \rangle}$.
    Panel \textbf{{(c)}} showcases the hierarchy of scales within the interacting flat band subspace, where $s$ is a small parameter related to the charge and Berry curvature concentration. 
    In Panel \textbf{{(d)}} we plot the decoupled moment entropy $S_0$ versus the flat band filling $\nu$, where $\pm 4$ denote full and empty respectively. For $U \gg T \gg Us^2$, we have $S\approx S_0$. Furthermore, we prove rigorously $S \geq S_0 - J_*/T$ at the dotted integer fillings, where $J_* \propto Us^2$ at small $s$. 
    Panels \textbf{{(e-g)}} show the spectral function $A(\bk,\omega)$ in the symmetric ``Mott regime" $U \gg T \gg Us^2$, calculated analytically. All spectral functions include a broadening $Us^2$, where $s=0.25$ here, reflecting the level of uncertainty in the calculation of the self energy $\Sigma$. 
    Panel \textbf{{(e)}} shows the Mott semimetal state at charge neutrality with zero bare dispersion; the spectral function peak looks like a single Dirac cone per spin per valley, despite $\Sigma(\bk) \propto |\bk|^2$ for $\bk \to 0$. The band touching at $\bk =0$ originates from the $\pm 2\pi$ Berry phase associated with the concentrated Berry curvature at the $\Gamma$ point. 
    Panel \textbf{{(f)}} shows the gapping of the Mott semimetal with a Zeeman field $B_{||}$; orbital effects are excluded for simplicity. The gap is proportional to the interaction scale and magnetization; the latter obeys a Curie law at small fields. The ``Curie gap", inversely proportional to temperature, highlights that the gaplessness of the Mott semimetal has its origins in the fluctuating nonlocal moments of the state. 
    In Panel \textbf{(g)} we plot the spectral function of the ``spectrally imbalanced" Mott state, here at $\nu = 1$. The inherent particle-hole asymmetry associated with $\nu \neq 0$ opens a spectral gap in the Mott semimetal. The gap opens in a nonstandard way, and preserves the eight-fold degeneracy at $\Gamma$ associated with the fluctuating moments. The eight-fold degeneracy is inherited by the holon band for $\nu > 0$, while the doublon band has quasiparticle residue $Z_\bk \propto |\bk|^2$ as $\bk \to 0$. 
    }
    \label{fig:summaryofresults}
\end{figure*}

In this paper, we will focus on the 
TBG flat bands in the strong coupling limit, with Coulomb interactions projected to the flat bands. In particular, we consider the hierarchy of scales $\Delta \gg U \gg h_b$ where $\Delta$ is the energy gap to the remote bands, $U$ is the Coulomb scale, and $h_b$ is the bare bandwidth. This hierarchy is depicted in Fig. \ref{fig:summaryofresults}a, on top of the BM band structure, which hosts two flat bands per spin per valley~\cite{Bistritzer12233}. In practice, whether this hierarchy of scales is realized or not depends on the dielectric constant $\epsilon$, which receives contributions from the experiment-dependent dielectric environment as well as screening from remote bands. However, the frequent observation of band gaps at full and empty filling of the flat bands suggest that the non-interacting band gaps are sufficiently large to not be eliminated by Coulomb interactions. We will discuss further experimental support for this hierarchy of scales in the discussion section. 

Close to the magic angle where the interaction scale dominates the non-interacting bandwidth, this pair of bands, for each spin and valley, is more conveniently expressed as a Chern $+1$ and a Chern $-1$ band related by $C_2 \T$ symmetry. Accounting for spin-valley degeneracy, we have four Chern $+1$ and four Chern $-1$ flat bands, as depicted in Fig. \ref{fig:summaryofresults}b. The interaction Hamiltonian turns out to have an approximate $\U(4) \times \U(4)$ symmetry associated with independent $\U(4)$ rotations in each Chern sector, which get broken down by smaller perturbations to the physical symmetry \cite{BultinckHidden, TBGIVGroundState, VafekRG}. The flat Chern bands and the $\U(4) \times \U(4)$ symmetry naturally lead to correlated insulator ground states at integer filling which can be understood as generalized quantum Hall ferromagnets \cite{BultinckHidden, ledwith2021strong}. However, both the BM model and experiments suggests an important difference between these Chern bands and those in quantum Hall systems: the concentration of real-space charge density and Berry curvature distributions. In the presence of small heterostrain, these properties lead to nematic semimetal and spiral states~\cite{huderElectronicSpectrumTwisted2018,MespleHeterostrain,bi_designing_2019,DaiEffectHeterostrain,WangUnusualMagnetotransport,STMNadjPerge,Nuckolls_2023,bocarsly2024imagingcoulombinteractionsmigrating,liuNematicTopologicalSemimetal2021,parkerStrainInducedQuantumPhase2021,soejimaEfficientSimulationMoire2020,KwanKekule,WangIKSDMRG}, which have no analog in the quantum Hall problem.

In Sec. \ref{sec:ModelIntro}, we introduce a minimal model for the TBG Chern bands. The model is constructed from three components that we argued to be essential (i) band topology, (ii) symmetry, and (iii) concentrated charge. A fundamental property of this model is a topological obstruction to arbitrarily localizing the charge at a single point, which for TBG is the AA point by symmetry. We will see that concentration of Berry curvature at $\Gamma$ is a natural consequence of this property, essentially because $\psi_{\Gamma}(\br = \rm{AA}) = 0$ manifests the obstruction. 

In Sec. \ref{sec:exactFM} we review the exact $T=0$ ferromagnetic states at integer filling without strain. Such ferromagnets have charge gaps and spin stiffness of the same order as the interaction scale $U$, and thus naively look stable to perturbations $\ll U$. Nonetheless, we later show explicitly that these ferromagnets are unstable to temperatures, $T$, and strain energy scales, $E_{\rm str}$, that are much smaller than $U$. This shows that the instability~\cite{liuNematicTopologicalSemimetal2021,KwanKekule,soejimaEfficientSimulationMoire2020,parkerStrainInducedQuantumPhase2021} of generalized ferromagnets in TBG to strain is not an accidential property depending on details, but is a direct consequence of obstructed charge concentration.

In Sec. \ref{sec:decoupledEntropy}, we demonstrate that the concentration of charge and Berry curvature gives rise to the remarkable phenomenon of weakly coupled nonlocal moments (Sec.~\ref{subsec:Wannier}). These moments correspond to electrons occupying Wannier states of the Chern bands. The band topology dictates that these states have asymptotic power law tails \cite{Thouless1984wannier,Li2024constraints} (Fig. \ref{fig:summaryofresults}b), resulting in an infinite localization length. Due to the infinite localization length, and the fact that the asymptotic tail of the state is non-perturbatively essential for producing the topology, we will refer to the flavor moments of such Wannier states as ``nonlocal moments."

Although the moments are nonlocal, we show that their exchange coupling 
is parametrically small, occurring at an energy scale of 
$Us^2\ll U$, where $s \ll 1$ is proportional to the width of both charge and Berry curvature distributions. The smallness of this coupling arises from the small overlap between Wannier states centered at different lattice sites, a consequence of charge concentration.

Using this information, in Sec.~\ref{subsec:entropy} we explicitly construct exponentially many states with energy $\sim Us^2$ relative to the ground state leading to a rigorous lower bound on the entropy at integer filling
\begin{equation}
    S \geq S_0 - NJ_*/T,  \qquad J_* \sim Us^2,
    \label{eq:entropyboundsummary}
\end{equation}
where $S_0$ is the entropy of classical, decoupled, moments at each lattice site and $N$ is the number of sites. For the hierarchy of scales of interest, $U \gg T \gg Us^2$, depicted in Fig. \ref{fig:summaryofresults}c, we expect $S \approx S_0$. We plot the $S_0$ of the entropy bound \eqref{eq:entropyboundsummary} in Fig. \ref{fig:summaryofresults}d, with and without strain. In Sec.~\ref{subsec:entropy_strain}, we show that for $U s^2 \ll T \ll E_{\rm{str}}$ strain quenches the entropy at the charge neutrality point by polarizing the system into a spin-valley singlet nematic semimetal with zero classical entropy. For $\nu \neq 0$, the entropy is only partially quenched, and is described by decoupled spin and valley moments. Our results are consistent with the filling and temperature dependence of entropy measured experimentally~\cite{IlaniEntropy} assuming a strain scale $U s^2 \ll E_{\rm{str}} \ll U$. A direct comparison with experimentally measured entropy \cite{IlaniEntropy} suggests a strain scale $E_{\rm{str}} \approx 50K$ with $Us^2 \approx U/16 \approx 10K$, which corresponds to an interaction scale $U \approx 15$ meV. In Sec.~\ref{sec:discussion}, we show that this interaction scale is consistent with most experimental measurements across different TBG samples.

In Sec. \ref{sec:Mott}, we discuss charged excitations within the fluctuating moment or ``Mott" regime $U \gg T \gg Us^2$. In this regime, charge is mostly frozen due to $U \gg T$ while there is a large entropy due to the nearly decoupled ($T \gg Us^2$) nonlocal moments at each lattice site. However, in topological bands it is 
impossible to have a complete separation between spin and charge degrees of freedom since the projected spin and density operators do not generally commute in a Chern band \cite{MoonSkyrmion}.
Indeed, spin textures with nontrivial winding number, skyrmions, carry electric charge proportional to the Chern number \cite{SondhiSkyrmions, MoonSkyrmion}. 

We will show how to calculate the self energy, and thus the single particle spectral function $A(\bk,\omega)$, in the Mott regime, in an analytically controlled manner in the limit $s \ll 1$. 
Although the limit $s \rightarrow 0$ is non-perturbative, we can make use of small phase space approximations controlled by the small parameter $s^2 \ll 1$ to evaluate the self energy to leading order in $U s^2$, either diagram-by-diagram or through the non-perturbative Schwinger-Dyson equations. Intuitively, the area of the Brillouin zone away from the $\Gamma$ point acts as a bath for electrons near the $\Gamma$ point, which occupies a phase space of order $s^2$. 

We now describe the spectral functions we obtain. In a typical Mott insulator, such as a Hubbard model with hopping $t=0$, the spectral function is given by a sum of delta functions at $\omega = \pm U$ for each $\bk$. We show this to be the case in our model as long as $\bk$ is far from the $\Gamma$ point. Near the $\Gamma$ point the spectral function exhibits a variety of features depending on the filling and which single particle terms are included. 

We show in Sec.~\ref{subsec:semimottal} that at charge neutrality, the spectral function is semimetallic with a Dirac like crossing as $\bk \to 0$, see Fig. \ref{fig:summaryofresults}e. The linear dispersion arises from a self energy which takes the form
\begin{equation}
    \Sigma^{\nu = 0}(\bk,\omega) = \frac{\abs{\bk}^2}{\abs{\bk}^2 + 2 s^2} \frac{U^2}{\omega},
    \label{eq:neutselfenergy_summary}
\end{equation}
in the first BZ, where $\bk$ is in units such that the area of the BZ is $2\pi$. Note that the self energy goes to zero quadratically in $\bk$. The $\omega^{-1}$ factor both opens the Mott gap away from $\Gamma$ and makes the crossing at $\Gamma$ linear. This can be seen by noting that the inverse Green's function is proportional to $\omega - \frac{U^2}{\omega}$ away from $\Gamma$ and $\omega - \frac{|\bk|^2}{2s^2} \frac{U^2}{\omega}$ close to $\Gamma$, leading to poles at $\omega = \pm U$ and $\omega = \pm \frac{|\bk|}{\sqrt{2}s} U$, respectively. We will refer to this state as a ``Mott semimetal." Single particle terms that are particle-hole symmetric and $\U(4) \times \U(4)$ symmetric cannot gap the Mott semimetal. We emphasize that due to the $1/\omega$ dependence of the self-energy, this state is sharply distinct from any weak coupling or mean-field state. Indeed, any symmetric mean-field state must have \emph{two} Dirac cones per spin per valley in the spectral function whereas the Mott semimetal only has one. We see that the conventional ties between Berry phase and energy spectrum require revision when the self-energy becomes strongly frequency-dependent.

The strong $\bk$ dependence of the self-energy \eqref{eq:neutselfenergy_summary} precludes a physical interpretation in terms of local moments and nearly decoupled conduction electrons. Indeed, a system consistent with the latter interpretation would have local moment correlators $\langle f_{\bR' \tau} f^\dag_{\bR 0} \rangle \propto \delta_{\bR \bR'}$; upon Fourier transformation this yields a $\bk$-independent spectrum for these states. We conclude that the absence of spectral weight at $(\bk = \Gamma, \omega = \pm U)$, together with the Mott gap at other $\bk$, is inconsistent with a local moment picture. Furthermore, this signature can be directly probed in future momentum resolved experiments. We now discuss how the response to perturbations of the $\Gamma$ point region also requires an interpretation in terms of \emph{nonlocal} moments.
 
To check that the gaplessness of the Mott semimetal is associated with decoupled nonlocal moments, we imagine coupling a Zeeman field $\H \to \H - B_{||} S_x$, where $S_x$ is a spin operator. Note that we do not include orbital effects of $B$, which cannot be realistically neglected in TBG but can be in alternating-twist trilayer graphene if the field is in-plane~\cite{ledwith2021tbtbcontrastingproperties,LakeSenthilPairing, LakeSenthilTTG, MacdonaldInPlane}. We find that the addition of this Zeeman field opens a gap proportional to the interaction scale and the Curie magnetization, $\sim U B_{||}/T$ at small $B_{||}$, see Fig. \ref{fig:summaryofresults}f. We expect the Mott semimetal to cross over into the gapped ferromagnet as $T \to 0$ through the spontaneous generation of a local exchange field $\propto Us^2$. This intuitive picture is consistent with the regime of validity $T \gg Us^2$ of our calculation. We emphasize that in a typical Mott insulator of local moments, spin and charge are entirely separated with charge degrees of freedom exponentially frozen at $T \ll U$; there is no analogous effect of gapping charge degrees of freedom through Curie magnetization.

In Sec.~\ref{subsec:SIMI}, we demonstrate that at integer fillings $\nu \neq 0$, the Mott part of the self energy still vanishes as $\bk \to 0$ leading to an eight-fold degeneracy if no other single particle terms are added. Remarkably, however, this does not imply gaplessness away from $\nu =0$. Instead, we find that there is now a gap as $\bk \to 0$, with a band separated by an order $U$ energy scale. For $\nu > 0$, the holon band has an eight fold degeneracy as $\bk \to \Gamma$, while the doublon band is separated by a band gap. To compensate for the eight-fold degeneracy of the $\bk = \Gamma$ holons, the doublon band has vanishing quasiparticle weight $Z_\bk \to 0$ as $\bk \to \Gamma$, as is visible in Fig. \ref{fig:summaryofresults}g. Thus, while the Mott state is gapped away from neutrality, the topology-induced spin-charge entanglement leads an unusual continuous shift of spectral weight between the holon and doublon bands close to the $\Gamma$ point.
In contrast, for $\nu < 0$ the eight-fold degeneracy is on the doublon side. The Mott semimetal lies in between at $\nu = 0$ and can thus be regarded as a kind of critical point for the spectrally imbalanced gap inversion.

\section{Model}\label{sec:ModelIntro}

\begin{figure*}
    \centering
    \includegraphics[width=\textwidth]{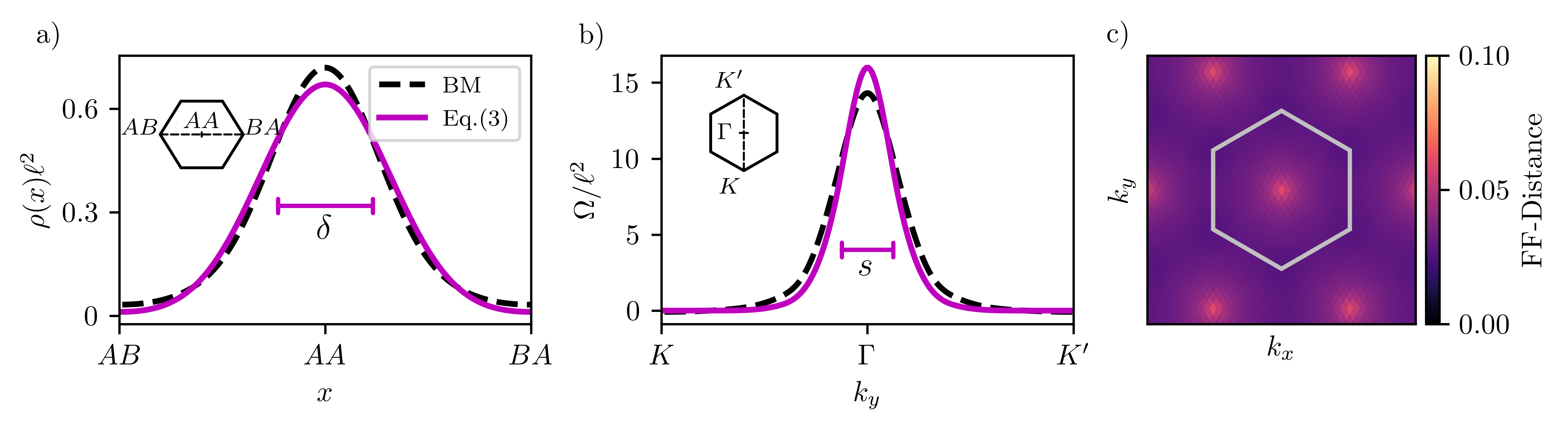}
    \caption{Comparison between the BM model and our obstructed charge localized model wavefunctions \eqref{eq:fullWF}. We use the BM parameters $w_0/w_1 = 0.7$ and $\theta = 1.06^\circ$, and match the charge density (panel \textbf{(a)}) with $\delta = 0.46$ and Berry curvature distribution (panel \textbf{(b)}) with $s = \beta^{-1} \ell^{-2} \delta = 0.25$, where we choose units $\ell = \sqrt{A_\uc/2\pi} \to 1$. The Berry curvature distribution of the BM model bands is obtained by diagonalizing the projected sublattice operator $\Gamma_{a b}(\bk) = \bra{u_{\bk a}} \sigma_z \ket{u_{\bk b}}$ and computing the Berry curvature of the resulting band with positive eigenvalue under $\Gamma$ \cite{BultinckHidden}. In panel \textbf{(c)} we plot the form factor distance defined in Eq. \eqref{eq:ffdistance} over the moir\'{e} Brillouin Zone; it takes a roughly constant $\sim 5\%$ value.}
    \label{fig:comparebmconc}
\end{figure*}

In the hierarchy of scales $\Delta \gg U \gg h_b$, the interacting problem is specified by projecting the Coulomb interaction onto the flat bands (see Sec. \ref{sec:exactFM} for the full Hamiltonian). Thus, the only input to the interacting problem are the flat band wavefunctions and their dispersion. Our approach is to construct a set of analytical wavefunctions that describe the eight TBG flat bands: four $C=1$ bands and four $C=-1$ bands with a $\U(4) \times \U(4)$ symmetry relating the wavefunctions in each Chern sector~\cite{BultinckHidden,VafekRG} and use these wavefunctions as input to the interacting problem. 
Our Chern band wavefunctions will have the topology and symmetry of TBG flat bands, with a charge distribution that is parametricallly localized at the AA sites. We will explicitly compare them to BM model wavefunctions and find excellent agreement. Their main advantage is that they contain a manifest small parameter, the charge concentration, that we will use to perform controlled, analytical, calculations in subsequent sections. 

To account for this charge concentration at AA, we write the following ansatz for a $C=1$ flat band wavefunction (the $C=-1$ bands follow from time reversal).
\begin{equation}
  \phi_\bk(\br) \to \psi_\bk(\br) = c_\bk \phi_\bk(\br) e^{- \lambda K(\br)}
\label{eq:concentrate}
\end{equation}
Here, $\phi_\bk(\br)$ are wavefunctions with the same topology and symmetry as those of the TBG $C=1$ bands, $K(\br)$ is a periodic function that attains its unique minimal values at the lattice sites, or $\br = 0$ in the first unit cell, and $c_\bk$ represents the gauge and normalization freedom. For each $\lambda$, \eqref{eq:concentrate} describes a band that has the right symmetry and topology and whose charge density has its maximum at AA, becoming more concentrated as $\lambda$ is increased. 

In the following, we will show that \eqref{eq:concentrate} reduces to a universal form independent of most details of $\phi_\bk(\br)$ and $K(\br)$ in the limit $\lambda \gg 1$ and that this universal form provides an excellent approximation to the flat band wavefunctions of the BM model at realistic parameters. However, with a view towards a systematic expansion, it is helpful to think of these as wavefunctions describing a family of models, with a widely tunable range of charge concentration. It is worth noting that bands of the type \eqref{eq:concentrate}, with arbitrarily tunable $\lambda$, can indeed be obtained as band structures of microscopic models. 
For instance, if we take $\phi_\bk(\br)$ to be the wavefunction for the LLL on the torus, we can directly obtain the wavefunction  (\ref{eq:concentrate}) for arbitrary $K(\br)$ from the Hamiltonian of a Dirac particle in a non-uniform magnetic field $\ell^{-2} + \lambda \nabla^2 K(\br)$~\cite{dong2022diracelectronperiodicmagnetic}. Since the wavefunctions of the chiral model are also related to those of a Dirac particle in magnetic field~\cite{LedwithFCI,JieWang_exactlldescription}, Eq.~(\ref{eq:concentrate}) provides a wavefunction that shares all the features of the chiral model wavefunction but with more concentrated charge and Berry curvature, fixing its main quantitative inaccuracy.  

For $\lambda \gg 1$, the values of $\phi_\bk(\br)$ away from $\br = 0$ are unimportant. If $\phi_\bk(0) \neq 0$ for all $\bk$, then multiplication by $e^{-\lambda K(\br)}$ will simply sample the wavefunction at $\br = 0$ leading to fully trivial band. However,  there is an obstruction to doing this here since  $C_3$ symmetry enforces $\psi_{\bk = 0}(\br = 0) = 0$. Indeed, this is required by the topology of the bands in the Chern basis; the $C_3$ eigenvalues must sum to the Chern number, $\pm 1$, modulo three \cite{FangGilbertBernevig, SymmetryIndicators}, so that $\psi_\Gamma$ must have a zero at $\br = 0$ if $\psi_{K,K'}$ do not. We must therefore expand $\phi_\bk(\br)$ to higher order in $\br$ around $\bk = 0$.

To derive an explicit form for the wavefunction, we first make the assumption that the spinor wavefunction $\psi_\bk(\br)$ or $\phi_\bk(\br)$ has the form of $\bk$-dependent part times a $\bk$-independent layer spinor. This condition is always satisfied in the chiral model and turns out to be approximately satisfied to very good accuracy for the realistic BM model. This will be verified later when we compare the form factors obtained from our wavefunction to the BM model form factors. Next, we expand $\lambda K(\br - \bR) \approx \lambda K_0 + \frac{1}{4}\lambda \nabla^2 K|_{\br = 0} \abs{\br - \bR}^2$. We see that a new, small, length scale $\delta$ has emerged, such that $\lambda \nabla^2 K|_{\br = 0} = \delta^{-2}$, which represents the charge density peak width at the AA sites. Away from the $\Gamma$ point, the wavefunction reduces to
\begin{equation}
  \begin{aligned}
  \psi_{\bk}(\br) & = e^{i \bk \cdot \br} u_c(\br), \text{Away from $\bk \approx \Gamma$.} \\
  u_c(\br) & = \frac{1}{\sqrt{2\pi \delta^2}} \sum_{\bR} e^{- \frac{\abs{\br - \bR}^2}{4 \delta^2} } \chi_0, \quad 
\end{aligned}
\label{eq:awayfromGamma}
\end{equation}
Here $\chi_0$ is a normalized layer-space spinor. 

\begin{table}
\begin{tabular}{ |c| c |c| }
\hline
 Parameters & Meaning\\ 
\hline 
 $s$ & Berry curvature concentration around $\Gamma$ \\
\hline
 $\delta$ & Charge density concentration around AA \\
\hline
 $\beta=\delta/s$ & $O(1)$ number \\
\hline
\end{tabular}
\caption{Parameters related to the wavefunction Eq.~\ref{eq:fullWF}. Comparing to the the BM model at $\theta=1.06^\circ$, $w_1=110\si{meV}$, $w_0/w_1=0.75$, we find that $s=0.25, \delta=0.46, \beta=1.85$ provides a good match (see Fig. \ref{fig:comparebmconc}). All quantities are measured in units where $\ell = \sqrt{A_{\uc}/2\pi} = 1$; a moiré lattice vector length of $L_M = 13 \si{nm}$ corresponds to $\ell = 4.8 \si{nm}$.}
\label{tab:parameters}
\end{table}

Near the $\Gamma$ point, the derivation of \eqref{eq:awayfromGamma} is obstructed due to $\phi_{\bk = 0}(\br = 0) = 0$. To obtain the wavefunction $\psi$ in the vicinity of $\Gamma$, we thus expand $\phi$ linearly around $\bk = 0$ and $\br = 0$. The $C_3$ symmetry and $C=1$ topology impose $\psi_{\bk \approx 0}(\br \approx 0) = c_\bk(z + i \beta \ell^2 k) u_c(\br)$, for $z = x+iy$ and $k = k_x + i k_y$ and a real dimensionless constant $\beta$. Here we have included a factor of $\ell^2$ on dimensional grounds, where $2 \pi \ell^2 $ is the area of the unit cell, as well as a factor of $i$ so that, for real $\beta$, the wavefunction is symmetric under the antiunitary $C_{2y} \T$ symmetry. The $C=-1$ bands are related by complex conjugation and $\bk \to -\bk$ as usual. 
Note that a momentum space scale, which we will call $s:= \beta^{-1} \ell^{-2} \delta$, has emerged: comparing $z \sim i \beta \ell^2 k$ defines a region of order $s$ in momentum space where the second term is significant. Here we used that $z \sim \delta$, otherwise the Gaussian piece makes the wavefunction exponentially small. The appearance of this scale will have important consequences as we will see later.

We will choose $c_\bk^{-1} = i\beta \ell^2 k \sqrt{1 + 2s^2/\abs{ k}^2}$ so that, in the first unit cell and first BZ,
\begin{equation}
  \psi_\bk(\br) = \frac{1 + \frac{z/\delta}{i k/s}}{\sqrt{1 + \frac{2s^2}{ \abs{k}^2}}} u_c(\br)
  \label{eq:fullWF}
\end{equation}
is normalized and reduces to the $\bk$-independent~\ref{eq:awayfromGamma} when $\abs{k} \gg s$ is not too close to $\Gamma$. Values of the wavefunction in other unit cells can be obtained by translating the wavefunction above properly with $\psi_{\bk}(\br + \bR) = e^{i \bk \cdot \bR} \psi_\bk(\br)$, and $\psi_{\bk + \bG} = \psi_\bk$. For plots and numerical calculations we will prefer to replace $k$ by a periodic function, which behaves as $k$ for small momenta, to ensure the periodicity of the resulting expression at all $s$. Such a choice of periodic function depends on the UV completion and only affects results at the $O(s^2)$ level; we will make the choice $k \mapsto \frac{1}{\zeta(\bk)}$ where $\zeta(\bk)$ is the modified Weierstrass $\zeta$ function~\cite{haldane2018modular}. From now on we will set $\ell \to 1$, and focus on the real-space scale $\delta$ and momentum space scale $s$, each regarded as a dimensionless number. The parameters are summarized in Table \ref{tab:parameters}. We note that a similar wavefunction appeared in the supplement of Ref. \cite{song2022TBGTHF}. Wavefunctions satisfying ideal band geometry~\cite{JieWang_exactlldescription,LedwithVishwanathParker22}, such as those of the chiral model~\cite{Tarnopolsky2019origin,LedwithFCI,wang2020chiral}, have $\beta = 1$.

We have fixed the gauge in \eqref{eq:fullWF} so that the wavefunction is approximately $\bk$-independent away from $\Gamma$. This gauge choice\footnote{Our gauge choice happens to satisfy gauge fixing conditions recently proposed for Wannier functions of Chern bands \cite{gunawardanaOptimallyLocalizedSingleband2024,xieChernBandsOptimally2024} related to optimizing a regularized localization length.} will lead to nearly decoupled nonlocal Wannier states in Sec. \ref{subsec:Wannier}. Note that a singularity in the gauge is necessary due to the $C=1$ topology; we have placed the singularity at the $\Gamma$ point.

Since the wavefunction is nearly $\bk$-independent away from $\Gamma$, all the Berry curvature must be concentrated near the $\Gamma$ point. Explicitly, we obtain 
\begin{equation}
  \Omega(\bk) = \frac{4 s^2  }{(2s^2 +  \abs{k}^2)^2},
  \label{eq:BerryC}
\end{equation}
We see that the Berry curvature is concentrated at a momentum region whose size is $\sim s^2$. Interestingly, our model reproduces the concentration of Berry curvature although we did not introduce it by hand. It is simply a consequence of UV/IR mixing of the Chern bands which implies that the introduction of a short length scale $\delta$ should lead to the emergence of a long length scale $1/s$ with $ s =  \ell^{-2} \beta^{-1} \delta $. The physical meaning of this length scale can be understood by considering a system with finite linear size $L$ which leads to a momentum space grid with spacing $|\Delta \bk| \sim 1/L$. If the Berry curvature is concentrated in a region of linear size $s \ll \Delta k$, it will be undetectable, suggesting that topology in our band is only `visible' at length scales $\gtrsim 1/s$. We note that similar expressions for Berry curvature distribution was 
previously obtained in the projected limit of the Song-Bernevig model~\cite{herzogarbeitman2024topologicalheavyfermionprinciple}. 

A comparison of the charge density, Berry curvature distribution, and $C=1$ form factors $\lambda_\bq(\bk) = \langle u_\bk|u_{\bk + \bq} \rangle$ (whose explicit form is provided in App.~\ref{app:formfactors_hartreefock}) obtained using Eq.~\eqref{eq:fullWF} with the ones obtained in the BM model is shown in Fig.~\ref{fig:comparebmconc}. The BM model quantities are computed with $\theta = 1.06^\circ, w_1 = 110 \si{meV}, w_0 / w_1 = 0.75$ and the parameters $s = 0.25$ and $\beta = 1.85$ are chosen to match the charge distribution and the Berry curvature distribution for these BM parameters. Note that in Fig.~\ref{fig:comparebmconc}a,b we are fitting two entire functions using just two parameters. 

Furthermore, in panel (c) we compare our form factors to the BM form factors through
\begin{equation}
    \rm{FF-Distance}(\bk) = \frac{\sum_\bq |\lambda_\bq(\bk) - \lambda^{\rm BM}_\bq(\bk)|^2}{\sum_\bq |\lambda^{\rm BM}_\bq(\bk)|^2},
    \label{eq:ffdistance}
\end{equation}
where $\lambda^{\rm BM}_\bq(\bk)$ is the form factor for a $C=1$ BM band, which we obtain in the standard way through the sublattice basis \cite{BultinckHidden}. Note that \eqref{eq:ffdistance} is zero if and only if the form factors are exactly equal, and its computation requires choosing a compatible gauge for the BM model wavefunctions. We find that \eqref{eq:ffdistance} is $\lesssim 5\%$ across the BZ. We note that there is a separate $\sim 5\%$ error associated with $\U(4)\times \U(4)$ symmetry breaking terms in the BM form factors, which \eqref{eq:ffdistance} does not include; we argue that this error should be treated separately and, if one wishes, addressed through adding corresponding symmetry breaking terms to our model, as discussed in Refs.~\cite{BultinckHidden, ledwith2021strong, khalaf2020soft}.
The excellent quality of the fit for the charge density, Berry curvature distribution, and the full form factor implies that our ansatz \eqref{eq:fullWF} captures the physics of TBG and provides justification for the analytical treatment we will develop in the next sections.

\section{Interacting Hamiltonian, Exact $T=0$ ferromagnets}\label{sec:exactFM}

\begin{figure*}
    \centering
    \includegraphics[width=\textwidth]{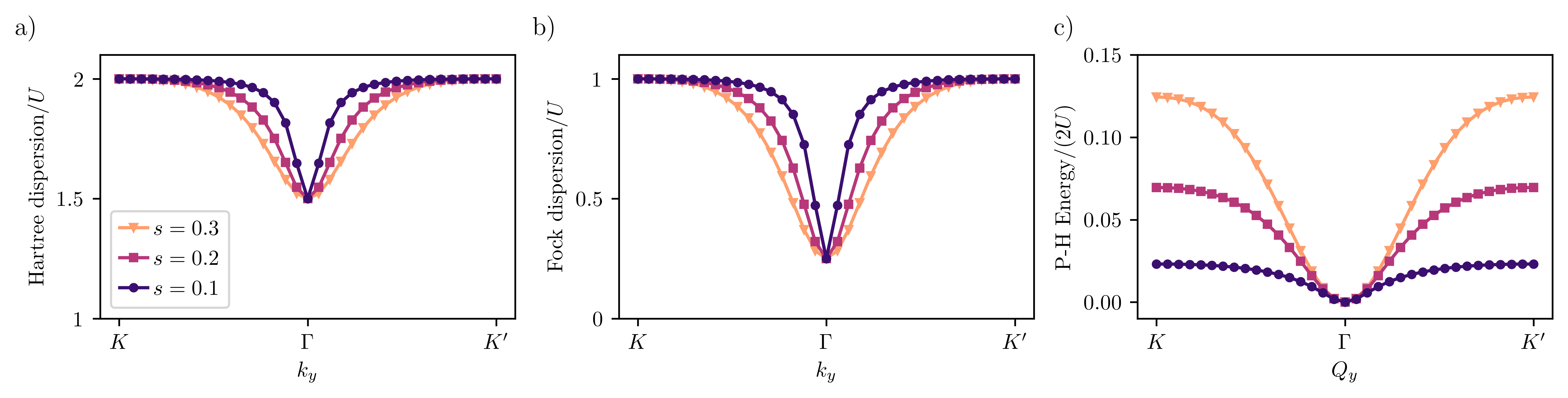}
    \caption{Panels \textbf{(a,b)}: Hartree dispersion $h_H(\bk)$ and Fock dispersion $h_F(\bk)$ for different values of the parameter $s$. Both Hartree and Fock dispersions have dips at the $\Gamma$ point of width $\sim s$. 
    Panel \textbf{(c)}: Energy of the Goldstone branch of particle-hole excitations. The overall bandwidth of the branch decreases with $s$, while the stiffness of the Goldstone mode itself remains $s$ independent due to the generation of a long length scale $s^{-1}$. }
    \label{fig:stiffness_HF}
\end{figure*}

In this section we will review the exact zero-temperature ferromagnetic eigenstates in the strong coupling limit, in the context of our model wavefunctions \eqref{eq:fullWF}. This section will also serve to set notation for subsequent sections, which will focus on $T>0$. The full Hamiltonian is 

\begin{equation}
    \H  = \sum_{\bk} c^\dag_\bk h_{\rm{BM}}(\bk) c_\bk + \frac{1}{2A}\sum_\bq V_\bq \delta \rho_\bq \delta \rho_{-\bq}
    \label{eq:Hamiltonian_forFMsection}
\end{equation}
where the bare BM dispersion $h_{\rm BM}(\bk)$, possibly renormalized by integrating out remote bands \cite{VafekRG}, is small sufficiently close to the magic angle and will be neglected in this section and the next. We will reintroduce it in the final section of the manuscript. $A = N A_{\uc}$ is the area of the sample in terms of the number of unit cells $N$ and the area of the unit cell $A_{\uc}$. We use the densities $\delta \rho$ which are measured relative to the charge neutrality density,
\begin{equation}
\begin{aligned}
    \delta \rho_\bq & = \rho_\bq - 4N \sum_\bG \delta_{\bq,\bG} \xi_\bG \\
    \rho_\bq & = \sum_\bk c^\dag_\bk \Lambda_\bq(\bk) c_{\bk + \bq}
    \qquad \xi_\bq=\frac{1}{N}\sum_\bk \Lambda_{\bq}(\bk),
\end{aligned}
\end{equation}
consistent with particle-hole symmetry. Here and throughout, we will use variables $\bk$, $\bk'$,$\dots$ to denotes BZ momenta, $\bq$, $\bq'$, $\dots$ to denote unrestricted momenta associated with the interaction $V_\bq$ and $\bG$, $\bG'$, $\dots$ to denote reciprocal lattice vectors. The form factor is an $8 \times 8$ matrix in the space of flat bands given by
\begin{equation}
    \Lambda_\bq(\bk) = \begin{pmatrix} \lambda_\bq(\bk) I_{4 \times 4} & 0 \\ 0 & \overline{\lambda_\bq(\bk)} I_{4 \times 4} \end{pmatrix}
    \label{eq:formfactorsblockmatrix}
\end{equation}
indicating four bands with $C=+1$ and four bands with $C=-1$. Here, we used $C_2 \T$ to relate the form factors in opposite Chern sectors, such that $\lambda_\bq^{-}(\bk) = \overline{\lambda^+_\bq(\bk)}$ where $\pm$ labels the Chern sector and $\lambda := \lambda^+$. $\xi_\bq$ is a scalar whose value for $\bq = \bG$ is equal to the Fourier components of the density of a fully filled band in either sector: time reversal relates the Chern sectors as
$\lambda^+_\bq(\bk) = \overline{\lambda^-_{-\bq}(-\bk)} = \lambda^-_{\bq}(-\bk-\bq)$, which implies $\sum_\bk \lambda_\bq^+(\bk) = \sum_{\bk} \lambda^-_\bq(\bk)$. $c^\dagger_{\bk}$ represents an eight-dimensional vector that is composed of electron creation operators corresponding to the four $C=+1$ and four $C=-1$ bands.
 
For $h_{\BM} = 0$, it is straightforward to reproduce the exact $T=0$ generalized ferromagnets at integer filling $\nu\in[-4,4]$ originally obtained in the strong coupling limit for the BM model \cite{BultinckHidden, TBGIVGroundState, KangStrongCoupling,KangVafekPRX,SeoStrongCoupling}. 
Consider a many-body state $\ket{\Psi}$ which fully fills some subset of the four Chern $1$ bands, and some subset of the four Chern $-1$ bands, described by a $\bk$-independent 
projector $P_{\alpha \beta} = \langle c^\dag_{\bk \beta} c_{\bk \alpha} \rangle$ that is diagonal in Chern sector. The state $\ket{\Psi}$ is then an exact eigenstate of $\rho_{\bq}$, and thus $\H$, because $\rho_\bq$ cannot scatter an electron within the same band due to Pauli blocking and there are no matrix elements between different bands due to the $\rm{U}(4) \times \rm{U}(4)$ symmetry of the form factors \eqref{eq:formfactorsblockmatrix}. Furthermore, single particle excitations on top of $\H$ are also exact eigenstates with energy bands $E_{\pm}(\bk) = \pm h_F(\bk) + \nu h_H(\bk)$, where $+$ corresponds to the conduction band and $-$ corresponds to the valence band~\cite{TBGVExcitations,KhalafBabySkyrmions}. Here $h_H(\bk)$ and $h_F(\bk)$ are the Hartree and Fock energies calculable from the form factors (see App \ref{app:formfactors_hartreefock}). 
\begin{equation}
\begin{aligned}
    h_H(\bk) & = \frac{1}{A_\uc} \sum_\bG V_\bG\lambda_\bG(\bk) \xi_\bG \\
    & = 2U \frac{\abs{\bk}^2}{\abs{\bk}^2 + 2 s^2} + 2(U-U_\Gamma) \frac{2s^2}{\abs{\bk}^2 + 2 s^2}, \\
    h_F(\bk) & = \frac{1}{2A}\sum_{\bq} V_\bq \abs{\lambda_\bq(\bk)}^2 \\
    & = U \frac{\abs{\bk}^2}{\abs{\bk}^2 + 2 s^2} + U_\Gamma \frac{2s^2}{\abs{\bk}^2 + 2 s^2}, 
\end{aligned}
\label{eq:hartreeFock_maintext}
\end{equation}
where
\begin{equation}
  \begin{aligned}
  U & = \frac{1}{2} \int \frac{d^2 \bq}{(2\pi)^2} V_\bq e^{-\abs{\bq}^2 \delta^2} \\
  & = \frac{1}{2} \int d^2 \br d^2 \br' V(\br - \br') \frac{e^{-\frac{\abs{\br}^2}{2 \delta^2}}}{2\pi \delta^2} \frac{e^{-\frac{\abs{\br'}^2}{2 \delta^2}}}{2 \pi \delta^2}
\end{aligned}
  \label{eq:hubbardU}
\end{equation}
is the on-site potential associated with two exponentially localized Gaussian orbitals with width $\delta$, and
\begin{equation}
  \begin{aligned}
    U_\Gamma & = \frac{1}{2} \int \frac{d^2 \bq}{(2\pi)^2} V_\bq \frac{1}{2}\abs{\bq}^2 \delta^2 e^{-\abs{\bq}^2 \delta^2},
\end{aligned}
  \label{eq:hubbardUGamma}
\end{equation}
is an interaction scale of order $U$ associated with scattering between the $\Gamma$ point electrons and the electrons away from $\Gamma$.
For the Coulomb potential $V_\bq = \frac{2\pi e^2}{\varepsilon \abs{\bq}}$ we have
\begin{equation}
  U = \frac{\sqrt{\pi}}{4} \frac{e^2}{\varepsilon \delta}, \qquad U_\Gamma = \frac{\sqrt{\pi}}{16} \frac{e^2}{\varepsilon \delta} = \frac{U}{4}.
  \label{eq:coulombscales_U_Ugam}
\end{equation}

From \eqref{eq:hartreeFock_maintext}, we see explicitly that charged excitations in this model have a characteristic energy scale of $U \sim e^2/\varepsilon \delta$. The Hartree and Fock dispersions are plotted in Fig.~\ref{fig:stiffness_HF}a,b, reproducing analytically the dip at the $\Gamma$ point~\cite{KangStrongCoupling, guineaElectrostaticEffectsBand2018, BultinckHidden, TBGVExcitations, KhalafBabySkyrmions}. Note that in our model, the minimum at $\Gamma$ is controlled by the scale $U_\Gamma$ which is of the same order as the Hubbard scale, yielding $\sim U$ gaps for the correlated insulators. Similar expressions have been obtained in the projected limit of the Song-Bernevig model~\cite{herzogarbeitman2024topologicalheavyfermionprinciple}.

In anticipation of the discussion of scales for spin and charge ordering in the next sections, we will now comment on the energy scale associated with particle-hole excitations on top of the generalized quantum Hall ferromagnets. Practically, the spectrum of these excitations can be solved for exactly by diagonalizing a particle-hole Hamiltonian $H_\bQ$ acting on the space of excitations $\sum_\bk  c^\dag_\bk \phi_\bQ(\bk) c_{\bk + \bQ}$ for each total-momentum $\bQ \in \rm{BZ}$~\cite{khalaf2020soft, TBGVExcitations}. 
Due to the $\U(4) \times \U(4)$ symmetry, excitations are split into two types: those in which the particle and hole are in the same Chern sector (intra-Chern) and those in which the particle and hole are in opposite Chern sectors (inter-Chern). Both spectra are split into some low-energy and high energy branches with the latter corresponding to far-separated particle hole pairs with energy of the order $2U$. The low energy branch of the intra-Chern modes corresponds to Goldstone modes of the broken $\U(4) \times \U(4)$, which generalize spin waves of conventional ferromagnets \cite{khalaf2020soft}. At small $\bQ$ this particle-hole band has the ferromagnetic Goldstone mode dispersion $E = \frac{1}{\pi}\rho_s |\ell \bQ|^2$, where we have reinstated $\ell$ temporarily to clarify that $\rho_s$ has the dimensions of energy. We find that the stiffness $\rho_s \sim U$, such that low $\bQ$ spin-wave excitations also seem to have an energy scale $\sim U$, see Fig. \ref{fig:stiffness_HF}c. In particular, $\rho_s$ is independent of the concentration scale (other than through the implicit dependence $U = U(\delta))$. Thus, the low lying excitations on top of the ferromagnet appear to be characterized by a single energy scale $U$.

However, in Fig. \ref{fig:stiffness_HF}c we see that the Goldstone energy begins to flatten at a small momentum scale $|\bQ| \sim s$, so that the top of the Goldstone energy band has energy decreasing with $s$. We will see in upcoming section that this is representative of a ferromagnetic exchange energy scale $J \sim U s^2 \ll U$ albeit over a long characteristic length scale $s^{-1}$. This state of affairs will lead to the conclusion that moments are \emph{mostly} decoupled, though at very low temperatures the relevant spin-waves, that exist in a small area of phase space $|\bQ| \sim s$, are described by a large stiffness $U \gg Us^2$. The parametric separation of the stiffness and exchange scales stands in stark contrast to lattice models with short-range interactions, e.g. Heisenberg model, where both are controlled by the same scale $J$.

\section{Nearly Decoupled nonlocal Moments}\label{sec:decoupledEntropy}

\begin{figure*}
    \centering
    \includegraphics{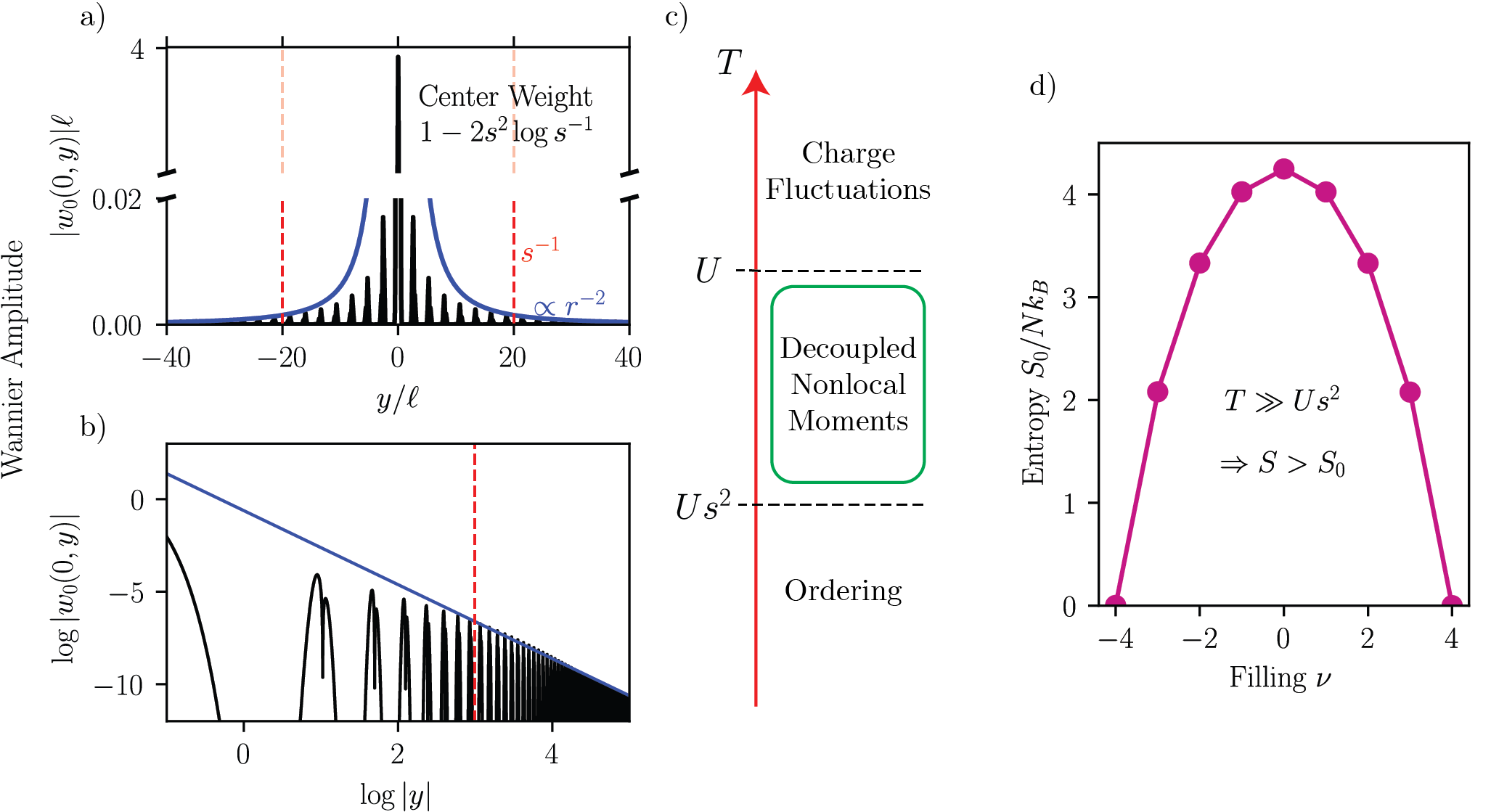}
    \caption{The Chern band Wannier states and their decoupled nonlocal moments. In panels \textbf{(a)} and \textbf{(b)} we plot the Wannier state centered at the origin in linear and log scale respectively, where each peak is an AA site. We use $s=0.05$ to better depict the multi-scale nature of the Wannier state. The central peak has almost all the weight of the state (note the $y$-axis discontinuity in panel \textbf{(a)}). Correspondingly, the long-distace tail, which cannot be suppressed at asymptotic distances, has its onset pushed to a long length scale $s^{-1}$. In panel \textbf{(c)}, we show the hierarchy of scales associated with the Wannier states. The on-site scale $U$ originates from the large repulsion associated with having two central peaks at the same site, whereas exchange is suppressed by $s^2$, since there is only $\sim s^2$ weight outside the central peak. Panel \textbf{(d)} plots the zero strain entropy of the decoupled moments, $S_0$. For $U \gg T \gg Us^2$, charge fluctuations are mostly frozen and the moments are mostly decoupled, up to order $s^2$ corrections, such that the total entropy $S \approx S_0$. This is consistent with the rigorous lower bound that we proved at integer fillings (dotted entropy values), $S \geq S_0 - N J_*/T$, where $J_* \sim Us^2$ at small $s$.}
    \label{fig:wannier_entropy_nostrain}
\end{figure*}

In the previous section, we have described the ferromagnetic ground state at charge neutrality and nearby integers. However, the treatment in the previous section is agnostic regarding nearby states, activated by $T>0$, consisting of fluctuating nonlocal moments. In this section we will describe such states and show that they have ground state energies $\sim Us^2$ above the ground state ferromagnets. This leads to an intermediate temperature regime $U \gg T \gg Us^2$ where the entropy is expected to be that of free decoupled moments. We will prove a rigorous lower bound on the entropy that demonstrates this fact, including when the ground state is not a ferromagnet due to strain, sufficiently large integer filling $\nu$, or both.

\subsection{Wannier functions}
\label{subsec:Wannier}

To describe the fluctuating moment state, we need to construct a basis of orthogonal orbitals associated with different lattice sites\footnote{An exponentially localized but overcomplete and non-orthogonal basis of ``coherent states" \cite{Li2024constraints,huang2020quantummagnetismwannierobstructedmott} can also be used to study direct exchange in Chern bands \cite{huang2020quantummagnetismwannierobstructedmott}.}. In a trivial band, this can be chosen to be a basis of exponentially localized Wannier orbitals, and the resulting moments will be local. Here, due to band topology, Wannier functions have a power law tail, but we can still use them to define decoupled nonlocal moments as follows. Consider the Wannier states defined via
\begin{equation}
    w_\bR^\gamma(\br) = \frac{1}{\sqrt{N}} \sum_\bk \psi_\bk^\gamma(\br) e^{-i \bk \cdot \bR},
\end{equation}
where $\gamma = \pm$ labels the Chern sector.
Due to band topology , the Wannier functions will be power law delocalized with infinite localization length. Their asymptotics depend only on the gauge singularity, whose existence is dictated by topology, and the normalized Bloch state at the singularity as shown in Ref.~\cite{Li2024constraints}:
\begin{equation}
\begin{aligned}
    w_{\bR = 0}^\gamma(\br \to \infty)  & = \frac{1}{\sqrt{N}}\sum_\bk e^{i\gamma \theta_\bk} e^{i \bk \cdot \br} \psi^\gamma_{\Gamma}(\br) \\
    & = - \frac{e^{-i \gamma \theta_\br}}{\abs{\br}^2} \sqrt{N}\psi^\gamma_\Gamma(\br)
\end{aligned}
    \label{eq:universaltail}
\end{equation}
Importantly, this power law tail cannot be made small. However, this universal power-law tail does not preclude parametrically small overlap between different Wannier states since the latter also depends on the length scale at which the tail set in. In fact, we find that in the gauge \eqref{eq:fullWF}, this overlap is parametrically small and leads to an exchange energy $Us^2$, as shown below.\footnote{We note that Bloch gauge transformations generally mix classical Wannier density-density and exchange terms; indeed, we choose the gauge \eqref{eq:fullWF} to minimize the exchange terms relative to the density-density terms.}

To understand how nonlocal Wannier states can nonetheless be nearly non-overlapping, we note that for most $\bk$ in the BZ, other than an order $s$ region surrounding the $\Gamma$ point, the wavefunctions \eqref{eq:fullWF} are approximately $\bk$-independent. As a result, almost all the weight of the Wannier state is concentrated in its central Gaussian peak, a fact which was noticed numerically in the BM model~\cite{zang2022realspacerepresentationtopological}. Indeed, we may directly calculate the weight of the Wannier function for $\br$ in the first unit cell (UC)
\begin{equation}
\begin{aligned}
    w_0(\br \in \uc) & \approx (1 - 2s^2 \log s^{-1}) \frac{e^{-\frac{\abs{\br}^2}{4s^2}}}{\sqrt{2\pi \delta^2}}\chi_0,
    \end{aligned}
\end{equation}
such that there is only $O(s^2)$ weight left for the peaks centered at other unit cells. For a detailed derivation of the asymptotic form of the Wannier states, see App.~\ref{app:wannier}.

At first sight the form of the Wannier state is surprising, because \eqref{eq:universaltail} appears to have order one weight when integrated. However, such  
tail only sets in when $\bR \gg 1/s$ is sufficiently large that $e^{i \bk \cdot \bR}$ probes momenta in the vicinity of the $\Gamma$-point, $\bk \ll s$. Thus, as $s$ is decreased towards zero, the onset of the tail \eqref{eq:universaltail} gets pushed to longer length scales, reducing the weight in the tail and concentrating it in the central peak. The Wannier states are plotted in Fig. 4(a)-(b) for a small $s=0.05$ to emphasize the aspects described above. Schematically speaking, $s^{-1}$ is the length scale at which the nontrivial band topology can be detected in the system. The emergence a long length scale $s^{-1}$ from a short length scale $\delta\sim s$ can be understood a consequence of UV/IR mixing, a generic feature topological bands due to the non-commutative geometry necessary for a nonzero Chern number, $C = 2\pi i \Tr [X,Y]$. Here $\Tr$ is a suitably regularized trace per unit volume and $X,Y$ are projected position operators \cite{kitaevAnyonsExactlySolved2006,prodanDisorderedTopologicalInsulators2011,bellissardNonCommutativeGeometryQuantum1994}.

\subsection{Rigorous bound on the entropy}
\label{subsec:entropy}

We will now derive a rigorous bound on the entropy which, when combined with the properties of the Wannier functions constructed and the hierarchy of energy scales in the model, implies the existence of temperature window $U s^2 \ll T \ll U$ where the entropy is $O(1)k_B$ per unit cell, reflecting the presence of decoupled moments. Our approach to establish the bound is to split the Hamiltonian for the density-density interaction in two pieces, $\H = \H_{\rm cl} + \H'$. The first part corresponds to the classical interaction energy associated with the local density operators constructed from the power-law localized Wannier functions whereas the second part contains the non-classical part of the interaction. We identify an exponentially degenerate manifold of ground states for $\H_{\rm cl}$. This space is spanned by $e^{S_0}$ states at an energy $E_{\rm cl}$ equal to that of the exact ferromagnetic eigenstate of $\H$, where $S_0 = N \log \begin{pmatrix} 8 \\ \nu + 4 \end{pmatrix}$ is the entropy of decopuled moments.

We show that $\H'$ (i) lifts the average energy of the $\H_{\rm cl}$ ground states by an amount $N J_{\rm ex}$ and (ii) cannot lower the ground state energy below $E_{\rm cl}$ by more than $N J_{\rm h}$; both terms are of order $ Us^2$ and are computed explicitly below. Points (i) and (ii) together imply that the total average energy $\langle \H \rangle$ in the degenerate set of trial states differs from the ground state energy by at most $N J_*$, where $J_* = J_{\rm ex} + J_{\rm h}$.
This fact is sufficient to rigorously bound the entropy by $S/N \geq S_0/N - \frac{J_*}{T}$.  The bound holds without any assumptions on the hierarchy of scales but is only non-trivial in the small $s$ limit where we can show that $J_* \sim U s^2$. The $\sim s^2$ dependence comes from the fact that $\H'$ acts nontrivially (and non-perturbatively) only on an $\sim s^2$ area of phase space. As a result, for $U s^2 \ll T \ll U$, this bound implies an entropy at least of order 1 per site confirming the existence of decoupled moments despite the power law tails of the associated Wannier functions. The hierarchy of scales and entropy bound are depicted in Fig. 4(c)-(d).

We begin by expanding the projected density operator $\rho_\bq$ in terms of the operators $c_{\bR,\alpha}^\dagger c_{\bR',\alpha}$ where $c_{\bR,\alpha}$ is the annihilation operator for the power-law localized Wannier function at site $\bR$ for band $\alpha$. Splitting the density in the parts with $\bR = \bR'$ and $\bR \neq \bR'$, we note that the first part yields a classical Hamiltonian $\H_{\rm cl}$ written purely in terms of the Wannier density operators $n_\bR = \sum_\alpha c^\dag_{\bR \alpha} c_{\bR \alpha}$:
\begin{equation}
  \begin{aligned}
 \H_{\rm{cl}} & =   \sum_{\bR, \bR'} U(\bR-\bR') (n_\bR - 4)(n_{\bR'} - 4), \\
  & U(\bR) = \frac{1}{2A} \sum_\bq V_\bq \abs{\xi_\bq}^2 e^{i \bq \cdot(\bR)}.
\end{aligned}
  \label{eq:classicalmodel}
\end{equation}
Since the Wannier orbitals are orthogonal, $[n_\bR, n_{\bR'}] = 0$ for $\bR \neq \bR'$, we can consider many-body states which are simultaneous eigenstates of all $n_\bR$. These are states with definite occupation number at each lattice site. This allows us to divide the many-body Hilbert space into sectors described by the occupations $\{n_\bR\}$. At any integer filling $\nu$, $\H_{\rm cl}$ is minimized for states with $n_\bR = 4 + \nu$ at every $\bR$. There are $\binom{8}{4+\nu}^N$ such states whose classical energy is $E_{\rm cl} = \sum_{\bR,\bR'} U(\bR-\bR')\nu^2$. Deviations from $n_\bR = 4 + \nu$ typically cost an energy of order $U$ and play the role of double occupancies in the Hubbard model. 

We begin with step (i). Projecting the remaining part of the Hamiltonian $\H' = \H - \H_{\rm cl}$ onto the ground state manifold of $\H_{\rm cl}$ yields an effective exchange Hamiltonian given by  
\begin{equation}
\begin{aligned}
    \H_{\rm ex} & = \frac{1}{2A} \sum_\bq V_\bq \sum_{\bR \neq \bR'} c^\dag_\bR \Lambda_\bq^{\bR \bR'} c_{\bR'} c^\dag_{\bR'}  \Lambda_{-\bq}^{\bR' \bR} c_{\bR} \\
    \Lambda_\bq^{\bR \bR'} & = \langle w_\bR| e^{-i \bq \cdot \br}| w_{\bR'} \rangle  =  \frac{e^{-i \bq \cdot \bR'}}{N} \sum_\bk e^{i \bk \cdot (\bR - \bR')} \Lambda_\bq(\bk) 
\end{aligned}
\end{equation}
Note that while $\H'$ generally contains various correlated hopping terms, these vanish upon projected to a space with fixed occupation $\{n_\bR \}$ leading to $\H_{\rm ex}$. The exchange Hamiltonian $\H_{\rm ex}$ splits the degeneracy of the exponentially many states which minimize $\H_{\rm cl}$ by favoring ferromagnets, which are the only states with vanishing expectation value under $\H_{\rm ex}$. Let $\rho_0$ be the equal-weight density matrix of ground states of $\H_{\rm cl}$. Then in App.~\ref{app:entropy} we compute
\begin{equation}
\begin{aligned}
    J_{\rm ex} & = \Tr \H' \rho_0 = \Tr \H_{\rm ex} \rho_0 \\
    & = (4-\frac{\nu^2}{4})\frac{1}{4NA} \sum_{\bk,\bq} V_\bq |\lambda_\bq(\bk) - \xi_\bq|^2,\\
    & \approx 2U_\Gamma s^2 (4-\frac{\nu^2}{4}) \log s^{-1}
\end{aligned}
    \label{eq:JexMainText}
\end{equation}
where we recall $\xi_\bq = \frac{1}{N} \sum_\bk \lambda_\bq(\bk)$ is the average form factor and $\lambda_\bq(\bk)$ is the form factor for one of the $C=1$ bands. The first two lines are exact whereas in the third line we used $s \ll 1$. 

We now briefly describe step (ii). For integer filling away from charge neutrality the exact ferromagnetic eigenstate may not be the ground state; in this case it is not sufficient to evaluate $\H'$ in the ground state subspace of $\H_{\rm cl}$. In App.~\ref{app:entropy} we prove
\begin{equation}
\begin{aligned}
    E_{\rm GS} & \geq E_{\rm cl} - J_H \\
    J_H & = 4\abs{\nu} \frac{1}{N}\sum_\bk |h_H(\bk) - h_H^{\rm avg}|, \\
    & \approx 16 \abs{\nu} U_\Gamma s^2 \log s^{-1}
\end{aligned}
\label{eq:JHmaintext}
\end{equation}
where $h_H(\bk)$ is the Hartree dispersion \eqref{eq:hartreeFock_maintext}.

We are now ready to discuss the entropy bound. We will lower bound the entropy through the Bogoliubov inequality $F[\rho_{\rm trial}] \geq F[\rho]$, where $\rho_{\rm trial}$ is any trial density matrix, $\rho \propto e^{-\beta \H}$ is the exact thermal density matrix, and $F[\rho] = \Tr \H \rho - T S[\rho]$ is the free energy functional. We will take $\rho_{\rm trial} = \rho_0$, the density matrix we used in step (i). This immediately translates to a \emph{lower} bound on the exact thermal entropy,
\begin{equation}
    S[\rho] \geq S_0 - \frac{\Tr \H \rho_0  - \Tr \H \rho}{T}.
\end{equation}
where $S_0 = N \log \begin{pmatrix} 8 \\ 4+\nu \end{pmatrix}$. Step (i) implies $\Tr \H \rho_0 = E_{\rm cl} + N J_{\rm ex}$, while step (ii) implies $\Tr \H \rho \geq E_{\rm GS} \geq E_{\rm cl} - J_H$. We therefore obtain 
\begin{equation}
\label{eq:entropybound}
    \frac{S[\rho]}{N} \geq \log \begin{pmatrix} 8 \\ 4+\nu \end{pmatrix} - \frac{J_*}{T}, \quad J_* = J_{\rm ex} + J_{\rm H}.
\end{equation}

We conclude that the bound \eqref{eq:entropybound} holds with $J_* = O(Us^2)$. We emphasize that no approximations were needed for this bound, with the sole exception of the concrete evaluation of the momentum-sums. In particular, while the bound is most useful and directly applicable for $T \gg Us^2$, in which it implies $S/N \gtrsim \log \begin{pmatrix} 8 \\ 4 +\nu \end{pmatrix}$, we did not use such a hierarchy to prove \eqref{eq:entropybound}. In fact, \eqref{eq:entropybound} implies a meaningful lower bound even when $T \sim Us^2$, and only becomes useless when $T \ll Us^2$. While the result above is only a bound, we can also argue that as long as $U\gg T\gg Us^2$, $S \approx S_0$ up to order $s^2$ corrections. We now briefly discuss this argument, which will be made more precise in the subsequent section.

Stepping back, an upshot of the above calculation is that even though the bands of twisted bilayer graphene are topological, with nonlocal Wannier states, the thermal entropy is similar to that of the trivial classical model \eqref{eq:classicalmodel}, which arises from formally setting $s=0$ and ignoring the $\Gamma$ point, provided $T \gg Us^2$. While for many quantities, such as Hall conductance and transport more generally, the $\Gamma$ point region is crucially important, for bulk thermodynamic quantities at not-too-small temperatures, small regions of phase space play a correspondingly small role. This intuition will be made more precise in the final technical section of the paper, where we will argue both diagrammatically and non-perturbatively that 1PI correlation functions with external legs away from $\Gamma$ mimic those of the classical model \eqref{eq:classicalmodel}, provided $T \gg Us^2$. Bulk thermodynamic quantities such as free energy, internal energy, entropy, and (generalized) magnetization, can all be broken down into such correlation functions to leading order in $s^2$, while transport quantities and the spectral function $A(\bk \approx \Gamma, \omega)$ are dominated by correlation functions with external momenta near the $\Gamma$ point and differ nonperturbatively from the classical model. The latter quantities must be computed carefully. However, we will show appropriately that the spectral function is analytically tractable through integrating out the momenta away from $\Gamma$, which serve as a classical bath.

Away from integer fillings, the classical Hamiltonian is minimized by states where $N (\lceil \nu \rceil - \nu)$ sites has filling $4 + \lfloor \nu \rfloor$ and $N (\nu - \lfloor \nu \rfloor)$ sites has filling $4 + \lceil \nu \rceil$. We therefore conclude that the entropy at non-integer filling will be equal to that of the classical model for $U \gg T \gg Us^2$. We note that in the Hubbard approximation $U(\bR-\bR') = U \delta_{\bR \bR'}$ the classical model \eqref{eq:classicalmodel} has a significant configurational charge entropy at non-integer $\nu$: here however we will assume that further-neighbor interactions (suppressed by a single power of $\delta$) are strong enough to quench the additional configurational entropy in practice. The classical entropy, for $T\ll U\delta$, is then equal to the sum of the flavor entropies at each site, such that $S_0(\nu)$ linearly interpolates between the nearest integers,
\begin{equation}
    S \approx S_0 = (\lceil \nu \rceil - \nu)S_0(\lfloor \nu \rfloor) + (\nu - \lfloor \nu \rfloor) S_0(\lceil \nu \rceil),
    \label{eq:entropyawayfromintegers}
\end{equation}
where we recall $S_0(\nu \in \mathbb{Z}) = N \log \begin{pmatrix} 8 \\ 4+\nu \end{pmatrix}$.

Likewise, again for $T \gg Us^2$, and for an applied (generalized) Zeeman field coupling to the $\SU(8)$ moments, we obtain generalized Curie laws for the magnetization. For example, consider an in-plane magnetic field in twisted trilayer graphene (for twisted bilayer graphene, there is an orbital coupling of the same order that should also be included \cite{ledwith2021tbtbcontrastingproperties, MacdonaldInPlane, LakeSenthilPairing, LakeSenthilTTG}; we neglect it here for simplicity) leads to the Zeeman coupling $-\mu_B B_{||} S_x$, where $S_x = \sum_\bR c^\dag_\bR s_x c_\bR$ is written in terms of the spin Pauli matrix $s_x = \pm 1$. The spin operator receives only an $O(s^2)$ contribution from the $\Gamma$ point electrons, such that we can compute its expectation value from the trivial classical model \eqref{eq:classicalmodel}. At an integer filling $\nu$, we can specialize to an single site partition function
\begin{equation}
    Z = \sum_{l=0}^{4+\nu} \binom{4}{l} \binom{4}{4+\nu - l} e^{ (2l-m) \beta \mu_B B_{||}}
    \label{eq:magnetizationcurie_partition}
\end{equation}
where we used that there are four states with each spin and $4+\nu$ electrons on the site; $l$ labels the number of $+x$ spins occupied. The full magnetization then follows from $\langle S_x \rangle = N \frac{\partial}{\partial (\beta \mu_B B_{||})} \log Z$, generalizing the usual Curie law $\tanh \frac{\mu_B B}{T}$ to our eight component on-site space. 

Above we have argued $S\approx S_0$ within the hierarchy $U \gg T \gg Us^2$. Let us briefly comment on corrections to this result. First, we expect negative corrections to $S$ of the order $Us^2/T$ due to exchange interactions. While we have not computed these explicitly, our rigorous entropy bound functions to upper bound their magnitude at integer filling. We can also have positive corrections to $S$ due to charged excitations. While in typical Mott insulators charge fluctuations are exponentially suppressed at $T \ll U$, we will find in Sec. \ref{sec:Mott} that the charge neutrality state is a Mott semimetal, with low energy charge excitations in an $\sim s^2$ region of the $\Gamma$ point. We therefore expect a power law positive correction to $S$, of the order $T^2 s^2/U^2$.

\subsection{Entropy with Strain}
\label{subsec:entropy_strain}

In this section we lower bound the entropy in the presence of strain, specifically uniaxial heterostrain which is the most important form of strain in TBG \cite{huderElectronicSpectrumTwisted2018, bi_designing_2019}. This rigorous lower bound is most relevant for $E_{\rm str} \gg T$, so that only the spin and valley moments present in the lower bound remain. As in the previous section, we will also report on a parametrically controlled entropy calculation within the hierarchy of scales $U \gg T \gg Us^2$; here the strain energy scale can be of the same order as $T$, and we will find close agreement with the curves in \cite{IlaniEntropy}. Finally, we will briefly discuss the ground state at charge neutrality in the presence of strain (the nematic semimetal), which is favored at parametrically small $E_{\rm str} \gtrsim Us^2$.

In App.~\ref{app:singleparticle_strain} we compute the projection of the strain part of the BM Hamiltonian on the wavefunctions \eqref{eq:fullWF}. We obtain a dispersion of the form
\begin{equation}
\begin{aligned}
    H_{\rm{str}} & = \sum_{\bk} h_{\rm str}(\bk) c^\dag_\bk \gamma_x c_\bk, \,\,\,\,
    h_{\rm str}(\bk) = E_{\rm str}\frac{\abs{\bk}^2}{\abs{\bk}^2 + 2s^2},
\end{aligned}
\end{equation}
where $E_{\rm str}$ is the energy scale associated with the projected strain dispersion.

We use a similar technique as in the strain-free case to lower bound the entropy. We split the Hamiltonian $\H = \H_{\rm cl} + \H'$, only now 
\begin{equation}
\begin{aligned}
    \H_{\rm cl} & =  \H_{\rm cl}|_{E_{\rm{str}} = 0} -  h_{\rm str}^{\rm av}\sum_\bR c^\dag_\bR \gamma_x c_\bR\\
    \H' & =  \H'|_{E_{\rm{str}} = 0} - \sum_{\bk} (h_{\rm str}(\bk) - h_{\rm str}^{\rm av}) c^\dag_\bk \gamma_x c_\bk.
\end{aligned}
\end{equation}
We see that the average value of the strain dispersion, $h_{\rm str}^{\rm av} = \frac{1}{N} \sum_\bk h_{\rm str}(\bk)$, acts as a classical $\U(8)$ Zeeman field along the $\gamma_x$ ``inter-Chern" direction. This leads to a single classical ground state at $\nu = 0$, defined by $c^\dag_\bR \gamma_x c_\bR = 4$ for all $\bR$. For $\nu > 0$ ($\nu < 0$), the four states per site with $\gamma_x = +1$ $(-1)$ are fractionally filled, leading again to exponentially many classical ground states. As before we define $\rho_0$ to be the equal-weight density matrix of $\H_{\rm cl}$ ground states. The entropy of $\rho_0$ is $N\log \binom{4}{\abs{\nu}}$, equal to the classical thermal entropy $S_0$ at temperatures $T \ll E_{\rm str}, U$.

We now carry out steps (i) and (ii) as before, beginning with the evaluation $\Tr \H' \rho_0$. The strain dependent term $\sum_{\bk} (h_{\rm str}(\bk) - h^{\rm av}_{\rm str}) \langle c^\dag_\bk \gamma_x c_\bk\rangle_{\rho_0}$ vanishes. Indeed, the $\bR$-independence of $\langle c^\dag_\bR \gamma_x c_\bR \rangle_{\rho_0}$ implies the $\bk$-independence of $\langle c^\dag_\bk \gamma_x c_\bk\rangle_{\rho_0}$. It suffices to evaluate the expectation value of $\H_{\rm ex}$ in $\rho_0$, which is carried out in App.~\ref{app:entropy}. The result is
\begin{equation}
\begin{aligned}
    J_{\rm ex} & = \frac{1}{N}\Tr \H' \rho_0 = \frac{1}{N}\Tr \H_{\rm ex} \rho_0 \\
    & = 4 (J_s-\tilde{J}_s) + 2 \abs{\nu}\tilde{J}_s - \frac{\nu^2}{4} (J_s+\tilde{J}_s)
    \end{aligned}
    \label{eq:Jex_strained_maintext}
\end{equation}
where
\begin{equation}
\begin{aligned}
    J_s & = \frac{1}{4NA} \sum_{\bk,\bq} V_\bq \abs{\lambda_\bq(\bk) - \xi_\bq}^2 \\
    & \approx 2 U_\Gamma s^2 \log s^{-1} \\
    \tilde{J}_s & = \frac{1}{4NA} \sum_{\bk,\bq} V_\bq \Re (\lambda_\bq(\bk) - \xi_\bq)^2 \\
     & \approx (\log(4) - 1) U s^2 
\end{aligned}
\label{eq:JandJtilde_maintext}
\end{equation}
are two exchange scales. The scale $J$ characterizes exchange between intra-Chern moments, while the scale $\tilde{J} < J$ couples inter-chern moments. These scales differ due to the $\U(8)$ anisotropy in \eqref{eq:formfactorsblockmatrix}; the form factors in the two Chern sectors are necessarily distinct $\lambda \neq \overline{\lambda}$. Step (ii) is modified by the second term in $\H'$, which we must bound. We obtain (App \ref{app:entropy}):
\begin{equation}
\begin{aligned}
    E_{\rm GS} & \geq E_{\rm cl} - NJ_H - NJ_{\rm str} \\
    J_{\rm str} & = \frac{4}{N}  \sum_\bk |h_{\rm str}(\bk) - h_{\rm str}^{\rm av}|\\
    & \approx 8 E_{\rm str} s^2 \log s^{-1}
\end{aligned} 
\label{eq:Jstr_Egslowerbound_main}
\end{equation}

We therefore have the bound
\begin{equation}
\begin{aligned}
    \frac{S[\rho]}{N} \geq \log \begin{pmatrix} 4 \\ \abs{\nu} \end{pmatrix} - \frac{J_*}{T}
\end{aligned}
\label{eq:strainbound}
\end{equation}
where
\begin{equation}
    J_* = J_{\rm ex}^{\rm str} + J_{\rm{H}} + J_{\rm str} \sim Us^2, E_{\rm str} s^2.
\end{equation}
We emphasize that the ground state could be a nematic semimetal, intervalley kekule spiral, or even a metal with a Fermi surface. The bound \eqref{eq:strainbound} still proves that the entropy will be bounded by that of decoupled moments as long as $T \gg Us^2, E_{\rm str} s^2$.

\begin{figure}
    \centering
    \includegraphics{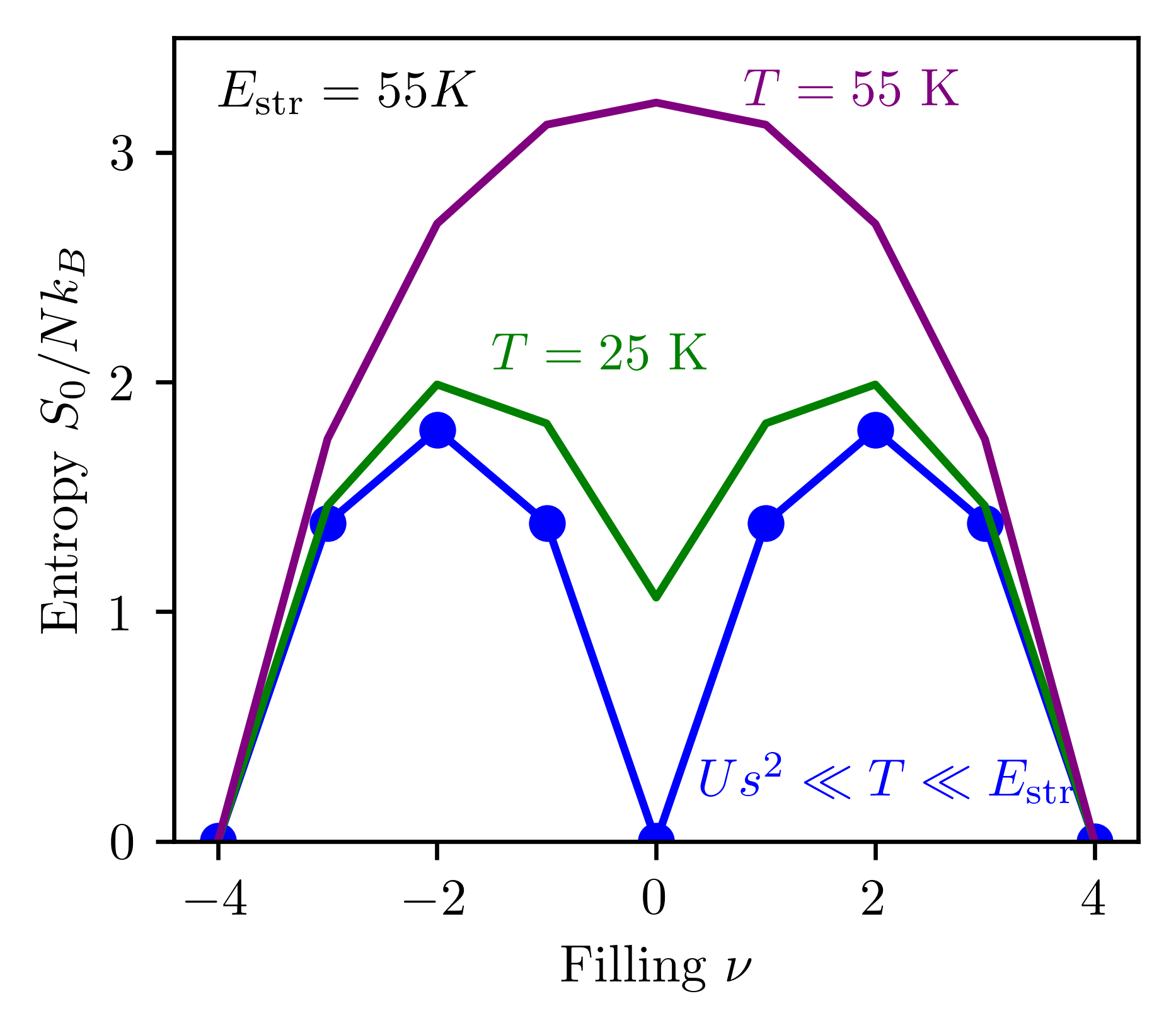}
    \caption{Entropy in the presence of strain at various temparatures. We have $S \approx S_0$ in the hierarchy of scales $U \gg T \gg Us^2$ when $T\gtrsim E_{\rm str}$. Below $E_{\rm str}$, the strain favors a valley-spin singlet nematic semimetal state at charge neutrality, creating a dip in the entropy.  We choose a strain scale $E_{\rm str} = 55 K$ and plot the entropy at the temperatures corresponding to Ref. \cite{IlaniEntropy}. We find that the curves are very closely matched. The dots correspond to the rigorous bound at integer fillings \eqref{eq:strainbound}.}
    \label{fig:strain_entropy}
\end{figure}

As discussed at the end of the previous subsection, we can also use the hierarchy of scales $U \gg T \gg Us^2$ to argue that the entropy of $\H$ is equivalent to that of $\H_{\rm cl}$. This reasoning enables us to compute $S \approx S_{0}$ away from the integers, as in the previous section. It also enables us to compute $S$ at temperatures $T \sim E_{\rm str}$, where the entropy crosses over between the strain-free and strainful limits. The entropy curves are plotted in Fig. \ref{fig:strain_entropy} at a variety of temperatures, relative to a strain scale $E_{\rm str} = 55$K. With this single fit parameter, we closely reproduce the curves in Ref. \cite{IlaniEntropy}.

Let us briefly comment on the ground state at charge neutrality with strain. As mentioned above, there is a non-degenerate ground state of $\H_{\rm cl}$ at this filling, $\ket{\Psi_{\nu = 0}}$, defined by $c^\dag_\bR \gamma_x c_\bR = 4$ for all $\bR$. Under the Hamiltonian $\H_{\rm cl}$, the state $\ket{\Psi_{\nu = 0}}$ describes a generalized ferromagnet in an ``inter-Chern" direction $\gamma_x$ with a gap of $U+E_{\rm{str}}$ to charge excitations. However, under the full topological Hamiltonian $\H$, an inter-Chern order parameter must have a $4\pi$ Berry phase winding around the BZ. The Berry phase winding implies that there are two vortices where the order parameter vanishes \cite{liuNematicTopologicalSemimetal2021,bultinckMechanismAnomalousHall2020,wang2024cherntexturedexcitoninsulatorsvalley,kwan2024texturedexcitoninsulators}. Indeed, the state $\ket{\Psi_{\nu=0}}$ describes the nematic semimetal (NSM) state, which has numerically been shown to be the ground state of the strained interacting BM model at charge neutrality\cite{KwanKekule,parkerStrainInducedQuantumPhase2021}. The NSM Fock dispersion renormalizes the strain dispersion $h_{\rm str}$ via $E_{\rm{str}} \to E_{\rm{str}} + U$, which has its two vortices at the $\Gamma$ point quadratic band touching. For most of the BZ, the classical spectral gap of $U + E_{\rm{str}}$ survives. 

The NSM becomes the ground state at parametrically small values of strain, $E_{\rm{str}} \gtrsim Us^2$. Indeed, the NSM has energy $-4 E_{\rm{str}} + N^{-1}\langle \H' \rangle = -4 E_{\rm{str}} + 4(J-\tilde{J})$ per site, where we used \eqref{eq:Jex_strained_maintext} at $\nu = 0$. This energy should be compared to $E_{\rm FM} = 0$, where $E_{\rm FM}$ is the ground state energy of the intra-Chern ferromagnets. We again see that $s>0$ can qualitatively alter excitations with $\abs{\bk} \sim s$ while having limited influence on bulk quantities such as ground state energy.

\section{Mott Semimetal and Insulators}\label{sec:Mott}
In this section, we study the properties of the fluctuating moment phase obtained at intermediate temperature $U s^2 \ll T \ll U$. At this temperature range, the moments are disordered and the entropy is large, as shown in the previous section. However, we will show that the spectral function of this phase still encodes non-trivial information about the topology of the underlying bands leading to an interesting ``Mott semimetal" at charge neutrality where the spectral gap closes near the $\Gamma$ point. Away from neutrality, we obtain an interesting class of "spectrally imbalanced" Mott states where the spectral weight continuously shifts between the lower and upper bands as we approach the $\Gamma$ point. These Mott spectra are calculable because the small parameter $s$, while non-perturbative, allows us to simplify the evaluation of {\it all}  diagrams that contribute to the self-energy. The simplification arises due to phase space considerations. This allows us to obtain the leading part of the spectral function in $s$ exactly.

Let us briefly discuss the model in a context and notation that will be useful for the spectral function calculation. A key observation is that all quantities of interest, including the form factors, Hartree, Fock, and strain dispersions, simplify to be essentially independent of the Bloch momentum $\bk$ as long as $\bk$ is not close to the $\Gamma$ point. We will use the symbol $``*"$ to indicate any BZ momentum $\bk$ that we can assume is ``not near the $\Gamma$ point." Then we have, from \eqref{eq:hartreeFock_maintext} and \eqref{eq:formfactor_calculation_supp}
\begin{equation}
    h_F(*) = U, \,\, h_H(*) = 2U, \,\, \lambda_\bq(*) = \xi_\bq.
\label{eq:staredquantities}
\end{equation}
These simplifications extend to the strain dispersion, which when evaluated away from $\Gamma$ has the generalized Zeeman form that we discused in the previous section, $h_{\rm str}(*) = -E_{\rm str} \gamma_x$.

Furthermore, these ``starred quantities" are exactly what one would compute in the classical model \eqref{eq:classicalmodel}. For these concrete computations, these simplifications are not too surprising: after all, the wavefunctions \eqref{eq:fullWF} look like Bloch states of the classical model as long as $\bk$ is not near the $\Gamma$ point.

We can also reintroduce the bare BM flat-band dispersion dispersion, $h_{\rm BM}(\bk)$, in this context. Note that $h_{\rm BM}$ nearly vanishes at the magic angle $\theta_M$ and increases linearly with small $(\theta-\theta_M)$. We will therefore capture its effect by expanding the full BM Hamiltonian $H_{\rm{BM}}^{\theta}$ around $\theta_M$ and projecting the term proportional to $(\theta-\theta_M)$ onto the wavefunctions \eqref{eq:fullWF}. We obtain (App. \ref{app:singleparticle_BM})
\begin{equation}
    h_{\rm BM}(\bk) = E_{\rm BM} \frac{2s^2}{\abs{\bk}^2 + 2 s^2} \begin{pmatrix} 0 & e^{-2 i \theta_\bk} \\ e^{2 i \theta_\bk} & 0 \end{pmatrix},
    \label{eq:BM_dispersion_maintext}
\end{equation}
where $E_{\rm BM} \propto (\theta - \theta_M)$ and $k_x + i k_y = \abs{\bk} e^{i \theta_\bk}$. Note that $h_{\rm BM}$ appears singular around $\bk = 0$, but this is simply because we chose a gauge that has its singularity at $\Gamma$. The Fourier transform of $h_{\rm BM}(\bk)$ can be interpreted as a long-range hopping between the non-local Wannier states; it decays as $\abs{\br}^{-2}$ due the gauge singularity (matching the Wannier tail amplitude). We have $h_{\rm BM}(*) = 0$ in this case, which is consistent with the fact that the Dirac points at $K$ and $K'$ have negligible velocity near the magic angle.

\subsection{Diagram set up}
\label{subsec:diagramsetup}

To set up the diagrammatic calculation for the Green's function and self-energy, we shift to a  Lagrangian path integral formulation. We start by rewriting the Hamiltonian as
\begin{equation}
    \H  = \sum_{\bk} c^\dag_\bk h_{T}(\bk) c_\bk + \sum_\bq V_\bq :\delta^\nu \rho_\bq \delta^\nu \rho_{-\bq}:
    \label{eq:Hamiltonianpreaction}
\end{equation}
where
\begin{equation}
    \delta^\nu \rho_\bq = \rho_\bq - (4+\nu) N\sum_\bG \delta_{\bq,\bG} \xi_\bG
\end{equation}
is the density relative to the average density at integer filling $\nu$, and we have normal ordered the interaction with respect to the fully empty $\nu = -4$ state. The rewriting of the interaction induces a shift in the two body term $h_{\rm BM} \to h_T$. In particular, the shift in average density leads to a two-body Hartree term, while the normal ordering produces a Fock term:
\begin{equation}
    h_T(\bk) = h_{\rm BM}(\bk) + h_F(\bk) + \nu h_H(\bk).
    \label{eq:totdispersion}
\end{equation}

It is important to note that all terms contributing to $h_T$ are nearly $\bk$-independent for most of the BZ, away from an $O(s^2)$ region of the $\Gamma$ point (see Eqs.~\eqref{eq:BM_dispersion_maintext},\eqref{eq:hartreeFock_maintext} and the discussion surrounding \eqref{eq:staredquantities}). The associated Green's function
\begin{equation}
G_0(\bk,i\omega_n) = (i \omega_n + \mu- h_T(\bk))^{-1}
\label{eq:nonintGreenFn}
\end{equation}
then satisfies $G(*,i \omega_n) = 1/(i\omega_n + \mu + (1+\nu)U)$ for $\bk \to *$ (away from $\Gamma$). Here $\omega_n = \frac{2\pi}{\beta} \left(n+\frac{1}{2}\right)$ is a fermionic Matsubara frequency. 

We emphasize that $G_0$ is not the physical Green's function in any physical limit, but it will play the role of the ``non-interacting" Green's function in the diagrammatics associated with \eqref{eq:Hamiltonianpreaction}. We also emphasize that the analysis of individual diagrams in the next section are not for the purpose of a perturbative calculation; rather, their purpose is to gain intuition for how diagrammatics simplify for small $s^2$. We will eventually consider all diagrams by computing suitable correlation functions.

Introducing a Hubbard-Stratonovich field to decouple the interactions, our path integral formulation is expressed as follows
\begin{equation}
\begin{aligned}
    Z  & = \int D\ov{\psi} D \psi D \phi e^{-S[\ov{\psi},\psi,\phi]} \\
    S & = \sum_{\bk, n} \ov{\psi}_{\bk, n} G_0^{-1}(\bk, i \omega_n) \psi_{\bk, n}  + \frac{i}{\beta A} \sum_{\bq, n,m} \phi_{\bq, m}  \delta^\nu \rho_{\bq m} \\
    & + \frac{1}{2\beta A} \sum_{\bq, n} \phi_{\bq, n} V^{-1}(\bq) \phi_{-\bq,-n} \\
    \delta^\nu \rho_{\bq m} & =  \sum_\bk \ov{\psi}_{\bk,n} \Lambda_{\bq}(\bk) \psi_{\bk+\bq,m+n} - (4+\nu)\delta_{[\bq],0}\delta_{m,0}\xi_\bq \\
    \end{aligned}
    \label{eq:partitionfn_maintext}
\end{equation}

We can write the full Green's function as
\begin{gather}
    G^{-1} = G_0^{-1} - \tilde{\Sigma} = -i \omega_n - h_{\rm BM}(\bk) - \Sigma \nonumber \\
    \Sigma = \tilde{\Sigma}(\bk,i\omega_n) + \nu h_H(\bk) + h_F(\bk)
    \label{eq:defnofFullSigma}
\end{gather}
We note that $\tilde{\Sigma}$ should be understood as the self energy relative to $h_T$ and the associated $G_0^{-1}$, and differs by some counter terms from total interaction-induced dispersion $\Sigma$. Indeed, $G_0^{-1}$ already contains the full Hartree self energy as well as a Fock counterterm associated with normal ordering with respect to $\nu = -4$. The physical self energy should be regarded as $\Sigma$, but our diagrammatics associated with \eqref{eq:partitionfn_maintext} will compute $\tilde{\Sigma}$. 

\subsection{Example diagrams}
\label{subsec:diagramexample}
\begin{figure}
     \begin{tikzpicture}
  \begin{feynman}[layered layout, every blob = {/tikz/fill=gray!0,/tikz/inner sep=2pt}]
    \vertex (i) at (0, 0) {};
    \vertex [right=of i] (qi) [dot] {};
    \vertex [right=1.5cm of qi] (fermleft) [];
    \vertex [right=0.65cm of fermleft] (rectanch) [];
    \vertex [above=1cm of fermleft] (photleft) [];
    \vertex [above=-0.5cm of rectanch] (rect) [blob,shape=rectangle,minimum height=2cm, minimum width = 1.3cm] {$\Xi$};
    \vertex [right=1.3cm of fermleft] (fermright) [];
    \vertex [above=1cm of fermright] (photright) [];
    \vertex [right=1.5cm of fermright] (qpf) [dot] {};
    \vertex [right= of qpf] (f) {};
    Draw the fermion lines
    \diagram* {
      (i) -- [fermion, edge label'=\(k\)] (qi)
      -- [fermion, edge label'=\(k - q\)] (fermleft),
      
      (fermright) -- [fermion, edge label'=\(k -  q'\)] (qpf)
      -- [fermion, edge label'=\(k\)] (f),

      (qi) -- [photon, out=90, in=180, looseness=1, edge label=\(q\)] (photleft),
      
      (photright) -- [photon, out=0, in=90, looseness=1, edge label=\(q'\)] (qpf)
    };
  \end{feynman}
\end{tikzpicture}
\includegraphics{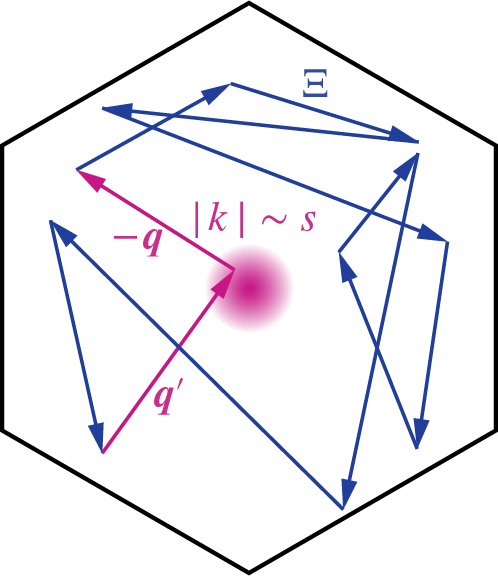}
\caption{Diagrammatic depiction of the general form of the self energy \eqref{eq:exactselfenergy} and depiction in terms of momentum-space scattering. The pink arrows represent the external vertices, which feel the $\Gamma$ point Berry phase, while the blue arrows labeled $\Xi$ represent the 1PI four point function box that can be evaluated in the trivial classical model \eqref{eq:classicalmodel}. An electron begins with a momentum $\bk$, which we allow to be the vicinity of $\Gamma$, but then scatters to a momentum $\bk - \bq$ which is generically not near $\Gamma$. The ensuing dynamics of the electron afterwards is equivalent to that of an electron in the classical model \eqref{eq:classicalmodel}, up to order $s^2$ corrections, until it scatters back into $\Gamma$ to end the 1PI diagram. The entire process can be regarded as $\Gamma$ point electrons feeling the rest of the Brillouin Zone as a bath.}
\label{fig:selfenergy_4ptfn_maintext}
\end{figure}

We begin by evaluating two example diagrams, which will serve to concretely explain the phase space approximations we can make at leading order in $s^2$. We begin with 
\begin{equation}
\begin{aligned}
& D_1(\bk,i\omega_n)  = 
\vcenter{\hbox{\begin{tikzpicture}
  \begin{feynman}
    \vertex (i) at (0, 0) {};
    \vertex [right=0.7cm of i] (qi) [dot] {};
    \vertex [right=1.3cm of qi] (qpi) [dot] {};
    \vertex [right=2cm of qpi] (qpf) [dot] {};
    \vertex [right=1.3cm of qpf] (qf) [dot] {};
    \vertex [right=0.7cm of qf] (f) {};

    \diagram* {
      (i) -- [fermion, edge label'=\(k\)] (qi)
      -- [fermion, edge label'=\(k - q\)] (qpi)
      -- [fermion, edge label'=\(k - q - q'\)] (qpf)
      -- [fermion, edge label'=\(k - q\)] (qf)
      -- [fermion, edge label'=\(k\)] (f),

      (qi) -- [photon, out=90, in=90, looseness=1, edge label=\(q\)] (qf),
      (qpi) -- [photon, out=90, in=90, looseness=1, edge label=\(q'\)] (qpf),
    };
  \end{feynman}
  \end{tikzpicture}}} \\
  & = \frac{1}{\beta^2 A^2}\sum_{\bq,\bq',m,m'}V_{\bq} V_{\bq'}\Lambda_{\bk,\bk-\bq} \\
  & \times \left[G_{0,\bk-\bq} \Lambda_{\bk-\bq,\bk-\bq-\bq'}G_{0,\bk-\bq-\bq'}   \Lambda_{\bk-\bq-\bq',\bk-\bq} G_{0,\bk-\bq} \right] \\
  & \times \Lambda_{\bk-\bq,\bk} \\
  \end{aligned}
\end{equation}
We have dropped the Matsubara frequency labels on the internal Green's functions; they may be read off from the momenta. 

The essential approximation is that internal lines within 1PI diagrams mostly have momenta away from the $\Gamma$ point. Indeed one sees explicitly that the momenta of all internal lines in $D_1$ vary over the BZ as $\bq$ and $\bq'$ are summed over. For most $\bq,\bq'$, these momenta do not lie in the $O(s^2)$ region around the $\Gamma$ point with nontrivial $\bk$-dependence. Furthermore, any potential divergences associated with the region around the $\Gamma$ point are regulated by $T>0$.  We may therefore replace the internal vertices and Green's functions by $\xi_\bq$ and $G_{0,*}$, which are their respective values away from $\Gamma$ (c.f. Eq.~\ref{eq:staredquantities}). 
Due to this great simplification, the momentum dependence may be extracted since it lies only in the external vertices:
\begin{equation}
\begin{aligned}
D_1(\bk,i\omega_n) & = \frac{1}{\beta^2 A^2}\sum_{\bq}V_{\bq} \Lambda_{\bk,\bk-\bq} \Lambda_{\bk-\bq,\bk} \\ 
& \times \sum_{\bq'mm'}V_{\bq'}\left[G_{0,*} \xi_{\bq'} G_{0,*}   \xi_{\bq'} G_{0,*} \right]  \\
& = \frac{\sum_{\bq} V_\bq \abs{\Lambda_{\bq}(\bk)}^2 }{\sum_{\bq} V_\bq \abs{\xi_\bq}^2} D(*,i\omega_n) \\
& = \frac{h_F(\bk)}{U} D(*,i\omega_n).
\end{aligned}
\label{eq:D1start}
\end{equation}
where we have identified the Fock dispersion \eqref{eq:hartreeFock_maintext}.

Furthermore, 1PI diagrams with external momenta far from the $\Gamma$ point, such as $D_1(*,i \omega_n)$, may as well have been evaluated in the effective classical model \eqref{eq:classicalmodel} with $\bk$-independent form factors $\Lambda_\bq(\bk) \to \Lambda_\bq(*) = \xi_\bq$. While our model of interest differs non-perturbatively from this trivial model, these distinctions lie in the external vertices where the momenta either not summed, or contracted with e.g. a current operator that has strong $\bk$-dependence near $\Gamma$. Indeed, even in the diagram $D_1$, the limit $s \to 0$ is singular due to the strong $\bk$ dependence of $h_F(\bk)$.

Most diagrams do not have momentum dependence that enters through $h_F(\bk)$, however. Let us consider the following diagram, and evaluate it in a similar way.
\begin{equation}
\begin{aligned} \\
& D_2(\bk,i\omega_n)  = \vcenter{\hbox{
\begin{tikzpicture}
  \begin{feynman}
    \vertex (i) at (0, 0) {};
    \vertex [right=0.7cm of i] (qi) [dot] {};
    \vertex [right=1.3cm of qi] (qpi) [dot] {};
    \vertex [right=2cm of qpi] (qf) [dot] {};
    \vertex [right=1.3cm of qf] (qpf) [dot] {};
    \vertex [right=0.7cm of qpf] (f) {};

    \diagram* {
      (i) -- [fermion, edge label'=\(k\)] (qi)
      -- [fermion, edge label'=\(k - q\)] (qpi)
      -- [fermion, edge label'=\(k - q - q'\)] (qf)
      -- [fermion, edge label'=\(k - q'\)] (qpf)
      -- [fermion, edge label'=\(k\)] (f),

      (qi) -- [photon, out=90, in=90, looseness=1, edge label=\(q\)] (qf),
      (qpi) -- [photon, out=-90, in=-90, looseness=1, edge label'=\(q'\)] (qpf),
    };
  \end{feynman}
  \end{tikzpicture}}}\\
& = \frac{\sum_{\bq'} V_{\bq'} \Lambda_{\bk,\bk-\bq'} \xi_{\bq'} \sum_{\bq} V_{\bq} \xi_\bq \Lambda_{\bk-\bq,\bk} }{\sum_{\bq} V_\bq \xi_\bq^2 \sum_{\bq'} V_{\bq'} \xi_{\bq'}^2} D_2(*,i\omega_n)
  \end{aligned}
  \label{eq:D2start}
  \end{equation}
We again obtain a $\bk$-dependent prefactor, which depends on $s$ non-perturbatively, multiplied by the value of the diagram in a $\bk$-independent model with form factors $\xi_\bq$. The prefactor may be evaluated similarly to the Fock dispersion below \eqref{eq:formfactor_calculation_supp}. We obtain
\begin{equation}
    D_2(\bk,i\omega_n) = \frac{\abs{\bk}^2}{\abs{\bk}^2 + 2 s^2} D(*,i \omega_n).
    \label{eq:D2end}
\end{equation}
The fact that $D_2 \to 0$ as $\bk \to 0$ may be understood by noting that the form factor $\Lambda_\bq(\bk)$ has a Berry phase of $2\pi$ as $\bk$ encircles the $\Gamma$ point, for generic $\bq$. In the diagram $D_2$, the average over $\bq$ and $\bq'$ are uncorrelated such that the scattering amplitude away from the $\Gamma$ point and the scattering amplitude into it separately vanish due to destructive interference as $\bk \to 0$. In contrast, in diagram $D_1$, the momentum transfers at the two external vertices are the same and the Berry phases cancel each other for each $\bq$. These scattering processes, with the external vertex processes highlighted as pink arrows, are depicted in Fig. \ref{fig:selfenergy_4ptfn_maintext}.

\subsection{All diagrams}
\label{subsec:diagramall}
We are now ready to explain the general calculation. We begin by deriving it diagrammatically, but we will later describe a corresponding non-perturbative Schwinger-Dyson calculation that yields the same result. We note that every diagram has the form depicted Fig. \ref{fig:selfenergy_4ptfn_maintext}, where the rectangular block $\Xi$ consists of all 1PI diagrams with the depicted external legs. This is nothing but the 1PI four point function, which is related to the usual four-point function by subtracting off the non-1PI part consisting of two connected three-point functions,
\begin{equation}
\begin{aligned}
\Xi_{\bq',\bq}(\bk) & = -\langle\T  \psi_{\bk-\bq'}\phi_{-\bq'} \phi_{\bq} \ov{\psi}_{\bk-\bq} \rangle_{\rm{1PI}} \\
& = -\langle  \T\psi_{\bk-\bq'}\phi_{-\bq'} \phi_{\bq} \ov{\psi}_{\bk-\bq} \rangle \\
& \,\,\,+ \langle \T \psi_{\bk-\bq'}\phi_{-\bq'} \ov{\psi}_{\bk} \rangle G^{-1}_\bk \langle \T \psi_{\bk} \phi_{\bq} \ov{\psi}_{\bk-\bq} \rangle.
\end{aligned}
\end{equation}
Note that the disconnected contributions to $\Xi$, which correspond to self-energy diagrams with no vertex corrections, are regarded as 1PI here. We then have (see Fig. \ref{fig:selfenergy_4ptfn_maintext})
\begin{equation}
    \tilde{\Sigma}(\bk,i\omega_n) = -\frac{1}{\beta^2A^2}\sum_{\bq m \bq'm'} \Lambda_{\bk,\bk-\bq'} \Xi_{\bq',\bq}(\bk)\Lambda_{\bk-\bq,\bk}
    \label{eq:exactselfenergy}
\end{equation}
The relationship \eqref{eq:exactselfenergy} is exact, with no approximations. Indeed, it can be derived as a Schwinger-Dyson equation for the self energy, where $\Xi$ is expressed in terms of third and fourth order 1PI vertices in the effective action (App.~\ref{app:SchwingerDyson}). 

We now use techniques equivalent to those used in the evaluation of $D_{1,2}$ to evaluate $\Xi$ at $T\gg Us^2$. From the diagrammatic viewpoint, because $\Xi$ is 1PI, any vertical cut through the diagram will intersect at least two lines which the momentum $\bk$ must be split between. Thus for most values of loop momenta, no internal line within $\Xi$ will have momentum $\bk \approx \Gamma$, and $\Xi$ may be evaluated within the $\bk$-independent classical model \eqref{eq:classicalmodel}. From the non-perturbative Schwinger-Dyson viewpoint (App.~\ref{app:SchwingerDyson}), $\Xi$ may be evaluated in terms of irreducible vertex functions $\Gamma^{(3)}(*), \Gamma^{(4)}(*),\ldots$ with $\bk$ arguments away from $\Gamma$. These irreducible vertex functions are related to each other self-consistently, with corrections from $\bk \approx \Gamma$ suppressed by $s^2$. A self-consistent solution to this set of Schwinger-Dyson equations is computable from the $\bk$-independent classical model. From either perspective, we arrive at the conclusion that, for $T \gg Us^2$, we can replace $\Xi(\bk) \to \Xi(*)$ in \eqref{eq:exactselfenergy} and evaluate $\Xi(*)$ in the classical model \eqref{eq:classicalmodel}.

We pause to emphasize that the same logic implies that any 1PI correlation function with external momenta away from $\Gamma$ may be calculated through the classical model. Many physical quantities, at $T \gg Us^2$, break down into such correlation functions and can be directly read off from the classical model. Entropy, as well as flavor magnetization and susceptibility, all take the values associated with the classical model, as we claimed at the end of Sec \ref{sec:decoupledEntropy}B. For these quantities, the small area of phase space near $\Gamma$ cannot compete with or reduce the large bulk contribution from the rest of the BZ. In contrast, quantities which either probe the $\Gamma$ point region directly, such as $G_{\bk,n}$ where $\abs{\bk} \sim s$, or transport quantities, which involve $\bk$-space derivatives that effectively magnify the $\Gamma$ region, often differ non-perturbatively from those of the classical model. For these quantities, the larger trivial region of the BZ serves as a bath for the active electrons near the $\Gamma$ point. 

The quantity $\Xi(*)$ is most readily evaluated in real space in terms of the original fermionic variables $c_{\bR,\alpha}$. Thus it can be written explicitly by integrating out the Hubbard-Stratonovich field and Fourier transforming to real space as
\begin{equation}
\begin{aligned}
    & \Xi_{\bq,\bq'}(*,i\omega_n)  = \beta V(\bq)G_{*,n-m} \delta_{\bq,\bq'}\delta_{m,m'} \\
    & - \frac{V_\bq \xi_\bq V_{\bq'} \xi_{\bq'}}{N}\sum_{\bR,\bR_1,\bR_2} e^{-i \bq \cdot (\bR - \bR_2)} e^{i \bq' (\bR-\bR_1)} \Xi_{\bR_1,\bR_2}(\bR), \\
    & \Xi_{\bR_1,\bR_2}(\bR)  = -\langle \T c_{\bR n-m' }\delta^{\nu} n_{\bR_1 m'} \delta^\nu n_{\bR_2 -m} c^\dag_{\bR n-m}\rangle \\
    & + \langle \T c_{\bR n-m' }\delta^{\nu} n_{\bR_1 m'} c^\dag_{\bR n} \rangle G_{*,n}^{-1} \langle \T c_{\bR,n} \delta^\nu n_{\bR_2 -m} c^\dag_{\bR n-m}\rangle
    \end{aligned}
    \label{eq:classicalXi}
\end{equation}
where $\T$ denotes imaginary time ordering and $\delta^\nu n_\bR = n_\bR - (\nu+4)$ is the occupation of site $\bR$ relative to its average at arbitrary filling $\nu$.

We have thus established a concrete, tractable, procedure to calculate the self energy in the TBG model, provided $T \gg Us^2$. One finds the classical ground state, evaluates \eqref{eq:classicalXi}, and plugs into \eqref{eq:exactselfenergy}. 

In this paper, we will restrict our attention to integer fillings where we can benefit from some convenient simplifications, but this restriction is not necessary. At integer fillings and $U \gg T$, the densities $\delta^\nu n_{\bR_{1,2}}$ annihilate the classical ground states. Thus, the densities must be evaluated at the same position as the fermion operators, $\bR = \bR_1 = \bR_2$, sit between them in the time ordering, and have sign $\pm1$ depending on whether the excitation is electron-like ($c$ has larger $\tau$) or hole-like ($c^\dag$ has larger $\tau$). We obtain\footnote{The Matsubara sum over $m$,$m'$ sets the time of each density equal to the corresponding fermion, leading to an ambiguity in time-ordering. One symmetric prescription for alleviating this ambiguity is $\T[c_\tau \delta n_\tau] \to \frac{1}{2}\{c_\tau, \delta n_\tau\}$, which can be separately justified through deriving the equation in a Hamiltonian picture without normal ordering. Because the path integral is derived through normal ordering the fermions, it also makes sense to normal order in the presence of this ambiguity; $\T[c_\tau \delta n_\tau] \to :c_\tau \delta n_\tau:$. Both of these prescriptions ultimately lead to the same result for $\Xi(i\omega_n)$. We thank Qingchen Li for a useful discussion on this point. }
\begin{equation}
\begin{aligned}
    \frac{1}{\beta^2}\sum_{m m'}& \Xi_{\bR_1,\bR_2}(\bR) = \frac{1}{4}\Xi(i\omega_n) \delta_{\bR_1,\bR} \delta_{\bR_2,\bR}, \\
    \Xi(i \omega_n) & = G_{*,n} - (G_{*,n}^+ - G_{*,n}^-) G_{*,n}^{-1}(G_{*,n}^+ - G_{*,n}^-), \\
    & =  2 \left( G_{*,n}^- G_{*,n}^{-1}G_{*,n}^+ +G_{*,n}^+G_{*,n}^{-1}G_{*,n}^-\right).
\end{aligned}
\label{eq:beyondmeanfieldXi}
\end{equation}
where $G_{*,n}^{\pm} = -\int_{-\beta/2}^{\beta/2} d\tau e^{i \omega_n \tau} \Theta(\pm \tau) \langle \T c_\tau c^\dag_0 \rangle$ are the electron-like and hole-like parts of the Green's function $G_{*,n} = G^+_{*,n} + G^-_{*,n}$. 

The self-energy \eqref{eq:exactselfenergy} splits into two parts when \eqref{eq:classicalXi} is inserted. The first part combines with the Fock counter term in $G_0^{-1}$, see \eqref{eq:totdispersion},\eqref{eq:nonintGreenFn} after using $\beta^{-1}\sum_m G_{*,n-m} = \langle c^\dag_{\bR} c_\bR \rangle = P_*$, which is a projector for a Slater determinant state but not otherwise. In the second term, the sums over $\bq,\bq'$ reproduce those of \eqref{eq:D2start},\eqref{eq:D2end}. 

We then have, for the full self-energy,
\begin{equation}
\begin{aligned}
\Sigma & = \tilde{\Sigma} + h_F(\bk) + \nu h_H(\bk) \\
    & = -\frac{1}{2A} \sum_\bq V_\bq \Lambda_{\bk,\bk-\bq} Q_* \Lambda_{\bk-\bq,\bk} \\
    & + \frac{\abs{\bk}^2}{\abs{\bk}^2 + 2 s^2}  U^2 \Xi(i \omega_n) + \nu h_H(\bk)
    \end{aligned}
    \label{eq:exactselfenergy_intfill}
\end{equation}
where $P_* = \frac{1}{2}(1+Q_*)$. 

The first part can be traced to diagrams of the form \eqref{eq:D1start} where the momenta $\bq$ and $\bq'$ in Fig.~\ref{fig:selfenergy_4ptfn_maintext} are set to be equal. These are diagrams where vertex corrections are not included. If the state is a Slater determinant with $Q^2 = 1$, this part simply reduces to the Fock exchange energy. The second term includes non-trivial vertex corrections and corresponds to diagrams of the form \eqref{eq:D2start},\eqref{eq:D2end} where $\bq$ and $\bq'$ in Fig.~\ref{fig:selfenergy_4ptfn_maintext} are independent. This term vanishes for a Slater determinant since $G^{+} = P G^+ P$, where $P$ projects onto occupied orbitals, while $G^{-} = (1-P) G^- (1-P)$, leading to $\Xi(i\omega_n) = 0$.

In contrast, for the symmetric state where the moments are disordered, the second term in \eqref{eq:exactselfenergy_intfill} is crucial: it opens the Mott gap away from the $\Gamma$ point. Indeed, a symmetric state at filling $\nu$ has $Q_* = \nu/4$, proportional to the identity matrix, so the first term gives a frequency and band independent dispersion that cannot open a Mott gap anywhere in the BZ. The second term should be computed from the Green's function of the classical model, which can be solved as a single site problem at integer filling:
\begin{equation}
    G_{*,n} = G^+_{*,n} + G^-_{*,n}, \quad G^{\pm}_{*,n} = \frac{\frac{1}{2} \mp \frac{\nu}{8}}{i\omega_n + \mu_\nu \mp U}.
    \label{eq:MottGreen}
\end{equation}
Here $\mu_\nu = \mu - 2\nu U$ is the chemical potential relative to the Hartree offset $h_H(*) = 2\nu U$. We should have $\abs{\mu_\nu} < U$ in order to be consistent with integer filling $\nu + O(s^2)$. The function \eqref{eq:MottGreen} leads to a nonzero, and ultimately self-consistent, expression for $\Xi(i\omega_n)$ as we now discuss.

\subsection{Mott Semimetal at Charge Neutrality}
\label{subsec:semimottal}
\begin{figure*}
    \includegraphics[width = \textwidth]{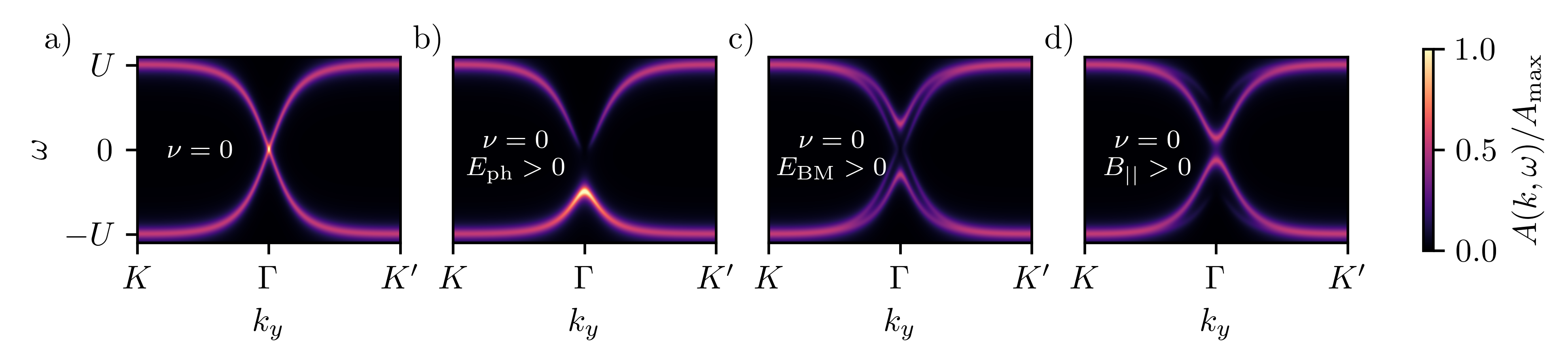}
    \caption{Mott states at charge neutrality ($\nu = 0$). A broadening term $\Sigma \to \Sigma - i U s^2$ is added to all spectral functions to model theoretical uncertainty. \textbf{(a)} Mott semimetal at charge neutrality with a single Dirac cone per spin per valley, described by the self energy \eqref{eq:neutselfenergy}. \textbf{(b)} Mott semimetal gapped by particle-hole breaking dispersion $E_{\rm ph} = U/2$, with $Z_\bk \sim |\bk|^2$ in the doublon band for $\bk \to 0$. The eight-fold at the Mott semimetal touching point is preserved, but it is inherited entirely by the holon band. \textbf{(c)} Mott semimetal with BM dispersion $E_{\rm BM} = 0.3U$. There is an exotic quadratic band touching at $\omega = 0$, with $Z_\bk \sim |\bk|^2$ as $\bk \to 0$; the latter makes the touching difficult to identify without analytical insight.\textbf{(d)} Curie-gapped Mott semimetal 
    with in-plane Zeeman field $\mu_B B_{||}/T = 1$.
    }
    \label{fig:semimottal}
\end{figure*}

We begin with the self energy and spectral function for the symmetric Mott state at charge neutrality. Using \eqref{eq:MottGreen}, with $\mu = \nu = 0$, we obtain 
\begin{equation}
    \Xi(i \omega_n) = \frac{1}{i \omega_n}
\end{equation}
such that
\begin{equation}
    \Sigma(\bk,i\omega_n) = \frac{\abs{\bk}^2}{\abs{\bk}^2 + 2 s^2} \frac{U^2}{i \omega_n}.
    \label{eq:neutselfenergy}
\end{equation}
We see that the self energy is equal to the self energy of the usual Hubbard model except modulated by a factor
\begin{equation}
    W_\bk = \frac{\abs{\bk}^2}{\abs{\bk}^2 + 2 s^2}
\end{equation}
The fact that $W_\bk \to 0$ quadratically with $\bk$ is reminiscent of a quadratic band touching and the $4\pi$ Berry phase typical in symmetric TBG band structures; Indeed, as we discussed in the example diagram $D_2$, the vanishing as $\bk \to 0$, arises from the Berry phase of the external form factors $\Lambda_\bq(\bk)$ when averaged over scattering directions $\bq$. Remarkably, however, we obtain a \emph{linear} dispersion in the spectral function. Indeed, for zero bare dispersion, a partial fraction decomposition leads to
\begin{equation}
\begin{aligned}
    G(\bk,i \omega_n) & = \frac{1}{i\omega_n - \Sigma(\bk,i\omega_n)} \\
    & = \frac{\frac{1}{2}}{i \omega_n - \sqrt{W_\bk} U} + \frac{\frac{1}{2}}{i \omega_n + \sqrt{W_\bk} U}
\end{aligned}
\label{eq:MottSemimetal_greensfunction}
\end{equation}
The spectral function associated to \eqref{eq:MottSemimetal_greensfunction} (Fig.~\ref{fig:semimottal}a) looks like a \emph{single} Dirac cone per spin and valley flavor near $\bk = 0$. Crucial to this apparent halving is the Mott-type $1/(i\omega_n)$ dependence of $\Sigma$, which leads to a square root of the Berry-phase induced prefactor of $\Sigma$. For $\bk \to *$ we reproduce $G_*$ as required by self-consistency. The Mott semimetal state we obtain is consistent with quantum Monte Carlo simulations at intermediate temperatures at charge neutrality~\cite{hofmannFermionicMonteCarlo2022,MengQMCSemimetal}.

The Mott semimetal is qualitatively distinct from all other TBG semimetals that we know of. The non-interacting or renormalized ``BM semimetal" has Dirac points at $K$ and $K'$ rather than $\Gamma$, as does the ``symmetric Kondo state" \cite{TBGKondoSDS,HuSymmKondo}. The nematic semimetal breaks $C_3$ and has either a quadratic band touching at $\Gamma$ or two Dirac cones in the vicinity of $\Gamma$. Another state was discussed in Ref. \cite{wang2024electronphononcouplingtopological} from the THF perspective and also dubbed Mott semimetal. This state appears to be describable by two decoupled systems, consisting of a traditional Mott insulator and eight Dirac bands, where the Dirac dispersion comes from the bare Hamiltonian. Our Mott semimetal possesses a few qualitative differences: (i) the semimetallic dispersion comes entirely from interactions; (ii) we have half the Dirac cone degeneracy, a fact which is enabled by the Mott-like self energy $\propto \abs{\bk}^2/\omega$; and (iii) our Mott semimetal responds to perturbations in a manner distinct from any non-interacting or mean-field system as we discuss below. 

We now discuss how a variety of different perturbations affect the Mott semimetal\eqref{eq:MottSemimetal_greensfunction} (we will use this abbreviated phrase to refer to the Mott semimetal): particle-hole breaking terms, BM dispersion, and Zeeman fields. We begin with the particle-hole breaking terms. We note that while such terms likely exist at some level microscopically, we are primarily discussing them here for conceptual purposes and as a simpler version of $\nu \neq 0$. We take 
\begin{equation}
    h_{\rm ph}(\bk) = -E_{\rm ph} \frac{2s^2}{\abs{\bk}^2 + 2 s^2} = -E_{\rm ph}(1-W_\bk),
    \label{eq:particleholedisp}
\end{equation}
for some scale $E_{\rm ph}$, which we choose to be positive for concreteness. Note that \eqref{eq:particleholedisp} has the same form as the non-constant part of the Hartree dispersion \eqref{eq:hartreeFock_maintext}. Any particle-hole breaking potential term $\Delta(\br)$ projected to the wavefunctions \eqref{eq:fullWF} produces a term of the form \eqref{eq:particleholedisp}, up to a constant that can be absorbed into the chemical potential. Since $h_{\rm ph}(*) = 0$, the self energy \eqref{eq:neutselfenergy} is unchanged. We have
\begin{equation}
    G(\bk,i \omega_n)  = \frac{1}{i \omega_n + E_{\rm ph} (1-W_\bk) - W_\bk \frac{U^2}{i \omega_n}}.
    \label{eq:particleholeGreen}
\end{equation}
The eight-fold degeneracy $\bk = 0$ is preserved and shifted to the energy $-E_{\rm ph}$. Despite this, we still open up a gap equal to $E_{\rm ph}$. To see this, we analytically continue $i \omega_n \to \omega$ and note that the denominator diverges when $\omega(\omega + E_{\rm ph}(1-W_\bk)) - W_\bk U^2 = 0$. This leads to the doublon ($+$) and holon ($-$) branches
\begin{equation}
    \omega_{\pm}(\bk) = -\frac{E_{\rm ph}(1-W_\bk)}{2} \pm \sqrt{E_{\rm ph}^2 + W_\bk U^2 },
\end{equation}
with quasiparticle residues
\begin{equation}
    Z_{\pm}(\bk) = \frac{|\omega_{\pm}(\bk)|}{\omega_{+}(\bk) - \omega_-(\bk)}.
\end{equation}
The holon branch has $\omega_-(\bk) \to -E_{\rm ph}$ as $\bk \to 0$. The doublon branch has $\omega_+(0) = 0$, and is separated by a gap $\geq E_{\rm ph}$ at all $\bk$. Both $\omega_\pm(\bk)$ are quadratic in $\bk$ in the vicinity of $\bk = 0$, with effective mass $\propto E_{\rm ph}^{-1}$. Note that $Z_-(\bk) \to 1$ as $\bk \to 0$ for each of the eight bands. The holon band thus has the full spectral weight at the $\Gamma$ point. The doublon branch correspondingly has $Z_{+}(\bk) \propto \abs{\bk}^2$ as $\bk \to 0$. As $\bk \to *$, we have $\omega_\pm(\bk) \to \pm U$ and $Z_\pm(\bk) \to \frac{1}{2}$, as required by self consistency. The corresponding spectral function is shown in Fig.~\ref{fig:semimottal}b.
\footnote{It may seem like the filling has deviated from $\nu = 0$ by an order $s^2$ amount, since spectral weight has shifted from the doublon to the holon band at $\Gamma$. However we are not including order $s^2$ corrections to the Green's function $G(*,i\omega_n)$, which describes the rest of the Brillouin Zone. As a result, the filling of \eqref{eq:particleholeGreen} is only controlled up to order $s^2$ corrections. Order $s^2$ corrections away from $\Gamma$ that shift spectral weight in the opposite direction, from the holon band to the doublon band, likely exist. We also expect an order $s^2$ broadening of the Mott bands which could lead to order $s^2$ weight at the Fermi level; we therefore do not claim that our Mott states with a spectral gap, which we define as the separation between peaks in the spectral function, are insulators.}  
We conclude that the particle hole breaking term \eqref{eq:particleholedisp} preserves the eight fold degeneracy of the Mott semimetal but still manages to open up a gap through making the $\Gamma$ point maximally spectrally imbalanced between the holon and doublon bands.

We now show that the BM dispersion leads to a quadratic touching at $\omega = \bk = 0$ with $Z_\bk \propto \abs{\bk}^2$. Since $h_{\BM}(*) = 0$, we may again use the self energy \eqref{eq:neutselfenergy}. We will also work in the BM energy band basis, which is straightforward since $\Sigma$ is a scalar. In this basis we have $h_{\BM} = \diag(h^+_\BM I_{4 \times 4},h^-_\BM I_{4 \times 4}) $, where
\begin{equation}
    h^{\pm}_{\BM}(\bk) =  \pm E_{\rm BM}(1-W_\bk).
\end{equation}
We relatedly have $G^{\pm} = \diag(G^+, I_{4 \times 4},G^- I_{4 \times 4})$ with
\begin{equation}
    G(\bk,i \omega_n)^\pm  = \frac{1}{i \omega_n \mp E_{\rm BM} (1-W_\bk) - W_\bk \frac{U^2}{i \omega_n}}.
    \label{eq:BMGreen}
\end{equation}
We see that in the $\mp$ BM-band we reproduce \eqref{eq:particleholeGreen} with $E_{\rm ph} = \pm E_{\rm BM}$. Thus, the doublon branch in the $-$ BM band and the holon branch in the $+$ BM band remain at $\omega = 0$ at $\bk = 0$. These branches touch quadratically as $\bk \to 0$ with quasiparticle residue $Z_\bk \propto \abs{\bk}^2$ (Fig.~\ref{fig:semimottal}c).

Next we discuss the inclusion of a Zeeman field $-\mu_B B_{||} S_x$. The Zeeman field affects states uniformly across the BZ, and we therefore need to obtain $G_*$ and $\Xi$ as a function of the field. In particular, we must use a nonzero $Q_{*\alpha \beta} = 2 \langle c^\dag_{\bR \beta}  c_{\bR \alpha} \rangle - 1$ in \eqref{eq:beyondmeanfieldXi} to capture the magnetization of the system. The Hermitian matrix $Q_*$ has eigenvalues $\in [-1,1]$, and in the diagonalized basis corresponds to a quarter of the filling of each diagonal component. The Green's function away from $\Gamma$ can still be computed in real space as a single site problem. For simplicity we will take $B_{||} \ll U$ since the most interesting effects come from comparing $B_{||}$ with $T$. Then, the holon and doublon band persist at $\pm U$ but become increasingly spin polarized:
\begin{equation}
    G(*,i \omega_n) = \frac{P_*}{i \omega_n - U} + \frac{1-P_*}{i \omega_n + U}.
    \label{eq:GreenFuncBfield}
\end{equation}
If $P_* = \frac{1}{2}(1+ Q_*)$ is a projector, the above Green's function reduces to the Slater determinant form $\frac{1}{i \omega_n + Q_*U}$ as expected. From \eqref{eq:GreenFuncBfield} we obtain
\begin{equation}
    \Xi(i \omega_n) = \frac{1-Q_*^2}{i \omega_n - Q_*U},
\end{equation}
which vanishes when the state is a Slater determinant, due to $Q_*^2 = 1$. From \eqref{eq:exactselfenergy_intfill},
\begin{equation}
\begin{aligned}
    \Sigma(\bk) & = -Q_* h_F(\bk) + W_\bk U^2 \frac{1-Q_*^2}{\omega - Q_*U} \\
    G(\bk,i \omega_n) & = \frac{1}{\omega + Q_*h_F(\bk) - W_\bk U^2 \frac{1-Q_*^2}{\omega-Q_*U}}
\end{aligned}
\label{eq:Greensfunction_curieBfield}
\end{equation}
where we used $[\Lambda,Q_*] = 0$ to simplify the first term in $\Sigma$. The calculation of the spectral function then proceeds in two steps: first we calculate the magnetiation through the single site partition function \eqref{eq:magnetizationcurie_partition} to obtain $Q_* = m s_x/4$, where $m = \langle S_x \rangle/N \in [-4,4]$ is the magnetization per site. Then we compute the poles and residues associated with \eqref{eq:Greensfunction_curieBfield}. The resulting spectral function is shown in Fig.~\ref{fig:semimottal}d. At $\bk = 0$ the Mott term drops and we obtain two poles from $\omega = -Q_* h_F(\bk=0)$, one for each eigenvalue of $Q_*$, with quasiparticle residue $Z_\bk = 1$. For $Q_* \neq 0,1$ there are also two other poles that have vanishing residue as $\bk \to 0$, with energy $\omega(\bk \to 0) = Q_* U$, coming from the denominator of the second term in the self energy \eqref{eq:Greensfunction_curieBfield}. We see that all four poles have energy proportional to $Q_*$ as $\bk \to 0$; a nonzero magnetization gaps the Mott semimetal.

\subsection{Spectrally Imbalanced Mott states}
\label{subsec:SIMI}
\begin{figure*}
    \centering
    \includegraphics[width=\textwidth]{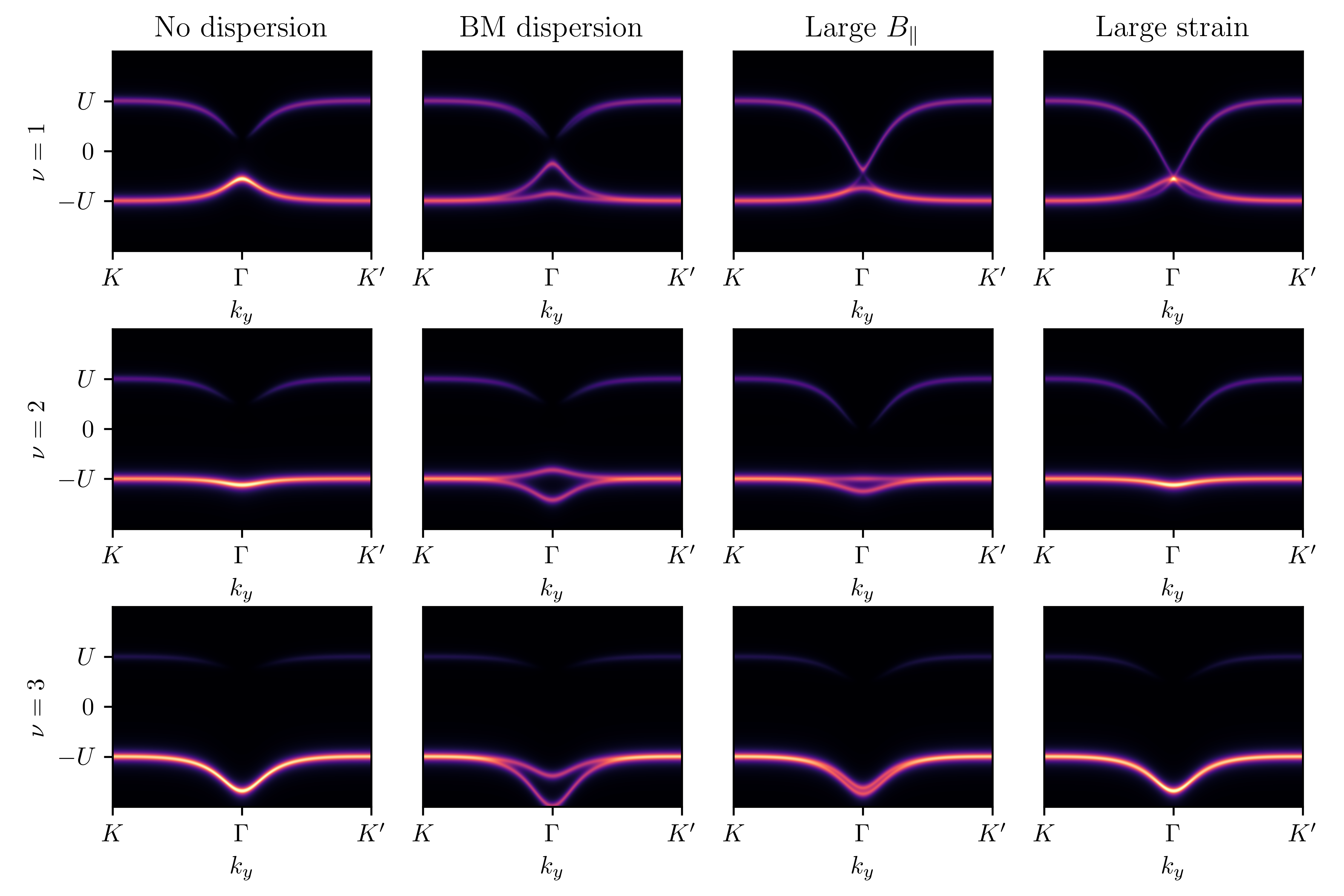}
    \caption{Mott states near fillings $\nu = 1,2,3$ (top, middle, and bottom rows) in response to, from the left column to the right column, no extra dispersion, BM dispersion $E_{\BM} = 0.3U$, large Zeeman magnetic field $B_{||} \gg T$, and large strain $E_{\rm str} \gg T$. All Mott states have large gaps $\gtrsim U$ in the absence of additional terms, larger than those of the zero temperature exact ferromagnets. The Zeeman magnetic field \emph{reduces} the spectrally-imbalanced $\Gamma$ point gap, especially at $\nu = 1$ where the gap is reduced to $\approx U/8$, rather than opening it as it does at $\nu = 0$. At intermediate temperatures $U \gg T \gg Us^2$, the Mott states on the leftmost column occupy all the $\Gamma$ point states available, reducing the quasiparticle weight $Z_\bk$ at the $\Gamma$ point of the doublon band.}
    \label{fig:imbalanced_spec}
\end{figure*}

We now discuss fillings away from neutrality. Here, the Hartree and Fock potentials break particle-hole symmetry and have a similar effect to the particle hole breaking dispersion considered earlier, Eq.~\eqref{eq:particleholedisp}. They gap the Mott semimetal and lead to spectrally imbalanced Mott states. 

The influence of symmetry breaking terms have a distinct effect, however. For example, at charge neutrality a large in-plane magnetic field is sufficient to turn the Mott state into a mean-field ferromagnet; the four Chern and valley states are in singlet representations at each site. But away from charge neutrality, spin polarization is not enough to quench the entropy; fluctuating Chern and valley moments with definite spin persist. 

The calculation of the spectral function is along the same lines as the previous section, with only a few differences. Now $Q_* = (\nu + m s_x)/4$, and $\abs{m} \leq 1-\frac{\abs{\nu}}{4}$. We also have a nonzero chemical potential $\mu \approx 2\nu U$,  and there is a Hartree dispersion $\nu h_F(\bk)$. We have
\begin{equation}
    G_* = \frac{P_*}{i \omega_n + \mu_\nu + U} + \frac{1-P_*}{i \omega_n + \mu_\nu - U}
\end{equation}
such that the self energy and full Green's function are given by
\begin{equation}
\begin{aligned}
    &\Sigma(\bk)  = \nu h_H(\bk) -Q_* h_F(\bk) + W_\bk U^2 \frac{1-Q_*^2}{i\omega_n + \mu_\nu - Q_*U} \\
    &G(\bk,i \omega_n)  = \frac{1}{i\omega_n + \mu  - \Sigma(\bk)}.
\end{aligned}
\label{eq:Greensfunction_curieBfield_nunonzero}
\end{equation}
We can verify self-consistency, $G(\bk \to *) = G_*$, after recalling $h_H(*) = 2U$ and $\mu_\nu = \mu - 2 \nu U$. 

We can also study the influence of strain in a similar manner to the in-plane Zeeman field; let us briefly discuss the modification necessary. Strain induces $Q_*$ to have a $\gamma_x$ component instead of a $s_x$ component. The first term in \eqref{eq:exactselfenergy_intfill} then evaluates to the nematic semimetal form $\frac{1}{2A}Q_* \sum_\bq V_\bq(\lambda_\bq(\bk))^2 = Q_* W_\bk U$ instead.

In Fig. \ref{fig:imbalanced_spec} we plot Mott spectra at fillings $\nu = 1,2,3$ in response to no extra dispersion, BM dispersion, large Zeeman field relative to temperature, and large strain relative to temperature. The Mott states with nothing added, left column of Fig. \ref{fig:imbalanced_spec} have large gaps $\gtrsim U$: even larger than those of the $\nu > 0$ exact ferromagnets. 
If one starts with a zero temperature generalized ferromagnetic ground state, the spectral gap will increase by an amount $\sim U$ when temperature is raised from zero to $\gtrsim Us^2$. The essential reason for the increased gap of the Mott states is their ability to have a $\bk$-dependent spectral imbalance between the holon and doublon bands. This enables the Mott states to entirely occupy the ``Hartree dip" at the $\Gamma$ point for $\nu > 0$ and lower the bandwidth of the conduction band. In contrast, the conduction band Hartree dip pierces the gap of the ferromagnet and rapidly lowers the gap as $\nu$ increases. We note that this points to an interesting connection to the intervalley Kekulé spiral state~\cite{Nuckolls_2023,liuNematicTopologicalSemimetal2021,parkerStrainInducedQuantumPhase2021,soejimaEfficientSimulationMoire2020,KwanKekule,WangIKSDMRG,wang2024cherntexturedexcitoninsulatorsvalley,kwan2024texturedexcitoninsulators}, which fully occupies the $\Gamma$ point electrons albeit through breaking translation symmetry\cite{KwanKekule,WangIKSDMRG}. The spectrally imbalanced Mott states we obtain thus seem similar in spirit to thermally disordered intervalley Kekulé spirals.

\begin{figure}
    \centering
    \includegraphics[width=0.5\textwidth]{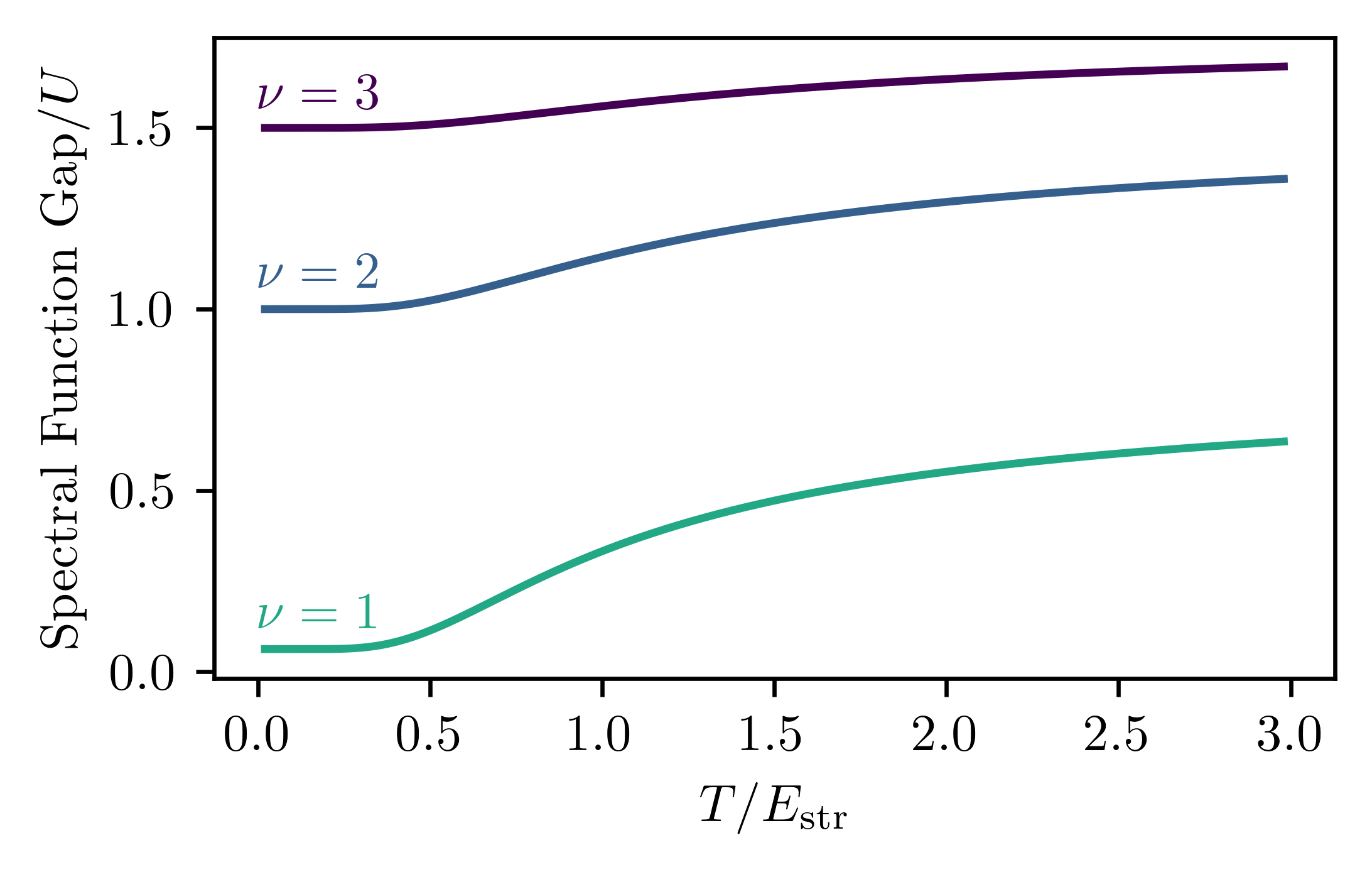}
    \caption{Spectral function gap as a function of temperature in the presence of heterostrain, $E_{\rm str} \neq 0$. In the Mott regime, $Us^2 \ll T \ll U$, the spectral function only depends on the ratio $T/E_{\rm str}$. We see that while the gap increases with temperature for all fillings $\nu = 1,2,3$, the increase for $\nu = 1$ is especially drastic. Indeed, the $\nu = 1$ gap essentially closes for $T/E_{\rm str} \to 0$. }
    \label{fig:pomeranchuk}
\end{figure}

Furthermore, both $B_{||}/T$ and $E_{\rm str}/T$ decrease the $\Gamma$ point Mott gap. This also implies a Pomeranchuk effect for the spectral gap upon increasing temperature, provided $E_{\rm str}$ is not too large. The difference in gaps is especially strong at $\nu = \pm  1$; here heterostrain essentially closes the Mott gap near $\Gamma$ (see the top right panel of Fig. \ref{fig:imbalanced_spec}). In Fig. \ref{fig:pomeranchuk} we plot the spectral gap as a function of $T/E_{\rm str}$ for $\abs{\nu} = 1,2,3$. We find it likely that the increase in spectral gap is largely responsible for the especially dramatic increase in resistance with temperature at $\nu = -1$ \cite{saitoIsospinPomeranchukEffect2021}. Of course for transport measurements it is difficult to disentangle the increase in the spectral gap from the temperature dependence of scattering off of activated moment fluctuations, though one may expect that the latter would not show a dramatic difference between $\abs{\nu} = 1$ and other integers. Measurements of the spectral gap as a function of temperature, through e.g. STM or Dirac cone spectroscopy in trilayer~\cite{shenDiracSpectroscopyStrongly2023,bocarsly2024imagingcoulombinteractionsmigrating}, should be able to test directly if the $\nu = -1$ spectral gap itself exhibits a Pomeranchuk effect.

\section{Discussion}
\label{sec:discussion}
We conclude by discussing the hierarchy of scales in TBG samples, the relevance of our results to TBG experiment, our relationship to other approaches, and promising future directions.

Studying TBG by projecting into the flat band subspace is, in our view, harmonious with the most exciting and unique aspect of the system: its realization of strongly interacting topology in a time-reversal invariant and beyond Landau level context. We also believe it is justified in reality, given the hierarchy of energy scales as we discuss now. To justify a projective treatment, we have taken the interaction scale $U$ to be much smaller than the gap to the remote bands $\Delta$. Whereas $\Delta$ is a band structure property, $U$ is sensitive to the dielectric environment which receives contributions both from the hBN substrate and the remote bands. The scale $U$ is thus difficult to determine precisely from first principles. On the other hand, since $U$ is the largest interacting scale in the problem, it should be possible to reliably extract it from experiment. There are several ways to do this. One is to identify the separation of the main peaks of the density of states at charge neutrality with $2U$. This is justified close to the magic angle and for not-too-large strain; for larger strain the separation is $2(U + E_{\rm str})$, which still yields an upper bound on $U$. This estimate does not depend on symmetry breaking except through subleading anisotropies.
The separation seen in various experiments \cite{xieSpectroscopicSignaturesManybody2019a, YazdaniCascade, Nuckolls_2023, EvaAndreiSTM2019, STMNadjPerge, BandFlattening} is consistently between 30 and 40 meV, leading to an estimate of $U \approx 15-20$ meV.  
Another estimate is obtained by the requirement to reproduce \cite{VafekCorrelatedChern} the measured change in chemical potential between $\nu = 0$ and $4$ \cite{IlaniCascade,saitoIsospinPomeranchukEffect2021} which leads to 
$U \approx 20$ meV (assuming our parameters). 
A third independent estimate can be obtained from recent nanoSQUID quantum oscillations in alternating twist trilayer graphene \cite{bocarsly2024imagingcoulombinteractionsmigrating}. The Fock scale and $K_M$-point gap scale obtained in that work are both $\sim20$ meV. Taking into account that all scales in trilayer are enhanced by a factor of $\sqrt{2}$~\cite{khalafMagicAngleHierarchy2019}, this gives a value $U \approx 15$ meV. Thus, rather remarkably, all these experiments across different devices yield a consistent estimate of $U \approx 15-20$ meV.

Phenomenologically determining the gap to the remote bands, $\Delta$, is more tricky. However, since $\Delta$ does not feature explicitly in our approach, we only need to justify the hierarchy $U \ll \Delta$ on phenomenological grounds. This can be addressed by noting that it is very difficult to reconcile $U \gtrsim \Delta/2$ with an insulating gap at full filling. To see this, we note that the last flat band electron added before $\nu = 4$ will be away from $\Gamma$ and feel a strong repulsion $2\nu U = 8U$. While half the remote bands are also AA localized and will feel the same repulsion, the other half have wavefunctions similar to the $\Gamma$ point flat bands \cite{song2022TBGTHF} which are not AA-localized; the latter feel a reduced repulsion. The reduction in repulsion is exactly the $\nu=4$ Hartree bandwidth, and can be estimated as $8U - 2\nu(U-U_\Gamma) \approx 2U$ for Coulomb interactions using Eq.~(\ref{eq:coulombscales_U_Ugam}). The non-interacting gap of $\Delta$ is therefore reduced by $\approx 2U$ at full filling which means that the gap is eliminated entirely if $U \gtrsim \Delta/2$. 
Taking a $\nu = 4$ gap of 15 meV from SET measurements \cite{IlaniCascade} yields $\Delta \approx 40 - 55$ meV, consistent with our hierarchy of scales $U \lesssim \Delta$. 
 
Although we derived the parameters of our model in Sec.~\ref{sec:ModelIntro} by fitting to the BM model, one of the advantages of our minimal approach is that it depends only on a few phenomenological parameters that can be extracted from experiments. In fact, once the hierarchy of scales $h_b \ll U \ll \Delta$ is established, we only need to determine the parameters $\delta$, $s$, $U$, in addition to the strain scale to explain most of the intermediate temperature physics. The interaction scale $U$ can be estimated from experiment as explained above. The AA-localization scale $\delta$ can be extracted by fitting the width of the charge distribution extracted from STM data~\cite{liObservationVanHove2010a,kerelskyMaximizedElectronInteractions2019,CorrelatedChernEvaAndrei,xieSpectroscopicSignaturesManybody2019a,YazdaniCascade,CorrelatedChernYazdani,Nuckolls_2023}. The strain scale, if present, can be extracted by identifying the temperature at which entropy gets quenched at charge neutrality. The parameter $s$ is more difficult to estimate at present, but will be obtainable once momentum-resolved probes like ARPES~\cite{lisiObservationFlatBands2021,satoObservationFlatBand2021,nunnARPESSignaturesFewLayer2023} and the quantum twisting microscope \cite{QTM} are sufficiently developed.

It is worth comparing and contrasting our approach to the topological heavy fermion (THF) approach \cite{song2022TBGTHF,calugaru2023TBGTHF2,yu2023TTGTHF,HuRKKY,herzogarbeitman2024topologicalheavyfermionprinciple, TBGKondo, THFKondo, TBGKondoSDS, SongLian, lau2024topologicalmixedvalencemodel, ValentiDMFT}. In this approach, the system is described using two flavors of fermions: the localized $f$ electrons that support local moments and are described by Hubbard physics, and the mobile $c$ electrons that are introduced through mixing with dispersing remote bands and support topology. We note that while the THF model can be studied in the projected limit, $U \ll \Delta$ \cite{herzogarbeitman2024topologicalheavyfermionprinciple}, it has been predominantly motivated by and studied in the opposite limit $U \gg \Delta$ \cite{song2022TBGTHF} where local moment states exist in the low-energy Hilbert space. However, as we have shown here, explicit local moments are not required to explain the extensive entropy seen in experiment. Furthermore, calculations done in the limit $U \gg \Delta$ do not yield a gapped band insulator at $\nu = -4$ \cite{HuRKKY}, consistent with the arguments given above.\footnote{Note that our definition of $U$ is smaller by a factor of $2$ compared to that of the THF papers; in our convention they quote $U \approx 30$meV~\cite{song2022TBGTHF}. In our convention, $U$ is half the Mott gap and $\Delta$ is half the hybridization gap, while in the THF papers $U$ is the full Mott gap.}We note that our argument does not preclude a gap at intermediate values $U \sim \Delta$, which can only be determined using unbiased numerics.
Finally, while both approaches are motivated by concentration of topology at the $\Gamma$ point, our approach formalizes this notion by introducing a small parameter $s$ and demonstrating that controlled analytic calculations of many-body observables can be achieved using phase space arguments. 

In particular, our approach has enabled the analytical calculation of electron spectral functions in TBG as reported here. Our analytic calculation for the spectral function explains some of the features observed numerically in DMFT for the THF model. For example, Fig. 1a of Ref. \cite{ValentiDMFT} looks qualitatively similar to our charge neutrality Mott semimetal with BM dispersion (Fig. \ref{fig:semimottal}c), and is generally consistent with our claim of a quadratic band touching at zero energy with vanishing residue; Figs. 1b,c of the same reference appear similar to our Fig.~\ref{fig:imbalanced_spec} at finite doping with BM dispersion by a particle-hole transformation. Furthermore, our spectral function for the Mott semimetal at neutrality agrees qualitatively with results obtained with QMC in the band-projected limit~\cite{hofmannFermionicMonteCarlo2022,MengQMCSemimetal}. A  quantitative comparison between our theory and numerical calculations would be valuable and is left to future work. It is also worth noting that some of the insights from our calculations may help uncover an analytically tractable regime of the THF model. In the THF model language, our $s\ll1$ projected limit amounts to taking large $c$-electron Fermi velocity $v_F k_\theta \gg U, \Delta$ in the projected limit $U \ll \Delta$. We anticipate large $v_F$, with $v_F k_\theta \gg U, \Delta$, will also yield analytical progress for the THF hierarchy of scales $U \gg \Delta$. This ensures that the $c$-electron phase space is parametrically small and one can use small phase space approximations similar to those that we have developed here. These approximations could be used to control the strength of RKKY interactions that couple the local $f$ moments~\cite{HuRKKY}, for example. We also anticipate $U/\Delta$ corrections to the projected limit to be tractable within $s \ll 1$. Such corrections could capture effects beyond any projected model, such as filling dependence of the active low-energy states.

We reiterate that our projected-limit spectral functions do not admit an interpretation in terms of local moments weakly coupled to a bath of conduction electrons. The presence of local moments imply a nearly $\bk$-independent branch of the spectral function; this is inconsistent with the lack of spectral weight at $(\bk = \Gamma, \omega = \pm U)$ in almost all of the spectra we obtain. Similarly, a weakly interacting bath of conduction electrons cannot explain the dispersion of our spectra near $\Gamma$ nor their response to perturbations.

We now discuss an outlook towards future experiments. A direct observation of the spectral function of the Mott semimetal and the spectrally imbalanced Mott state is a clear goal. This is in principle possible with momentum-resolved techniques under active development \cite{lisiObservationFlatBands2021,satoObservationFlatBand2021,nunnARPESSignaturesFewLayer2023,QTM}. We have also predicted a Pomeranchuk effect for the spectral gap, especially at $\abs{\nu} = 1$ when strain $E_{\rm str} \ll U$ is not too large. Further measurements on entropy over a wide temperature range as a function of in-plane field direction would would provide a further test of our results. Indeed, because the in-plane field couples orbitally in TBG with a strength that is comparable to the Zeeman splitting, we expect the residual degeneracy to depend on whether the in-plane field is parallel or perpendicular to the axis of strain. Trilayer entropy measurements as a function of in-plane field would also be useful, since here the orbital effect can be neglected \cite{ledwith2021tbtbcontrastingproperties, MacdonaldInPlane, LakeSenthilPairing, LakeSenthilTTG}.

There are also a number of theoretical calculations that now appear within reach. For example, the intervalley Kekulé spiral (IKS) state \cite{Nuckolls_2023,liuNematicTopologicalSemimetal2021,parkerStrainInducedQuantumPhase2021,soejimaEfficientSimulationMoire2020,KwanKekule,WangIKSDMRG,wang2024cherntexturedexcitoninsulatorsvalley,kwan2024texturedexcitoninsulators} first obtained within Hartree-Fock~\cite{KwanKekule} and later verified in DMRG~\cite{WangIKSDMRG}, has thus far evaded strong coupling arguments that would analytically establish it as the ground state in the presence of strain. We are optimistic that our $s^2 \ll 1$ phase space approximations will both enable such a treatment and shed light on some puzzling features of the state, including the nearly flat landscape of spiral wavevectors. We also speculate that a theory of  superconducting pairing could be analytically controlled with $s \ll 1$. Initial steps could be to calculate the Drude weight for the $T\gg Us^2$ normal state, or the total flat-band optical spectral weight\cite{mendezvalderrama2023theorylowenergyopticalsumrule}; the latter is an upper bound for the superconducting stiffness \cite{mao2023upperboundssuperconductingexcitonic,MaoDiamagneticResponse,Verma_2021}.

More broadly, we propose that our techniques lay the groundwork for a new analytical approach towards low-temperature TBG. We have worked in the hierarchy of scales $U \gg T \gg Us^2 \sim 1$meV. There are several other effects we have neglected such as asymmetric form factors, dispersion-induced superexchange, and phonons; all are on the $1$meV scale. While the zero temperature symmetry breaking pattern depends sensitively on these anisotropies \cite{BultinckHidden,kwanElectronphononCouplingCompeting2024}, for us their effect is suppressed by temperature. 
Beginning with the symmetric state and incorporating the various $\sim 1$meV effects perturbatively relative to temperature will of course lead to more accurate $T>0$ spectral functions (e.g. through incorporating nonzero broadening, or splitting from asymmetric form factors\cite{bultinckMechanismAnomalousHall2020}). However, the process of initially ignoring them, and then gradually incorporating them perturbatively relative to temperature, could also yield a more universal perspective on the doped state that is obscured by doping a very particular zero temperature insulator. 

Furthermore our findings invite some deeper conceptual questions. In this work, we have described thermal Mott states where charge is mostly frozen but the nonlocal moments are disordered. A natural conceptual question is to ask about states where nonlocal moments are quantum disordered at zero temperature. For the $\nu \neq 0$ states, it seems likely that the state will be a gapped spin-liquid, and that the main signature of the non-interacting topology will be the $Z_\bk \to 0$ imbalance in the spectral function. For $\nu = 0$, however, the situation is more interesting. Let us consider a simpler case of a single $C=1$ spinful band for concreteness and a chiral symmetry that pins the filling at zero (which can be regarded as a combination of particle-hole and time reversal). Then the system may be regarded as the surface of a three-dimensional class AIII topological insulator at level $k=2$. It was shown that $\SU(2)$ symmetric surface states in this class do not admit gapped ground states, even in the presence of topological order \cite{WangSenthilAIII}. A quantum disordered version of our Mott semimetal, with a spectral function equivalent to \emph{a single} Dirac cone, then seems like a natural possibility. Note that it has half the spectral weight of the minimal non-interacting surface state.

In closing, we point out potential generalizations of the model we have presented here beyond the $|C| = 1$ bands relevant to twisted bilayer graphene. For example, what types of Chern bands host decoupled nonlocal moments when $|C|>1$? We anticipate that the concentration of topology will still be an essential feature, but for $|C|>1$ we anticipate a wider range of possibilities with distinct physics and Mott spectra. Finally, an exciting question raised by recent experiments \cite{holleis2024isospinpomeranchukeffectfinite} which observed enhanced entropy and isospin fluctuations in rhombohedral graphene, is whether the concepts presented here can be extended to flat band systems even in the {\em absence} of a moiré lattice.

\section*{Acknowledgements}

We thank Qingchen Li, Nisarga Paul, Henry Shackleton and Ophelia Sommer for related collaborations. We acknowledge Nick Bultinck, Debanjan Chowdhury, Piers Coleman, Leonid Glazman, Hart Goldman, Sid Parameswaran, Dan Parker, T. Senthil, Alex Thomson, and Oskar Vafek for fruitful discussions.
This research was supported in part by grant NSF PHY-2309135 to the Kavli Institute for Theoretical Physics (KITP). A.V. is supported
by a Simons Investigator award, the Simons Collaboration on Ultra-Quantum Matter,
which is a grant from the Simons Foundation (651440,
A.V.)  and by NSF-DMR
2220703.  

\bibliographystyle{unsrt}
\bibliography{references}

\appendix
\renewcommand{\thefigure}{\thesection\arabic{figure}}
\setcounter{figure}{0}

\section{Form factors and Hartree Fock Dispersion}\label{app:formfactors_hartreefock}

In this section we compute the form factors $\Lambda_\bq(\bk)$ and Hartree Fock dispersion. It suffices to calculate the form factor associated with the $C=+1$ bands, $\lambda_\bq(\bk)$, because the form factors, as well as the Hartree and Fock dispersions, are diagonal in Chern sector with the two Chern sectors related by $C_2 \T$. For notational simplicity we will drop the Chern sector label $\gamma = +$ throughout this section. Using that we are in a periodic gauge $\psi_\bk = \psi_{\bk + \bG}$, such that the periodic wavefunction satisfies $u_{\bk+\bG} = e^{- i \bG \cdot \br} u_\bk$, we have
\begin{equation}
\begin{aligned}
& \braket{u_{\bk} | u_{\bk + \bq}}  = \bra{u_{\bk}} e^{-i \bG \cdot \br} \ket{ u_{[\bk + \bq]}} \\
  & =  \frac{1}{N_\bk N_{\bk+\bq}}\int_\uc d^2 \br \left(1 + i \frac{\ov{z}/\delta}{\ov{k}/s} \right) \left(1 - i \frac{z/\delta}{[k+q]/s} \right) \\
  & \qquad \qquad \qquad \times e^{-i \bG \cdot \br} \frac{e^{-\frac{\abs{z}^2}{2 \delta ^2}}}{2\pi \delta^2}
  \end{aligned}
    \end{equation}
where
\begin{equation}
    N_\bk = \frac{1}{\sqrt{1+\frac{2s^2}{\abs{k}^2}}}
\end{equation}
is a normalization factor. We have used the notation $[k+q]$ to denote the part of $k+q = [k+q] + G$ in the first BZ, and also take $k = [k]$ and $z$ in the first unit cell. Since the integral is dominated by $\abs{z} \sim \delta$, we can remove the integration bound restriction to the first unit cell and integrate over the whole plane, as is standard in saddle point approximations. Furthermore we can replace $e^{-i \bG \cdot \br} \approx e^{-i \bq \cdot \br}$: the difference $\bq - \bG$ is BZ scale and exponential $e^{-i \bq \cdot \br}$ is only sensitive to momenta $\sim \delta^{-1}$ much larger than the BZ scale. The resulting Gaussian integral can be explicitly evaluated, though the calculation is made slighlty easier with the trick of replacing factors of $z$ and $\ov{z}$ with derivatives with respect to $\ov{q}$ and $q$ respectively:
\begin{widetext}
\begin{equation}
  \begin{aligned}
\braket{u_{\bk} | u_{\bk + \bq}} 
 & = \frac{1}{\sqrt{1+\frac{2s^2}{\abs{k}^2}} \sqrt{1+\frac{2s^2}{\abs{[k+q]}^2}}  }\int_\uc d^2 \br \left(1 + i \frac{\ov{z}/\delta}{\ov{k}/s} \right) \left(1 - i \frac{z/\delta}{[k+q]/s} \right) e^{-i \bq \cdot \br} \frac{e^{-\frac{\abs{z}^2}{2 \delta ^2}}}{2\pi \delta^2} \\
 & = \frac{1}{\sqrt{1+\frac{2s^2}{\abs{k}^2}} \sqrt{1+\frac{2s^2}{\abs{[k+q]}^2}}  }\left(1 -  \frac{2 \delta^{-1}\partial_q}{\ov{k}/s} \right) \left(1 +  \frac{2 \delta^{-1} \partial_{\ov{q}}}{[k+q]/s} \right) \int_\uc d^2 \br e^{-\frac{i}{2}(\ov{q}z + q \ov{z})} \frac{e^{-\frac{\abs{z}^2}{2 \delta ^2}}}{2\pi \delta^2}  \\
     & = \frac{1}{\sqrt{1+\frac{2s^2}{\abs{k}^2}} \sqrt{1+\frac{2s^2}{\abs{[k+q]}^2}}  }\left(1 -  \frac{2 \delta^{-1}\partial_q}{\ov{k}/s} \right) \left(1 +  \frac{2 \delta^{-1} \partial_{\ov{q}}}{[k+q]/s} \right) e^{-q^2 \delta^2/2} \\
     & = \frac{\left(1 +  \frac{\ov{q} \delta}{\ov{k}/s} \right) \left(1 -  \frac{q \delta}{[k+q]/s} \right) + \frac{2 s^2}{\ov{k}[k + q]}}{ \sqrt{1+\frac{2s^2}{\abs{k}^2}} \sqrt{1+\frac{2s^2}{\abs{[k+q]}^2}}}e^{-\abs{q}^2 \delta^2/2}
\end{aligned} 
\label{eq:formfactor_calculation_supp}
\end{equation}
\end{widetext}

The Fock dispersion is given by
\begin{equation}
\begin{aligned}
  h_F(\bk) & = \frac{1}{2A} \sum_\bq V_\bq \abs{\lambda_\bq(\bk)}^2 \\
  & = \frac{1}{2} \int \frac{d^2 \bq}{(2\pi)^2} V_\bq \abs{ \frac{1 + \frac{\ov{q} \delta}{\ov{k}/s}}{\sqrt{1 + \frac{2 s^2}{\abs{\bk}^2}}}}^2 e^{-\abs{q}^2 \delta^2}
  \end{aligned}
  \label{eq:fockstart}
\end{equation}
where we used that while $\bk$ may be near $\Gamma$, for most of the $\bq$-integral $\abs{[\bk + \bq]} \gg s$. Expanding out the suare and dropping terms that vanish after the $\bq$ integral by circular symmetry, we obtain
\begin{equation}
  \begin{aligned}
  h_F(\bk)  & =  \frac{1}{2} \int \frac{d^2 \bq}{(2\pi)^2} V_\bq  \frac{ \abs{\bk}^2 + \abs{\bq}^2 \delta^2 s^2 }{\abs{\bk}^2 + 2 s^2} e^{-\abs{q}^2 \delta^2} \\
    & = \frac{1}{2} \int \frac{d^2 \bq}{(2\pi)^2} V_\bq  \frac{ \abs{\bk}^2 + \abs{\bq}^2 \delta^2 s^2 }{\abs{\bk}^2 + 2 s^2} e^{-\abs{q}^2 \delta^2}\\
      & = U \frac{ \abs{\bk}^2}{\abs{\bk}^2 + 2 s^2}  + U_\Gamma \frac{2 s^2}{\abs{\bk}^2 + 2 s^2}
  \end{aligned}
  \label{eq:fockstart_fin}
\end{equation}
where 
\begin{equation}
  U  = \frac{1}{2} \int \frac{d^2 \bq}{(2\pi)^2} V_\bq e^{-\abs{\bq}^2 \delta^2}
  \label{eq:hubbardU_supp}
\end{equation}
and
\begin{equation}
  \begin{aligned}
    U_\Gamma & = \frac{1}{2} \int \frac{d^2 \bq}{(2\pi)^2} V_\bq \frac{1}{2}\abs{\bq}^2 \delta^2 e^{-\abs{\bq}^2 \delta^2}.
\end{aligned}
  \label{eq:hubbardUGamma_supp}
\end{equation}

The Hartree dispersion is
\begin{equation}
  \begin{aligned}
    h_H(\bk) & = \frac{1}{ A_{\uc}} \sum_{\bG} V_\bG \lambda_\bG(\bk) \xi_{-\bG} \\
      & = \frac{1}{A_\uc} \sum_\bG V_\bG \frac{\left(1 +  \frac{\ov{G} \delta}{\ov{k}/s} \right) \left(1 -  \frac{G \delta}{k/s} \right) + \frac{2 s^2}{\abs{k}^2}}{ 1+\frac{2s^2}{\abs{k}^2} }e^{-\abs{G}^2 \delta^2} \\
      & = \int \frac{d^2 \bq}{(2\pi)^2} V_\bq \frac{\abs{\bk}^2 + 2 s^2  - \abs{\bq}^2 \delta^2 s^2}{ \bk^2 + 2 s^2} e^{-\abs{q}^2 \delta^2}\\
      & = 2U \frac{\bk^2}{\bk^2 + 2 s^2} + 2(U-U_\Gamma) \frac{2 s^2}{\bk^2 + 2s^2}
\end{aligned}
  \label{hartree}
\end{equation}
where we used the fact that the characteristic scale of the summand is $\bG \sim \delta^{-1} \gg \ell^{-1}$, such that we can replace the sum by an integral (and relabel $\bG \to \bq$).

\section{Single particle terms}\label{app:singleparticle}
In addition to the form factors and Hartree and Fock dispersions derived in the previous section, in the main text we have occasionally referenced the non-interacting BM dispersion as well as the strain dispersion. In this section we derive these expressions. We note that in this section, since we will be interfacing with the microscopic Hamiltonian, we will \emph{not} adopt the $\ell=1$ units in which $A_\uc = (2\pi)^2/A_\bz = 2\pi \ell^2 \to 2\pi$ until the end. The model parameters $\delta$ and $s = \beta^{-1} \ell^{-2} \delta$ then have units of length and inverse length respectively.

\subsection{BM dispersion}\label{app:singleparticle_BM}

We begin with the BM dispersion. We note that the BM Hamiltonian has very small bare bandwidth at the magic angle, even away from the chiral limit. However, the bandwidth increases quickly upon deviating slightly from the magic angle, such that $h_{\rm BM} \propto (\theta - \theta_M)$, where $\theta_M$ is the magic angle. We will therefore approximate the BM dispersion through this linearization in the vicinity of the magic angle. The BM Hamiltonian in the $K$ valley of graphene is given by
\begin{equation}
    H_{\BM}(\theta) = \begin{pmatrix} -i v \bsigma \cdot \bnabla & T(\br,k_\theta) \\ T^\dag(\br,k_\theta) & -i v \bsigma \cdot \bnabla \end{pmatrix},
    \label{eq:BMHamiltonian}
\end{equation}
where we neglected the twist angle dependence of the kinetic terms (their effect goes as $\theta^2 \ll \theta$). The Hamiltonian in the $K'$ graphene valley is related by time reversal. The $\theta$ dependence enters in the moiré scale modulation of the tunneling terms through the wavevector $k_\theta = 2 k_D \sin \frac{\theta}{2} \approx k_D \theta$ where $k_D$ is the distance between the grpahene $K$-points (side length of the graphene BZ hexagon). Explicitly we have
\begin{equation}
    T(\br,k_\theta) = w_{\rm AB} \begin{pmatrix} \kappa U_0(\br,k_\theta) & U_{-1}(\br,k_\theta) \\ U_1(\br,k_\theta) & U_0(\br,k_\theta) \end{pmatrix},
\end{equation}
where $U_l(\br,k_\theta) = \sum_{n=0}^2 e^{\frac{2\pi i}{3} l n} e^{-i\boldsymbol{q}_n \cdot \br}$ and $\bq_n = \bq_n(k_\theta)$ is given by $q_{n,x} + i q_{n,y} = -i k_\theta e^{2\pi i n/3}$. It is convenient to use the fact that $T(\br,k_\theta) = T(\br k_\theta,1) = \tilde{T}(\tilde{\br})$ only depends on the combination $\br k_\theta$ to re-scale \eqref{eq:BMHamiltonian} so that it depends on $\theta$ more explicitly and not through the unit cell size. We obtain
\begin{equation}
\begin{aligned}
    H_{\BM}(\theta) & = \begin{pmatrix} -i v k_\theta \bsigma \cdot \tilde{\bnabla} & \tilde{T}(\tilde{\br}) \\ \tilde{T}(\tilde{\br}) & -i v k_\theta \bsigma \cdot \tilde{\bnabla} \end{pmatrix} \\
    & = H_{\BM}(\theta_M) + (\theta - \theta_M)k_D I^{\rm layer}_{2 \times 2} (-i \bsigma \cdot \tilde{\bnabla}),
    \end{aligned}
    \label{eq:BMHamiltoniansplit}
\end{equation}
where we have separated the ``magic angle" Hamiltonian $H_{\BM}(\theta_M)$ from part corresponding to the deviation from the magic angle, where the second part has a simple, layer-independent, from. We now wish to project \eqref{eq:BMHamiltoniansplit} onto our explicit model wavefunctions \eqref{eq:fullWF}. We neglect the first term by hypothesis (the BM dispersion is known to be small at the magic angle) but compute the second term explicitly. Undoing the scale transformation leads to $\frac{(\theta - \theta_M)}{\theta}  (-i \bsigma \cdot \bnabla)$ such that we must evaluate
\begin{equation}
   h_{\BM}^{\gamma \gamma'}(\bk) = \frac{\theta - \theta_M}{\theta} v \bra{\psi_\bk^\gamma}  (-i \bsigma \cdot \bnabla) \ket{\psi_\bk^{\gamma'}}.
   \label{eq:bmdisp}
\end{equation}
We recall that our wavefunctions are sublattice polarized (we have neglected the corresponding off-diagonal part of the form factor) and the Chern sector is given by $\gamma_z = \sigma_z \tau_z$, where $\sigma_z$ is the sublattice matrix within each valley and $\tau_z = \pm 1 = K,K'$ is the graphene valley and we are focusing on the $K$ valley. Thus, in \eqref{eq:bmdisp} we may identify $\gamma$ and $\sigma$. Moreover, $-i \bsigma \cdot \bnabla$ anticommutes with $\sigma_z$ while the wavefunctions are eigenstates of $\sigma_z$: this observation implies that the diagonal terms, with $\gamma = \gamma'$, vanish. We can then focus on $\gamma'=-\gamma = +$ and obtain $\gamma=-\gamma=-$ through $C_2 \T$, which acts as $\sigma_x K$ and $\br \to -\br$. Since $\bra{\sigma = -} (-i\bsigma \cdot \bnabla) \ket{\sigma +} = -2i \overline{\partial}$, where $\overline{\partial} = \frac{1}{2}( \partial_x + i \partial_y$) is the derivative with respect to $\overline{z}$, we obtain
\begin{equation}
\begin{aligned}
    & \bra{\psi_\bk^-}  (-i \bsigma \cdot \bnabla) \ket{\psi_\bk^{+}} \\
    & = \frac{1}{N_\bk^2}\int_{\uc} \frac{d^2 \br}{2\pi \delta^2} \ov{\left(1 + \frac{\ov{z}/\delta}{i \ov{k}/s}\right)}\left(1 + \frac{z/\delta}{i k/s}\right) \\
    \,\, & \times
    e^{-\frac{\abs{\br}^2}{4 \delta^2}} \chi_0^\dag (-2i \ov{\partial})\left[e^{-\frac{\abs{\br}^2}{4 \delta^2}} \chi_0(\br) \right].
    \end{aligned}
\end{equation}
Here we plugged in the wavefunctions \eqref{eq:fullWF} with $\psi^-_\bk(\br) = \overline{\psi_{\bk}^+(-\br)}$ by $C_2 \T$. We also used that we can move the derivative $\ov{\partial}$ past $z$. The above expression is sufficient to extract the $\bk$ dependence, up to an overall coefficient with a layer-spinor dependent integral that can be in principle evaluated. Note that, for this calculation, we have not replaced $\chi_0(\br) \to \chi_0(\br=0)$ as in the main text because otherwise the integral above vanishes. In practice the overall scale of the BM dispersion can be matched to numerics on the BM model or symmetry-preserving generalizations, so we will focus on extracting the $\bk$-dependence. For this purpose it is sufficient to use $C_3$ symmetry under which $z$ and $\ov{\partial}$ have angular momentum one and the other parts of the wavefunction have angular momentum zero. The only terms that survives the $C_3$ symmetric integral over the UC is then the angular momentum $3 \equiv 0$ part with two factors of $z$. We then obtain
\begin{equation}
\begin{aligned}
    h^{-+}_\BM & = E_{\BM} \frac{k^{-2}}{1 + 2s^2 \abs{\bk}^{-2}} \\ 
    E_\BM & = \frac{\theta - \theta_M}{\theta}  \int_\uc \frac{d^2 \br}{2\pi \delta^2} \frac{z^2}{\delta^2} e^{-\frac{\abs{\br}^2}{4 \delta^2}} \chi_0^\dag (-2iv \ov{\partial})\left[e^{-\frac{\abs{\br}^2}{4 \delta^2}} \chi_0(\br) \right].
\end{aligned}
\end{equation}
After using $h^{+-}(\bk) = \overline{h^{-+}(\bk)}$ by $C_2 \T$ we arrive at the expression quoted in the main text.

\subsection{Strain dispersion}\label{app:singleparticle_strain}
We now move on to the strain dispersion. To leading order, strain couples to the BM Hamiltonian \eqref{eq:BMHamiltonian} as $-i \bnabla \to -i\bnabla - \mu_z \bA $ where $\mu_z = \pm$ labels the top and bottom graphene layer and $\bA = \frac{\beta \sqrt{3}}{2a} (\epsilon_{xx} - \epsilon_{yy}, -2 \epsilon_{xy})$\cite{manesSymmetrybasedApproachElectronphonon2007,guineaGaugeFieldInduced2008,kimGrapheneElectronicMembrane2008,pereiraStrainEngineeringGraphene2009,vozmedianoGaugeFieldsGraphene2010} where the tensor $\epsilon$ is the dimensionless strain tensor, $a$ is the graphene lattice constant, and $\beta \approx 3.14$ is a dimensionless parameter associated with the graphene hopping as a function of distance\cite{bi_designing_2019,namLatticeRelaxationEnergy2017,huderElectronicSpectrumTwisted2018}. We note that there is a layer dependent distortion matrix as well, described in other works\cite{bi_designing_2019,soejimaEfficientSimulationMoire2020,parkerStrainInducedQuantumPhase2021}, which captures the energetic effects associated with distorting the moiré unit cell. Its effect is of order $\theta \epsilon \ll \epsilon$ or $\epsilon^2$, and so we regard it as subleading here. It may be included in a very similar calculation, however.

It follows that we must project $v\mu_z \bsigma \cdot \bA$ onto the wavefunctions \eqref{eq:fullWF}. As with the BM dispersion, only the off-diagonal in Chern sector terms are nonzero, and we may begin with the $\gamma' = - \gamma = +$ component.
\begin{equation}
\begin{aligned}
    h^{+-}_{\rm strain} & = (\bsigma \cdot \bA)_{+-} \chi_0^\dag \mu_z \chi_0 \frac{1}{N_\bk^2} \\
    & \times \int_\uc d^2 \br \ov{\left(1 + \frac{\ov{z}/\delta}{i \ov{k}/s}\right)}\left(1 + \frac{z/\delta}{i k/s}\right) \frac{e^{-\frac{\abs{\br}^2}{2\delta^2}}}{2\pi \delta^2}
    \end{aligned}
\end{equation}
Here, the terms proportional to $z$ vanish under the unit cell integral and we obtain
\begin{equation}
    h_{\rm strain} = v \bA \cdot \v{\gamma} (\chi_0^\dag \mu_z) \chi_0 \frac{\abs{\bk}^2}{\abs{\bk}^2 + 2s^2}.
\end{equation}
where we used time reversal symmetry to deduce that the final result is the same in both valleys, and thus replaced $\bsigma \to \v{\gamma}$. This is the same form we quoted in the main text.

Note that the orbital coupling of an in-plane magnetic field in TBG has the same form as strain in the $K$ valley, but is odd under time reversal symmetry such that it couples oppositely to strain in the $K'$ valley. We therefore obtain the same result except with an extra factor of $\tau_z$ that accounts for the valley sign change. This orbital coupling can generally not be ignored in TBG, since the g-factor is roughly the same as Zeeman. However, in alternating twist trilayer graphene, an in-plane field is odd under the $M_z$ symmetry that exchanges the top and bottom layer; as a result it does not directly project to the active flat bands and the orbital coupling dominates.

\section{Wannier State Asymptotic Forms}
\label{app:wannier}
In this section we calculate Wannier state wavefunctions analytially for the various regimes: $\bR = 0$, $\abs{\bR} \ll s^{-1}$, and $\abs{\bR} \gg s^{-1}$. The Wannier states are related by translation, $w_\bR^\gamma(\br) =  w_{\bR = 0}^\gamma(\br - \bR)$, so we will focus on the Wannier state centered at $\bR = 0$. Additionally, time reversal implies that we may choose $w_\bR^-(\br) = \overline{w_\bR^+(\br)}$, so we will focus only on $\gamma = +$ and drop the $\gamma$ label for the rest of the section. Finally since $\psi_\bk(\br)$ is exponentially suppressed away from AA sites, we may evaluate it in the neighborhood of each AA site and use the Bloch periodicity $\psi_\bk(\br \approx \bR) = e^{i \bk \cdot \bR} \psi_\bk(\br \approx 0)$. We therefore have
\begin{equation}
  \begin{aligned}
    w_{\bR=0}(\br) & = \int \frac{d^2 \bk}{2\pi} \psi_\bk(\br) \\
    & = \sum_\bR \int \frac{d^2 \bk}{2\pi}  e^{i \bk \cdot \bR} \frac{1 + \frac{(z - R)/\delta}{i k/s} }{\sqrt{1 + \frac{2s^2}{\abs{k}^2}}} \frac{e^{-\frac{\abs{z-R}^2}{4 \delta^2}}}{\sqrt{2\pi \delta^2}} \chi_0 \\
    & = \sum_{\bR} \left(c_1(\bR) + c_z(\bR) \frac{z-R}{ \sqrt{2} \delta } \right) \frac{e^{-\frac{\abs{z-R}^2}{4 s^2}}}{\sqrt{2\pi \delta^2}} \chi_0
  \end{aligned}
  \label{eq:wan_supp}
\end{equation}
where 
\begin{equation}
  \begin{aligned}
  c_1(\bR) & = \int \frac{d^2 \bk}{2\pi} \frac{e^{i \bk \cdot \bR}}{\sqrt{1 + \frac{2s^2}{\abs{k}^2}}} \\
  c_z(\bR) & = \int \frac{d^2 \bk}{2\pi} \frac{\sqrt{2}s}{ik}\frac{e^{i \bk \cdot \bR}}{\sqrt{1 + \frac{2s^2}{\abs{k}^2}}} 
\end{aligned}
\label{coef1z_defn_supp}
\end{equation}
Note that we have used $\sum_\bk = A_\uc \int \frac{d^2 \bk}{(2\pi)^2}$ with $A_\uc = 2\pi \ell^2 = 2\pi$. For $\bR = 0$ we have $c_z = 0$ by symmetry whereas
\begin{equation}
  \begin{aligned}
    c_1(\bR = 0) & = 1 - \int \frac{d^2 \bk}{2\pi} \left( 1 - \frac{\abs{k}}{\sqrt{\abs{k}^2 + 2 s^2}} \right) \\
    & = 1 - 2 s^2 \log s^{-1},
  \end{aligned}
  \label{eq:requalzerointegral}
\end{equation}
where for the first term we used that the area of the Brillouin Zone is $2\pi$. For $\bR \neq 0$, the integrals are all dominated by the $k \approx 0$ region and have a circular symmetry. We can perform the integral over $\theta_\bk$, leading to Bessel functions $J_{0,1}(u)$ for $u = \abs{\bk}\abs{\bR}$. 
\begin{equation}
  \begin{aligned}
    c_1(\bR \neq 0)  & = \frac{1}{\abs{\bR}^2}\int_{0}^\infty du  \frac{u^2 J_0(u)}{\sqrt{u^2 + 2 s^2 \abs{\bR}^2}}, \\
    & = -2s^2 \int_{0}^\infty du  \frac{u J_1(u)}{(u^2 + 2 s^2 \abs{\bR}^2)^{3/2}}, \\
    c_z(\bR) & = \frac{\sqrt{2} s e^{-i \theta_{\bR}}}{\abs{\bR}} \int_0^\infty du \frac{u J_1(u)}{\sqrt{u^2 + 2 s^2 \abs{\bR}^2}}.
\end{aligned}
  \label{eq:besselform}
\end{equation}
To go from the first to second line above we integrated by parts and used $\frac{d}{du}(u J_1(u)) = u J_0(u)$. We can now evaluate the asymptotoics of $c_1$ and $c_z$ by replacing $(u^2 + 2 s^2 \abs{\bR}^2)$ with $u^2$ and $s^2 \abs{\bR}^2$, for $\abs{\bR} \ll s^{-1}$ and $\abs{\bR} \gg s^{-1}$ respectively. The resulting Bessel function integrals can mostly be computed straightforwardly through integration by parts, $\frac{d}{du} J_0(u) = - J_1(u)$, and $\int_0^\infty J_m(u) du = 1$. For $c_1(\abs{\bR} \ll s^{-1})$, however, this procedure leads to $-2s^2\int J_1(u)/u^2$. This integral is logarithmically divergent as $u \to 0$ because $J_1(u) \approx u/2$ for small $u$. The divergence is cut off at $u \sim s \abs{\bR} \ll 1$ leading to $-s^2 \log (s \abs{\bR})^{-1}$. We therefore conclude

\begin{equation}
  \begin{aligned}
  c_1(\bR) & = \begin{cases} 1-2 s^2 \log s^{-1} , \quad R=0 \\ -s^2 \log (s\abs{\bR})^{-1}, \quad \abs{\bR} \ll s^{-1} \\ -\frac{1}{\sqrt{2} s \abs{\bR}^3}, \quad R \gg s^{-1} \end{cases} \\
  c_z(\bR) & = e^{-i \theta_\bR} \begin{cases} \frac{\sqrt{2} s }{\abs{\bR}}, \quad \abs{\bR} \ll s^{-1} \\  \frac{1}{\abs{\bR}^2}, \quad \abs{\bR} \gg s^{-1} \end{cases}
  \end{aligned}
\label{coef1z_asymp_supp}
\end{equation}

The near complete weight of the Wannier state at its center implies that the density operator is nearly-diagonal in the Wannier basis, since the Wannier state overlap is parametrically small. An explicit calculation yields
\begin{equation}
  \begin{aligned}
    \bra{\psi_{\bR}} & e^{-i \bq \cdot \br} \ket{\psi_{\bR'}} = e^{i \bq \cdot \bR'} \int \frac{d^2 \bk}{2\pi} \braket{u_k|u_{k+q}} e^{i \bk \cdot (\bR'-\bR)}\\
    & = e^{-\abs{q}^2 \delta^2/2}
    \begin{cases} 
    e^{i \bq \cdot \bR} + O(s^2) \text{ for } \bR = \bR' \\
    (c_1(\bR-\bR') - \frac{i q \delta}{\sqrt{2}})e^{i \bq \cdot \bR}\\
    + (c.c, \bR \leftrightarrow \bR', \bq \leftrightarrow -\bq) \text{ for } \bR \neq \bR'
    \end{cases}
\end{aligned}
  \label{eq:fourierformfact_supp}
\end{equation}
such that off diagonal elements are suppressed by $O(s^2)$ relative to diagonal ones. We will now use this fact to bound the entropy at any filling.

\section{Entropy Bound Calculations}\label{app:entropy}

In this section we describe the calculations associated with the entropy bounds \eqref{eq:entropybound} and \eqref{eq:strainbound}. We begin by explicitly computing the exchange energy scales without strain, Eq. \eqref{eq:JexMainText}, and with strain, Eq. \eqref{eq:Jex_strained_maintext}. Later we will address the ground state energy lower bounds \eqref{eq:JHmaintext}, \eqref{eq:Jstr_Egslowerbound_main}.

\subsection{Exchange Energy Scale}

We copy the exchange Hamiltonian below for readability
\begin{equation}
\begin{aligned}
    \H_{\rm ex} & = \frac{1}{2A} \sum_\bq V_\bq \sum_{\bR \neq \bR'} c^\dag_\bR \Lambda_\bq^{\bR \bR'} c_{\bR'} c^\dag_{\bR'}  \Lambda_{-\bq}^{\bR' \bR} c_{\bR} \\
    \Lambda_\bq^{\bR \bR'} & = \langle w_\bR| e^{-i \bq \cdot \br}| w_{\bR'} \rangle  =  \frac{e^{-i \bq \cdot \bR'}}{N} \sum_\bk e^{i \bk \cdot (\bR - \bR')} \Lambda_\bq(\bk).
\end{aligned}
\end{equation}
Before taking expectation values, we use the standard exchange-interaction procedure of grouping the fermion operators at the same site together and using the Fierz identity for $\SU(8)$ generators $r^\mu_{\alpha \beta}$, where $\mu$ labels the generator and $\alpha,\beta$ labels the band index. The Fierz identity reads $r^\mu_{\alpha \beta} r^\mu_{\gamma \delta} + \frac{1}{8} \delta_{\alpha \beta} \delta_{\gamma \delta} = \delta_{\beta \gamma} \delta_{\alpha \delta} $, where the generators satisfy $\tr r^\mu r^\nu = \delta^{\mu \nu}$. To obtain interactions in terms of the flavor operators $R^\mu(\bR) = c^\dag_{\bR} r^\mu c_{\bR}$. We obtain
\begin{equation}
\begin{aligned}
    \H_{\rm ex} & = \frac{1}{2A} \sum_{\bq}V_\bq \sum_{\bR \neq \bR'} \abs{\lambda_\bq^{\bR,\bR'}}^2 n_\bR(1-\frac{n_{\bR'}}{8}) \\
    & \,\,- \tr(r^\mu \Lambda_\bq^{\bR \bR'} r^\nu \Lambda^{\dag \bR \bR'}_\bq) R^\mu(\bR) R^\nu(\bR').\\
    \label{eq:ExHam_afterfierz_supp}
\end{aligned}
\end{equation}
Note that we cannot set $\mu = \nu$ because $\Lambda$ is not the identity matrix (it is different between the Chern sectors for $\bR \neq \bR'$.)

\subsubsection{Strain-free}

We now compute the zero-strain $N J_{\rm ex} \Tr \H_{\rm ex} \rho_0 = \langle\H_{\rm ex} \rangle_{\rho_0}$ where we recall $\rho_0$ is the equal weight density matrix of states $\ket{\Psi}$ that satisfy $n_\bR \ket{\Psi} = (n+\nu) \ket{\Psi}$. Products of such flavor operators at distinct sites have zero expectation value under $\rho_0$ because $\rho_0$ is symmetric under \emph{local} $\SU(8)$ rotations. Thus, the last line in \eqref{eq:ExHam_afterfierz_supp} has zero expectation value. We therefore have
\begin{equation}
\begin{aligned}
    J_{\rm ex} = \langle \H_{\rm ex} \rangle_{\rho_0} & =\frac{N}{8}(4-\frac{\nu^2}{4}) \sum_{\bR\neq0} J(\bR)
\end{aligned}
\end{equation}
where we have identified
\begin{equation}
    J(\bR\neq0) = \frac{2}{A} \sum_\bq V_\bq \abs{\lambda_\bq^{\bR,0}}^2 
\end{equation}
as the $\SU(8)$ symmetric part of the exchange coupling at distance $\bR$ (we will counter other parts in the strain calculation).  We can further evaluate, using the Plancherel theorem
\begin{equation}
\begin{aligned}
    J_{\rm ex} & = \frac{1}{8}(4-\frac{\nu^2}{4})\sum_{\bR \neq 0} J(\bR) \\
    & =\frac{1}{4N}(4-\frac{\nu^2}{4})\sum_{\bk,\bq} V_\bq |\lambda_\bq(\bk) - \xi_\bq|^2,
\end{aligned}
\end{equation}
which is the general form that we quoted in \eqref{eq:JexMainText}. Note we have not yet had to assume $s \ll 1$. 

For $s \ll 1$ we can show $J_{\rm ex} \propto s^2$ up to a log. To see this, note that 
there are only two order $s^2$ areas of phase space, $\bk \approx \Gamma$ and $\bk + \bq \approx \Gamma$, where $\lambda_\bq(\bk)$ is appreciably distinct from $\xi_\bq$. We can explicitly evaluate the expression by evaluating $\lambda_\bq(\bk)$ in these two areas, using \eqref{eq:formfactor_calculation_supp}, and integrating.
\begin{equation}
\begin{aligned}
& \frac{1}{4N}\sum_{\bk,\bq} V_\bq |\lambda_\bq(\bk) - \xi_\bq|^2 \\
    & = \frac{1}{4NA}\sum_{\bq,\bk}V_\bq \xi_\bq^2 \left( \abs{\frac{1+\frac{\ov{q}\delta}{\ov{k}/s}}{\sqrt{1 + \frac{2s^2}{ \abs{k}^2}} } - 1}^2 + \abs{\frac{1-\frac{q\delta}{([k+q])/s}}{\sqrt{1 + \frac{2s^2}{ \abs{[k+q]}^2}} } - 1}^2 \right) \\
    & = \frac{1}{2NA}\sum_{\bq,\bk}V_\bq \xi_\bq^2  \abs{\frac{1+\frac{\ov{q}\delta}{\ov{k}/s}}{\sqrt{1 + \frac{2s^2}{ \abs{k}^2}} } - 1}^2  \\
    & = 2 s^2 \log s^{-1} \frac{1}{A}\sum_\bq V_\bq \frac{1}{2} \abs{\bq}^2 \delta^2 e^{-\abs{\bq^2} \delta^2} \\
    & = 2 U_\Gamma s^2 \log s^{-1},
    \end{aligned}
    \label{eq:Jstareval}
\end{equation}
Between the second and third line above we did the change of variables $[k+q] \to k$ in the summation over $k$ (done before the integration over $\bq$). We also dropped terms subleading to $\log s^{-1}$, and used the circular symmetry to drop terms that would vanish under the $\bq$ integration. We therefore conclude
\begin{equation}
    J_{\rm ex} = 2(4-\frac{\nu^2}{4}) U_\Gamma s^2 \log s^{-1},
\end{equation}
as claimed in the main text \eqref{eq:JexMainText}.

\subsubsection{With Strain}
We now derive the strainful case \eqref{eq:Jex_strained_maintext}. Again beginning with \eqref{eq:ExHam_afterfierz_supp}, we take the expectation value in the density matrix $\rho_0$ where now $\rho_0$ is an equal weight density matrix of states $\ket{\Psi}$ that satisfy $n_\bR \ket{\Psi} = (4 + \nu) \ket{\Psi}$ as well as $c^\dag_\bR \gamma_x c_\bR \ket{\Psi} = (4 - \abs{\nu})\ket{\Psi}$. 

The expectation value is more complicated now because some of the terms in the second line of \eqref{eq:ExHam_afterfierz_supp}, involving $r^\mu \propto \gamma_x$, have nonzero expectation value in $\rho$ (the first line is equivalent to the strain-free case). We therefore split the $\SU(8)$ generator index $\mu \to (i,a)$ and write $r^\mu = \frac{\gamma^i}{\sqrt{2}} t^a$. Here $\gamma^i$ are the $2 \times 2$ Chern-sector Pauli matrices, while $t^a$ are the generators of $\SU(4)$ rotations in spin-valley space. 

We must first evaluate the expectation value, for $\bR \neq \bR'$. It is zero unless $i = j$ and $a=b=0$ where $t^0 = \frac{1}{\sqrt{4}}I_{4\times4}$ by the symmetry of $\rho_0$. For these indices we have
\begin{equation}
    \langle R^{x 0}(\bR) R^{x 0}(\bR') \rangle_{\rho_0} = \frac{1}{2} \frac{1}{4}\left(4 - \abs{\nu}\right)^2
\end{equation}
Next we need to evaluate the trace $\tr(r^{i a} \Lambda_\bq^{\bR \bR'} r^{j b} \Lambda^{\dag \bR \bR'}_\bq)$, where we can replace $i, j \to x$ and $a,b \to 0$ since otherwise the expectation value vanishes. We obtain
\begin{equation}
    \tr(r^{x 0} \Lambda_\bq^{\bR \bR'} r^{x 0} \Lambda^{\dag \bR \bR'}_\bq) = \Re (\lambda^{\bR \bR'})^2,
\end{equation}
where we used that $\Lambda = (\lambda I_{4\times 4}, \overline{\lambda} I_{4\times4})$. If the form factors were $\U(8)$ symmetric we would have obtained the modulus squared instead of the real part as in the strain-free case. We identify the inter-chern exchange coupling
\begin{equation}
    J(\bR\neq0) = \frac{2}{A} \sum_\bq V_\bq \Re(\lambda_\bq^{\bR,0}\lambda_\bq^{0,\bR})^2.
\end{equation}

Putting these sub-computations together, and computing the first term in $\H_{\rm ex}$ identically to the strain free case, we have
\begin{equation}
\begin{aligned}
    J_{\rm ex} & = \frac{1}{N} \langle \H_{\rm ex} \rangle_{\rho_0}  \\
    & = \frac{1}{8}(4 - \frac{\nu^2}{4}) \sum_{\bR \neq 0} J(\bR) \\
    & - \frac{1}{8} \frac{1}{4}(4 - \abs{\nu})^2 \sum_{\bR \neq 0} \tilde{J}(\bR) \\
    & = 4(J_s - \tilde{J}_s) + 2 \abs{\nu} \tilde{J}_s - \frac{\nu^2}{4}(J_s + \tilde{J}_s),
\end{aligned}
\end{equation}
where
\begin{equation}
\begin{aligned}
    J_s & = \frac{1}{8} \sum_{\bR \neq 0} J(\bR) = \frac{1}{4NA} \sum_{\bk,\bq} \abs{\lambda_\bq(\bk) - \xi_\bq}^2, \\
    \tilde{J}_s & = \frac{1}{8} \sum_{\bR \neq 0} \tilde{J}(\bR)  = \frac{1}{4NA} \sum_{\bk \bq} \Re (\lambda_\bq(\bk) - \xi_\bq)^2.
\end{aligned}
\end{equation}
We have already evaluated $J_s$ in \eqref{eq:Jstareval}. We can similarly integrate
\begin{equation}
    \begin{aligned}
\tilde{J}_s & = \frac{1}{4NA}\sum_{\bq,\bk}V_\bq \xi_\bq^2  \\
& \times \Re \left( \left[\frac{1+\frac{\ov{q}\delta}{\ov{k}/s}}{\sqrt{1 + \frac{2s^2}{ \abs{k}^2}} } - 1\right]^2 + \left[\frac{1-\frac{q\delta}{([k+q])/s}}{\sqrt{1 + \frac{2s^2}{ \abs{[k+q]}^2}} } - 1\right]^2 \right) \\
& = \frac{1}{2NA} \sum_{\bq ,\bk} V_\bq \xi_\bq^2 \left( \frac{1}{\sqrt{1 + \frac{2s^2}{|k|^2}}}  - 1\right)^2 \\
& = \frac{1}{2A} \sum_{\bq} V_\bq \xi_\bq^2 (\log(4) - 1) s^2\\
& = (\log(4) - 1) U s^2 
    \end{aligned}
\end{equation}
where we redefined $\bk$ in the second big bracket term to map it to the first and used that the integration over $q$ leads to many terms dropping by circular symmetry.

\subsection{Ground State Energy Lower Bounds}

We now discuss the lower bounds of ground state energy \eqref{eq:JHmaintext} \eqref{eq:Jstr_Egslowerbound_main}. It suffices to prove the latter strainful bound, since it reduces to the former for $E_{\rm str} \to 0$. 

For readability we restate the interacting Hamiltonian in the presence of strain
\begin{equation}
\begin{aligned}
    \H  & = \sum_{\bk} c^\dag_\bk h_{\rm{str}}(\bk) c_\bk + \frac{1}{2A}\sum_\bq V_\bq \delta \rho_\bq \delta \rho_{-\bq}\\
    \delta \rho_\bq & = \rho_\bq - 4 \sum_\bG \delta_{\bq,\bG} \xi_\bG \\
    \rho_\bq & = \sum_\bk c^\dag_\bk \Lambda_\bq(\bk) c_{\bk + \bq}
    \qquad \xi_\bq=\frac{1}{N}\sum_\bk \Lambda_{\bq}(\bk).
\end{aligned}
\end{equation}

For the purpose of lower bounding the ground state energy, relative to $E_{\rm cl}$ we will rewrite the Hamiltonian by separating out the reciprocal lattice vector wavevectors from the $\bq$ sum, since they alone contribute to the $E_{\rm cl}$ at integer filling. Among the reciprocal lattice vector density operators, we will also separate out the part corresponding to the classical energy by splitting the form factor $\Lambda_\bG(\bk) = \xi_\bG + (\Lambda_\bG(\bk) - \xi_\bG)$
and
\begin{equation}
    \tilde{\rho}_\bG  = \sum_{\bk} c^\dag_{\bk \alpha} (\Lambda_\bG(\bk) - \xi_\bG) c_{\bk}.
\end{equation}
Finally, we will find it convenient to work in a space with fixed \emph{total} particle number $N \nu$ so that we can replace $\sum_\bk (c^\dag_\bk c_\bk - 4) \to N_\nu$ in various Hamiltonian terms. For example
\begin{equation}
    \delta \rho_\bG = \sum_\bk \xi_\bG (c^\dag_\bk  c_\bk - 4) + \tilde{\rho}_\bG \to N \nu + \tilde{\rho}_\bG.
\end{equation}

With this splitting, we obtain
\begin{equation}
\begin{aligned}
    \H & = \frac{1}{2A} \sum_{[\bq] \neq 0} V_\bq \rho_\bq \rho_{-\bq} \\
    & + \frac{N}{2A_{\rm uc}} \sum_\bG V_\bG \abs{\xi_\bG}^2 \nu^2  \\
    & + \tilde{H}_{\rm H} \\
    & + \frac{1}{2A} \sum_\bG V_\bG \tilde{\rho}_{\bG}\tilde{\rho}_{-\bG}, \\
    & + \sum_{\bR} h_{\rm str}^{\rm av} c^\dag_\bR \gamma_x c_\bR, \\
    & + \tilde{H}_{\rm str}.
    \end{aligned}
    \label{eq:Hlowerboundsupp}
\end{equation}
We define and discuss the terms in \eqref{eq:Hlowerboundsupp} line by line. In the first term, which is positive semidefinite, we sum over all $\bq$ that are not reciprocal lattice vectors. The terms in the sum that did include reciprocal lattice vectors were split between the second, third, and fourth lines. The second term is exactly the interaction contribution to $E_{\rm cl}$,
\begin{equation}
    \frac{N}{2A_{\rm uc}} \sum_\bG V_\bG \abs{\xi_\bG}^2 \nu^2 = \sum_{\bR \bR'} U(\bR - \bR')\nu^2,
\end{equation}
where we recall $U(\bR-\bR') = \frac{1}{2A} \sum_\bq V_\bq \abs{\xi_\bq}^2 e^{i \bq \cdot (\bR - \bR')}$.
The third term contains the cross terms and can be expressed in terms of the dispersive part of Hartree
\begin{equation}
\begin{aligned}
    \tilde{H}_{\rm H} & = \frac{\nu}{A_{\uc}} \sum_\bG V_\bG  \xi_{\bG} \tilde{\rho}_{-\bG} \\
    & = \frac{\nu}{A_\uc} \sum_\bG V_\bG \xi_\bG \sum_{\bk} c^\dag_{\bk} (\Lambda_\bG(\bk) - \xi_\bG) c_{\bk}, \\
    & = \nu \sum_\bk (h_H(\bk) - h_H^{\rm{av}})  (c^\dag_{\bk} c_{\bk} - 4),
\end{aligned}
\end{equation}
and the fourth term is positive semidefinite. The fifth term is the classical part of the strain Hamiltonian, while the sixth is the dispersive part of strain
\begin{equation}
\begin{aligned}
    \tilde{H}_{\rm str} & = \sum_{\bk} (h_{\rm str}(\bk) - h_{\rm str}^{\rm av} ) c^\dag_\bk  \gamma_x c_\bk
\end{aligned}
\end{equation}

We are now ready to bound the ground state energy from below. Let us take the expectation value of \eqref{eq:Hlowerboundsupp} in the ground state. Then we have
\begin{equation}
    E_\GS = \bra{\Psi_\GS} \H \ket{\Psi_\GS} \geq E_{\rm cl} + \bra{\Psi_\GS} \tilde{H}_{\rm H} + \tilde{H}_{\rm str} \ket{\Psi_\GS}.
    \label{eq:groundstatelowerbound}
\end{equation}
Here we used that the second and fifth terms in \eqref{eq:Hlowerboundsupp} are equal and minimized respectively by their respective contributions to $E_{\rm cl}$, while the first and fourth terms are positive semidefinite. 

It suffices to lower bound the expectation value on the right hand side of \eqref{eq:groundstatelowerbound}. We begin with the Hartree term
\begin{equation}
\begin{aligned}
    \langle \tilde{H}_{\rm H} \rangle & = \nu \sum_\bk (h_H(\bk) - h_H^{\rm{avg}})  \langle c^\dag_{\bk \alpha} c_{\bk \alpha} - 4 \rangle  \\
    & \geq - 4 \abs{\nu} \sum_\bk \abs{h_H(\bk) - h_H^{\rm{avg}}} = -NJ_H.
    \end{aligned}
\end{equation}
Similarly
\begin{equation}
\begin{aligned}
    \langle \tilde{H}_{\rm str} \rangle & = \sum_{\bk} (h_{\rm str}(\bk) - h_{\rm str}^{\rm av} )\langle c^\dag_\bk  \gamma_x c_\bk \rangle \\
    & \geq -4 \sum_{\bk} \abs{(h_{\rm str}(\bk) - h_{\rm str}^{\rm av} ) } = -N J_{\rm str}.
\end{aligned}
\end{equation}
Combining the above two bounds with \eqref{eq:groundstatelowerbound} we obtain \eqref{eq:Jstr_Egslowerbound_main} as claimed.

In the small $s$ limit we have the following expressions for $J_H$ and $J_{\rm str}$ that we quoted in the main text:
\begin{equation}
\begin{aligned}
    J_H & =  4\abs{\nu} \int \frac{d^2 \bk}{2\pi} \abs{2U - \frac{2 U_\Gamma}{\abs{\bk}^2 + 2 s^2} - 2U} \\
    & = 8  \abs{\nu} U_\Gamma \int_0^{\Lambda \sim 1} \frac{\abs{\bk} d \abs{\bk}}{1 + \frac{\abs{\bk}^2}{2 s^2}} \\
    & = 16  \abs{\nu} U_\Gamma  s^2 \log s^{-1},
    \end{aligned}
\end{equation}
and similarly
\begin{equation}
    J_{\rm str} =  4 E_{\rm str} \int_0^{\Lambda \sim 1} \frac{\abs{\bk} d \abs{\bk}}{1 + \frac{\abs{\bk}^2}{2 s^2}} = 8 E_{\rm str} s^2 \log s^{-1}.
\end{equation}

\section{Schwinger-Dyson Calculation of Spectral Function}\label{app:SchwingerDyson}

In this section we show how 1PI correlation functions, such as the self energy, can be calculated non-perturbatively (i.e. without resorting to ``all orders in perturbation theory in $U$'' arguments given in the main text) through the Schwinger-Dyson equations. In particular we will prove
\begin{equation}
  \tilde{\Sigma}(\bk,i\omega_n) = -\frac{1}{\beta^2A^2}\sum_{\bq m \bq'm'} \Lambda_{\bk,\bk-\bq'} \Xi_{\bq',\bq}(\bk)\Lambda_{\bk-\bq,\bk},
    \label{eq:exactselfenergy_supp}
\end{equation}
as argued for in the main text diagrammatically in the text surrounding \eqref{eq:exactselfenergy}. Also explained in the main text diagrammatically is the approximation $\Xi_{\bq',\bq}(\bk) \to \Xi_{\bq',\bq}(*)$ for $T \gg U s^2$ and the prescription that $\Xi_{\bq', \bq}(*)$ can be evaluated in real space in a classical, topologically trivial, model \eqref{eq:classicalmodel}. In this appendix, we will also explain this approximation and prescription from the non-perturbative Schwinger-Dyson perspective.

\subsection{Effective Action and Schwinger Dyson Formalism}

We begin by reviewing the quantum effective action approach and deriving the Schwinger-Dyson equations that we will use. The formulas in this subsection are more or less common to any fermionic system interacting via density-density interactions. We will reserve the usage and explanation of $s \ll 1$ approximations for the next subsection. We begin by restating the partition functions and coupling it to source fields.
\begin{equation}
\begin{aligned}
    Z[\eta,\ov{\eta}, J] & = \int D\ov{\psi} D \psi D \phi e^{-S[\ov{\psi},\psi,\phi] - S_{\rm src}}  \\
    S & = \sum_{\bk, n} \ov{\psi}_{\bk, n} G_0^{-1}(\bk, i \omega_n) \psi_{\bk, n} \\
    & + \frac{i}{\beta A} \sum_{\bq, n,m} \phi_{\bq, m}  \delta^\nu \rho_{\bq m} \\
    & + \frac{1}{2\beta A} \sum_{\bq, n} \phi_{\bq, n} V^{-1}(\bq) \phi_{-\bq,-n} \\
    \delta^\nu \rho_{\bq m} & =  \sum_\bk \ov{\psi}_{\bk,n} \Lambda_{\bq}(\bk) \psi_{\bk+\bq,m+n} - (4+\nu)\delta_{[\bq],0}\delta_{m,0}\xi_\bq \\
     S_{\rm src} & = \sum_{\bk,n}\ov{\eta}_{\bk, n} \psi_{\bk, n} + \ov{\psi}_{\bk,n} \eta_{\bk,n} + \sum_{\bq,m} J_{\bq,m} \phi_{\bq,m}  \\
    \end{aligned}
    \label{eq:partitionfn_withsources_supp}
\end{equation}

We additionally use the quantum effective action $\Gamma[\ov{\psi},\psi,\phi]$, also referred to as the 1PI effective action as its power series expansion coefficients correspond to the irredicible vertices. In other words, $\Gamma$ can be defined as the action under which every correlation function can be evaluated at tree level; its terms can in principle be computed by evaluating such correlation functions through other means. It is also commonly defined, non-perturbatively, as the Legendre transform of the generator of connected correlation functions $W[\eta,\ov{\eta},J]$, where $e^{i W} = Z$ is the partition function \eqref{eq:partitionfn_withsources_supp}. Generally, the quantum effective action contains infinitely many terms and is not local in space or time. However, it may be expanded in powers of the fields, and since it is exact at tree level only a few terms (irreducible vertices) enter any given correlation function. The first few terms are
\begin{equation}
\begin{aligned}
    \Gamma & = \sum_{\bk,n} \ov{\psi}_{\bk n} G^{-1}(\bk, i \omega_n) \psi_{\bk n}\\
    & + \frac{1}{2\beta A} \sum_\bq \phi_{-\bq,-m} D^{-1}(\bq, i\omega_m) \phi_{\bq,m} \\
    & + \frac{i}{\beta A} \sum_{\bk \bq m n} \phi_{\bq,m} \\
    & \,\,\, \times \left( \ov{\psi}_{\bk,n} \Gamma^{(3), \bk,n}_{\bq m} \psi_{\bk + \bq, m+n} - (4+\nu) \delta_{[\bq],0}\delta_{m,0} \xi_\bq \right) \\
    & - \frac{1}{3! \beta^2 A^2 } \sum_{\bq m \bq'm' \bq''m''} \tilde{\Gamma}^{(3)}_{\bq m \bq' m' \bq'' m''} \\
    & \,\,\, \times\phi_{\bq m} \phi_{\bq' m'}  \phi_{\bq'' m''} \delta_{[\bq + \bq' + \bq''],0} \delta_{m+m'+m'',0} \\
    & - \frac{1}{2 \beta^2 A^2} \sum_{\bk \bq \bq' m m' n}  \phi_{-\bq',-m'}\phi_{\bq m} \\
    & \quad \times\left(\ov{\psi}_{\bk - \bq',n-m'}  \Gamma^{(4), \bk,n}_{\bq m \bq' m'} \psi_{\bk - \bq n-m} - (4+ \nu) \delta_{[\bq],[\bq']} \xi_\bq \right) \\
    & + \cdots .
\end{aligned}
\label{eq:schematiceffaction_supp}
\end{equation}

Of course, the coefficients of the effective action are generally unknown and, generically, too complex to determine by symmetry. They are in principle determined through the Schwinger Dyson equations (SDEs). The SDEs are an infinite hierarchy that relate lower order correlations to successively higher order correlation functions (we will review their derivation soon). When the correlation functions in the SDEs are evaluated using the effective action $\Gamma$, one obtains related SDEs which relate lower order vertices to successively higher order vertices. Typically, this hierarchy is still impossible to solve, though here we will find it amenable to phase space approximations for $s \ll 1$ as described in the next subsection.

We begin by deriving the SD equations that we need. Shift $\ov{\psi} \to \ov{\psi} + \ov{\varepsilon}$ in the path integral, take a derivative with respect to $\ov{\varepsilon}$, and set $\ov{\varepsilon} \to 0$. Since the partition function does not change under such a change of variables we have
\begin{equation}
    0 = \int D\ov{\psi} D \psi D \phi \left(\frac{\delta}{\delta \ov{\epsilon}_{\bk,n}} S[\ov{\psi}+ \ov{\varepsilon},\psi , \phi] + \eta_{\bk, n} \right) e^{-S- S_{\rm{src}}}
\end{equation}
Using
\begin{equation}
\begin{aligned}
    & \frac{\delta}{\delta \ov{\varepsilon}_{\bk,n}} S[\ov{\psi} + \ov{\varepsilon},\psi, \phi] \\
    & =  G_0^{-1}(\bk,i \omega_n) \psi_{\bk,n} + \frac{i}{\beta A}\sum_{\bq,m} \phi_{\bq,m} \Lambda^\dag_{\bq}(\bk) \psi_{\bk+\bq,m+n},
    \end{aligned}
\end{equation}
and the fact that we can replace factors of $\psi$ and $\phi$ with derivatives with respect to $\ov{\eta}$ and $J$ respectively, we arrive at
\begin{equation}
\begin{aligned}
    \bigg( &  G_0^{-1}(\bk)_{\alpha \beta} \frac{\delta}{\delta \ov{\eta}_{\bk,n}^\beta}\\
    & + \frac{i}{\beta A} \sum_{\bq,m}  \Lambda_\bq(\bk)_{\alpha \beta} \frac{\delta^2}{\delta J_{\bq,m} \delta \ov{\eta}_{\bk+\bq,m+n}^\beta} \\
    & + \eta_{\bk,n}^\alpha  \bigg) Z[\eta,\ov \eta,J] = 0
    \end{aligned}
    \label{eq:SDstart_supp_diffbar}
\end{equation}

We could have also varied $\psi$ instead of $\ov{\psi}$; following the same steps leads to
\begin{equation}
\begin{aligned}
    \bigg(&  - \frac{\delta}{\delta \eta_{\bk,n}}  G_0^{-1}(\bk) \\   & -\frac{i}{\beta A}\sum_{\bq,m}   \frac{\delta^2}{\delta J_{\bq,m} \delta \eta_{\bk-\bq,n-m}}\Lambda_\bq(\bk-\bq) \\
    & + \ov{\eta}_{\bk,n}  \bigg) Z[\eta,\ov \eta,J] = 0
    \end{aligned}
    \label{eq:SDstart_supp_diff}
\end{equation}
where we suppressed the matrix-vector multiplications in the flavor indicies for concision.

Now we differentiate \eqref{eq:SDstart_supp_diff} respect to $\ov{\eta}_{\bk,n}$ and set all the sources to zero. We then obtain
\begin{equation}
\begin{aligned}
    \delta_{\alpha \beta} & = [G(\bk,i \omega_n)G_0^{-1}(\bk,i \omega_n)]_{\alpha \beta} \\
    & -  \frac{i}{\beta A}\sum_{\bq,m}\langle \phi_{\bq,m} \psi_{\bk,n}^\alpha [\ov{\psi}_{\bk-\bq,n-m} \Lambda_{\bq}(\bk-\bq) ]^{\beta}\rangle 
    \end{aligned}
    \label{eq:gapeqn}
\end{equation}
where $G_{\alpha \beta}(\bk, i\omega_n) = -\langle \psi_{\bk, n, \alpha} \ov{\psi}_{\bk, n, \beta} \rangle$ is the full Green's function. The expression $\psi \ov{\psi}$ will, from now on, refer to the outer product in flavor indices $\psi_\alpha \ov{\psi}_\beta$ as indicated explicitly above.  We could have also differentiated \eqref{eq:SDstart_supp_diffbar} instead, which exchanges the arguments of $\psi$ and $\ov{\psi}$ and switches $\Lambda$ to the right side. This yields two expressions for the self energy \begin{equation}
\begin{aligned}
    G^{-1} & = G_0^{-1} - \tilde{\Sigma} \\
    \tilde{\Sigma}_{\bk,n} & = \frac{i}{\beta A} \sum_{\bq m} G^{-1}_{\bk,n}\langle  \psi_{\bk,n} \ov{\psi}_{\bk-\bq,n-m}  \phi_{\bq,m}\rangle \Lambda_{\bq}(\bk-\bq)  \\
    & = \frac{i}{\beta A} \sum_{\bq m} \Lambda_{-\bq}(\bk)\langle \phi_{-\bq,-m} \psi_{\bk-\bq,n-m} \ov{\psi}_{\bk,n}  \rangle G^{-1}_{\bk,n}.  \\
    \end{aligned}
\end{equation}
The correlation functions can then be evaluated using the 1PI effective action $\Gamma$, and for the above correlation functions there is only a single diagram with vertex $\Gamma^{(3)}$. We then obtain two equivalent expressions for the self energy, which may be represented diagrammatically. 
\begin{equation}
    \begin{aligned}
        \tilde{\Sigma}_{\bk,n} & = -\frac{1}{\beta A}\sum_{\bq m} D_{\bq,m} \Gamma^{(3)}_{-\bq -m}(\bk,i\omega_n) G_{\bk - \bq,n} \Lambda_{\bq}(\bk - \bq) \\
        & = \vcenter{\hbox{\begin{tikzpicture}
  \begin{feynman}[layered layout,every blob={/tikz/fill=gray!30,/tikz/inner sep=2pt}]
    \vertex (i) at (0, 0) {};
    \vertex [right=0.7of i] (qi) [dot] {};
    \vertex [right=2cm of qi] (qf) [blob] {$\Gamma^{(3)}$};
    \vertex [right=1cm of qf] (f) {};

    \diagram* {
      (i) -- [fermion, edge label=\(k\)] (qi)
      -- [fermion, style = ultra thick, edge label=\(k - q\)] (qf)
      -- [fermion, edge label=\(k\)] (f),

      (qi) -- [photon,style=ultra thick, out=90, in=90, looseness=1, edge label=\(q\)] (qf),
    };
  \end{feynman}
  \end{tikzpicture}}} \\
        & = -\frac{1}{\beta A}\sum_{\bq m} D_{\bq,m} \Lambda_{-\bq}(\bk) G_{\bk - \bq,n} \Gamma^{(3)}_{\bq m}(\bk-\bq,i\omega_n) \\
        & = \vcenter{\hbox{\begin{tikzpicture}
  \begin{feynman}[layered layout, every blob={/tikz/fill=gray!30,/tikz/inner sep=2pt}]
    \vertex (i) at (0, 0) {};
    \vertex [right=1cmof i] (qi) [blob] {$\Gamma^{(3)}$};
    \vertex [right=2cm of qi] (qf) [dot] {};
    \vertex [right=0.7cm of qf] (f) {};

    \diagram* {
      (i) -- [fermion, edge label=\(k\)] (qi)
      -- [fermion, style = ultra thick, edge label=\(k - q\)] (qf)
      -- [fermion, edge label=\(k\)] (f),

      (qi) -- [photon,style=ultra thick, out=90, in=90, looseness=1, edge label=\(q\)] (qf),
    };
  \end{feynman}
  \end{tikzpicture}}}
    \end{aligned}
    \label{eq:two_gap_eqns}
\end{equation}
We will primarily use the first expression for the self-energy, for concreteness. But \eqref{eq:SDstart_supp_diffbar} will be useful in what follows since we will want to shift the argument of $\ov{\psi}$ away from the external momentum $\bk$, which can be close to $\Gamma$.

We must now solve for $\Gamma^{(3)}$; the above expressions are not yet suitable for the $s \ll 1$ approximation since the $\Gamma^{(3)}$ vertices above each have a fermion line at $\bk$, which we do not want to assume is away from $\Gamma$. We therefore need to go one higher order in the hierarchy and relate the three point function to the four point function. We differentiate \eqref{eq:SDstart_supp_diffbar} with respect to both $\eta_{\bk-\bq,n-m}$ and $J_{\bq,m}$, such that the first term becomes $G_0^{-1}$ multiplied by the three-point function in \eqref{eq:gapeqn}. We also set the sources to zero, and relabel the summand $\bq \to -\bq'$ so that $\bq$ is reserved for the momentum in the three-point function. We then obtain
\begin{equation}
\begin{aligned}
    &-G_0^{-1}(\bk,i \omega_n)\langle \phi_{\bq,m} \psi_{\bk,n} \ov{\psi}_{\bk - \bq, n-m}  \rangle  \\
    & +  \frac{i}{\beta A}\sum_{\bq',m'}\Lambda_{-\bq'}(\bk) C_{\bq \bq'}(\bk)  \\
    & = 0
    \end{aligned}
    \label{eq:threeptSDE_supp}
\end{equation}
where 
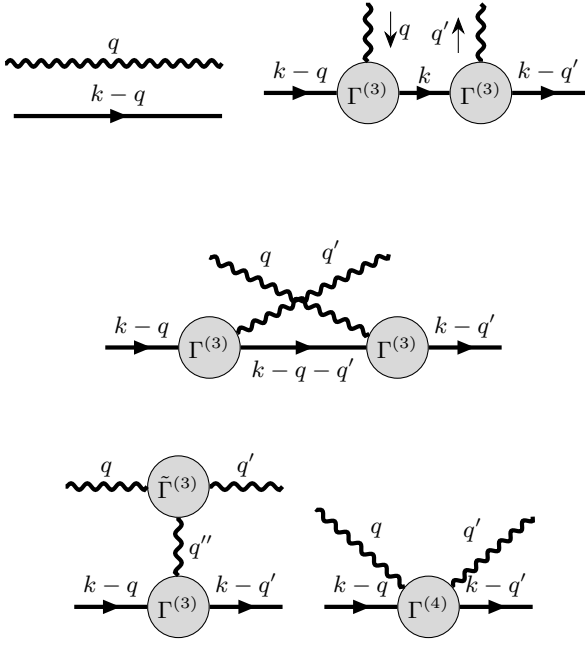
\begin{figure}
    \centering

\begin{tikzpicture}
  \begin{feynman}[layered layout, every blob = {/tikz/fill=gray!30,/tikz/inner sep=2pt}]
    \vertex (i) at (0, 0) {};
    \vertex [right=3 cm of i] (f) {};
    \vertex [above=0.7cm of i] (qi);
    \vertex [above=0.7cm of f] (qf) {};
    
    \diagram*{
      (i) -- [fermion, style = ultra thick, edge label=\(k-q\)] (f),

      (qi) -- [photon, ultra thick,  edge label=\(q\)] (qf),
    };
  \end{feynman}
\end{tikzpicture}    
\vspace{0.7cm}
\begin{tikzpicture}
  \begin{feynman}[layered layout, every blob = {/tikz/fill=gray!30,/tikz/inner sep=2pt}]
    \vertex (i) at (0, 0) {};
    \vertex [right=1.5cm of i] (qf) [blob] {\(\Gamma^{(3)}\) };
    \vertex [right=1.5cm of qf] (qpf) [blob] {$\Gamma^{(3)}$};
    \vertex [right=1.5cm of qpf] (f) {};
    \vertex [above=1.2cm of qf] (qi);
    \vertex [above=1.3cm of qpf] (qpi) {};
    
    \diagram* {
      (i) -- [fermion, style = ultra thick, edge label=\(k-q\)] (qf)
      -- [fermion, style = ultra thick, edge label=\(k\)] (qpf)
      -- [fermion, style = ultra thick, edge label=\(k - q'\)] (f),

      (qi) -- [photon, ultra thick,  momentum=\(q\)] (qf),
      (qpf) -- [photon, ultra thick, momentum=\(q'\)] (qpi),
    };
  \end{feynman}
\end{tikzpicture}
\vspace{0.7cm}

\begin{tikzpicture}
  \begin{feynman}[layered layout, every blob = {/tikz/fill=gray!30,/tikz/inner sep=2pt}]
    \vertex (i) at (0, 0) {};
    \vertex [right=1.5cm of i] (qf) [blob] {\(\Gamma^{(3)}\) };
    \vertex [right=2.5cm of qf] (qpf) [blob] {$\Gamma^{(3)}$};
    \vertex [right=1.5cm of qpf] (f) {};
    \vertex [above=1.2cm of qf] (qi);
    \vertex [above=1.3cm of qpf] (qpi) {};
    
    \diagram* {
      (i) -- [fermion, style = ultra thick, edge label=\(k-q\)] (qf)
      -- [fermion, style = ultra thick, edge label'=\(k-q-q'\)] (qpf)
      -- [fermion, style = ultra thick, edge label=\(k - q'\)] (f),

      (qi) -- [photon, ultra thick,  edge label=\(q\),near start] (qpf),
      (qf) -- [photon, ultra thick, edge label = \(q'\), near end] (qpi),
    };
  \end{feynman}
\end{tikzpicture}

\vspace{0.7cm}
\begin{tikzpicture}
  \begin{feynman}[layered layout, every blob = {/tikz/fill=gray!30,/tikz/inner sep=2pt}]
    \vertex (i) at (0, 0) {};
    \vertex [right=1.5cm of i] (qqpf) [blob] {\(\Gamma^{(3)}\) };
    \vertex [right=1.5cm of qqpf] (f) {};
    \vertex [above=1.6cm of i] (qi);
    \vertex [above=1.6cm of qqpf] (qqpi) [blob] {\(\tilde{\Gamma}^{(3)}\)};
    \vertex [above=1.6cm of f] (qpf) {};
    
    \diagram* {
      (i) -- [fermion, style = ultra thick, edge label=\(k-q\)] (qqpf)
      -- [fermion, style = ultra thick, edge label=\(k - q'\)] (f),

      (qi) -- [photon, ultra thick,  edge label=\(q\)] (qqpi) -- [photon, ultra thick, edge label = \(q'\)] (qpf),
      (qqpi) -- [photon, ultra thick, edge label = \(q''\)] (qqpf)
    };
  \end{feynman}
\end{tikzpicture}
\vspace{0.7cm}
\begin{tikzpicture}
  \begin{feynman}[layered layout, every blob = {/tikz/fill=gray!30,/tikz/inner sep=2pt}]
    \vertex (i) at (0, 0) {};
    \vertex [right=1.5cm of i] (qqpf) [blob] {\(\Gamma^{(4)}\) };
    \vertex [right=1.5cm of qqpf] (f) {};
    \vertex [above=1.3cm of i] (qi);
    \vertex [above=1.3cm of f] (qpf) {};
    
    \diagram* {
      (i) -- [fermion, style = ultra thick, edge label=\(k-q\)] (qqpf)
      -- [fermion, style = ultra thick, edge label=\(k - q'\)] (f),

      (qi) -- [photon, ultra thick,  edge label=\(q\)] (qqpf) -- [photon, ultra thick, edge label = \(q'\)]  (qpf)
    };
  \end{feynman}
\end{tikzpicture}
\caption{Diagrams corresponding to the evaluation of the four-point function \eqref{eq:fourpointdiagrams_supp}. The second diagram depicted is not 1PI, and corresponds to a product of three point functions together with an extra $G^{-1}$ to correct for the overcounting of the middle propagator as written in Eq. \eqref{eq:fourpointdecompose1PI}. All the other diagrams contribute to $\Xi$, the 1PI correlation function. }
\label{fig:fourpointfunctiondiagrams}
\end{figure}
\begin{equation}
\begin{aligned}
    C_{\bq m \bq' m'}^{\bk,n} & = -\big\langle \phi_{-\bq',-m'}  \psi_{\bk - \bq',n-m'}  \ov{\psi}_{\bk - \bq,n-m} \phi_{\bq,m} \big\rangle\\
    & = D(\bq  ,i\omega_m) G(\bk-\bq,i\omega_n) \beta A \delta_{\bq,\bq'}\delta_{mm'} \\
    &  -G_{\bk-\bq'} D_{\bq'}\Gamma^{(3),\bk-\bq'}_{\bq'} G_{\bk} \Gamma^{(3), \bk}_{-\bq} D_\bq G_{\bk-\bq} \\
    &  -G_{\bk-\bq'} D_{\bq'}\Gamma^{(3),\bk-\bq'}_{-\bq} G_{\bk-\bq-\bq'} \Gamma^{(3), \bk-\bq-\bq'}_{\bq'} D_\bq G_{\bk-\bq} \\
    & -iG_{\bk-\bq'} D_{\bq'} L_{\bq \bq'} D_\bq G_{\bk-\bq}\\
    & + G_{\bk-\bq'} D_{\bq'} \Gamma^{(4) \bk}_{-\bq, -\bq'} D_\bq G_{\bk-\bq}.
\end{aligned}
\label{eq:fourpointdiagrams_supp}
\end{equation}
For concision we suppressed the Matsubara indices; they may be recovered by looking at the associated momenta. We also used the short-hand
\begin{equation}
  L_{\bq m \bq' m'} = \sum_{\bq''} \Gamma^{(3)}_{-\bq''} D_{\bq''} \tilde{\Gamma}^{(3)}_{-\bq -m,\bq' m',\bq'' m-m'} \delta_{[\bq''+\bq'-\bq],0}.
  \label{eq:tildevertexshorthand}
\end{equation}
The associated diagrams are depicted in Fig. \ref{fig:fourpointfunctiondiagrams}. Note that we associated the factors $\beta^{-1}A^{-1}$ to the sums over $\bq,m$, not the diagrammatic vertices, though we do associated the factors of $(-i)$ with the diagrammatic $\Gamma^{(3)}$ vertices leading to the $-1$ prefactors in the second and third diagrams.
\begin{figure}
    \centering
\begin{tikzpicture}
  \begin{feynman}[layered layout, every blob = {/tikz/fill=gray!30,/tikz/inner sep=2pt}]
    \vertex (i) at (0, 0) {};
    \vertex [right=1.5cm of i] (qf) [dot] {};
    \vertex [right=1.5cm of qf] (f) {};
    \vertex [above=1.5cm of qf] (qi);
    blahblahblah
    \diagram* {
      (i) -- [fermion,  edge label=\(k-q\)] (qf) --  [fermion,  edge label=\(k\)] (f),
      (qi) -- [photon,  edge label=\(q\)] (qf),
    };
  \end{feynman}
\end{tikzpicture} 
\vspace{0.7cm}
\begin{tikzpicture}
  \begin{feynman}[layered layout, every blob = {/tikz/fill=gray!30,/tikz/inner sep=2pt}]
    \vertex (i) at (0, 0) {};
    \vertex [right= of i] (qpi) [blob] {$\Gamma^{(3)}$};
    \vertex [right=2cm of qpi] (qf) [blob] {$\Gamma^{(3)}$};
    \vertex [right= of qf] (qpf) [dot] {};
    \vertex [right= of qpf] (f) {};
    \vertex [above=2cm of qf] (qi) {};
    \diagram* {
      (i) -- [fermion,  edge label=\(k-q\)] (qpi) -- [fermion, style = ultra thick, edge label=\(k-q-q'\)] (qf) -- [fermion, ultra thick, edge label=\(k-q'\)] (qpf) -- [fermion, edge label=\(k\)] (f),
      (qpi) -- [photon, style = ultra thick, out=90, in=90, looseness=1, edge label=\(q'\), near start] (qpf),
      (qi) -- [photon, edge label=\(q\), near start] (qf),
    };
  \end{feynman}
\end{tikzpicture}    
\vspace{0.7cm}
\begin{tikzpicture}
  \begin{feynman}[layered layout, every blob = {/tikz/fill=gray!30,/tikz/inner sep=2pt}]
    \vertex (i) at (0, 0) {};
    \vertex [right= of i] (qqpf) [blob] {$\Gamma^{(3)}$};
    \vertex [right= of qqpf] (qpf) [dot] {};
    \vertex [right= of qpf] (f) {};
    \vertex [above= of qqpf] (qqpi) [blob] {$\tilde{\Gamma}^{(3)}$};
    \vertex [above= of qqpi] (qi) {};
    \diagram* {
      (i) -- [fermion,  edge label=\(k-q\)] (qqpf) -- [fermion, style = ultra thick, edge label=\(k-q'\)] (qpf) -- [fermion, edge label=\(k\)] (f),
      (qqpi) -- [photon, style = ultra thick, out=0, in=90, looseness=1, edge label=\(q'\)] (qpf),
      (qqpi) -- [photon, style = ultra thick, edge label=\(q''\)] (qqpf),
      (qi) -- [photon, edge label=\(q\)] (qqpi),
    };
  \end{feynman}
\end{tikzpicture}

\vspace{0.7cm}

\begin{tikzpicture}
  \begin{feynman}[layered layout, every blob = {/tikz/fill=gray!30,/tikz/inner sep=2pt}]
    \vertex (i) at (0, 0) {};
    \vertex [right=of i] (qqpf) [blob] {\(\Gamma^{(4)}\) };
    \vertex [right=2cm of qqpf] (qpf) [dot] {};
    \vertex [right=of qpf] (f) {};
    \vertex [above= of qqpf] (qi);
    
    \diagram* {
      (i) -- [fermion, edge label=\(k-q\)] (qqpf)
      -- [fermion, style = ultra thick, edge label=\(k - q'\)] (qpf) -- [fermion, edge label = $k$] (f),
      (qi) -- [photon,  edge label=\(q\)] (qqpf) -- [photon, style = ultra thick, out=45, in=90, looseness=1.3, edge label=\(q'\)]  (qpf)
    };
  \end{feynman}
\end{tikzpicture}
\caption{Diagrams associated with the SDE \eqref{eq:vertexSDE} for the $\Gamma^{(3)}$ vertex.}
\label{fig:dressedvertex_supp}
\end{figure}
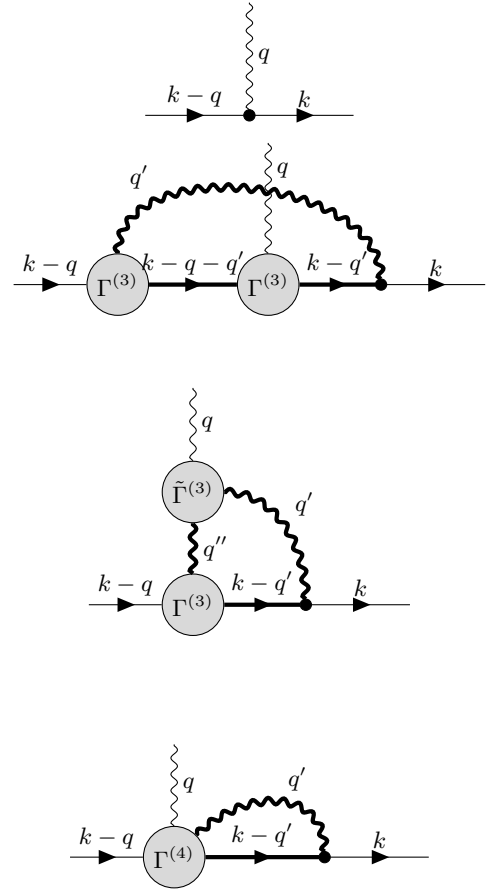

Note that we may group the diagrams in $C$ into non-1PI (first connected diagram in Fig. \ref{fig:fourpointfunctiondiagrams} and third line in \eqref{eq:fourpointdiagrams_supp}) and 1PI parts (everything else).
\begin{equation}
  \begin{aligned}
  & C_{\bq m \bq' m'}^{\bk,n}  = \Xi_{\bq m \bq' m'}^{\bk,n} \\
  & + 
\big\langle  \psi_{\bk - \bq',n-m'}  \phi_{-\bq',-m'} \ov{\psi}_{\bk,n}  \big\rangle G_{\bk,n}^{-1} \big\langle  \psi_{\bk,n}  \phi_{\bq,m} \ov{\psi}_{\bk - \bq,n-m} \big\rangle
\end{aligned}
  \label{eq:fourpointdecompose1PI}
\end{equation}
Here we again used that the three point functions above can be written in terms of $\Gamma^{(3)}$, $G$, and $D$. The factor of $G^{-1}$ comes from ``double counting'' the middle propagator of the non-1PI diagram  between the two three-point functions.

We can solve for $\Gamma^{(3)}$ through substituting into \eqref{eq:threeptSDE_supp}. The first, disconnected, diagram in Fig. \ref{fig:fourpointfunctiondiagrams}, and the corresponding first line in the evaluation of \eqref{eq:fourpointdiagrams_supp}, will lead to the bare vertex contribution to $\Gamma^{(3)}$. The second diagram appears to lead to a non-1PI contribution to $\Gamma^{(3)}$, which is not allowed; instead, it combines with the term in \eqref{eq:threeptSDE_supp} with a $G_0^{-1}$ prefactor and converts it to $G^{-1}$. Indeed, its left half is the three point function whereas its right half, when contracted with $\Lambda_{-\bq'}(\bk)$, leads to $\tilde{\Sigma}$ (see the second diagram in \eqref{eq:two_gap_eqns} and corresponding expression). The final three diagrams then lead to the dressed vertex. The vertex is then
\begin{equation}
\begin{aligned}
    -i&\Gamma^{(3)}_{-\bq}(\bk)  = -i\Lambda_{-\bq}(\bk) \\
    & + \frac{i}{\beta A}\sum_{\bq'}\Lambda_{-\bq'}(\bk) G_{\bk-\bq'} D_{\bq'}  \Gamma^{(3),\bk-\bq'}_{-\bq} G_{\bk-\bq-\bq'} \Gamma^{(3), \bk-\bq-\bq'}_{\bq'} \\
    & -\frac{1}{\beta A}\sum_{\bq'}\Lambda_{-\bq'}(\bk) G_{\bk-\bq'} D_{\bq'} L_{\bq \bq'} \\
    & -\frac{i}{\beta A}\sum_{\bq'}\Lambda_{-\bq'}(\bk) G_{\bk-\bq'} D_{\bq'} \Gamma^{(4) \bk}_{-\bq, -\bq'} .
    \end{aligned}
    \label{eq:vertexSDE}
\end{equation}
The corresponding diagrams are shown in Fig. \ref{fig:dressedvertex_supp}. Note that we derived the equation such that the right-most external vertex is bare, and all 1PI effective vertices have incoming momenta that are not near $\Gamma$ for generic $\bq,\bq'$. 

We now plug back into the self energy, specifically the first expression for $\tilde{\Sigma}$ in \eqref{eq:two_gap_eqns}. Using \eqref{eq:fourpointdecompose1PI}, and the associated expressions associated with $\Xi$ in \eqref{eq:fourpointdiagrams_supp}, we arrive at
\begin{equation}
  \begin{aligned}
    \tilde{\Sigma}(\bk,i\omega_n) & = -\frac{1}{\beta^2 A^2}\sum_{\bq,\bq'} \Lambda_{\bk,\bk-\bq'} \\
  & \times \bigg( D_\bq G_{\bk-\bq} \beta A \delta_{\bq \bq'} \delta_{mm'} \\
    & - G_{\bk-\bq'} D_{\bq'} \Gamma^{(3),\bk-\bq'}_{-\bq} G_{\bk-\bq-\bq'}\Gamma^{(3), \bk-\bq-\bq'}_{\bq'}  D_{\bq} G_{\bk-\bq}\\
    & - i G_{\bk-\bq'} D_{\bq'} L_{\bq \bq'} D_{\bq}  G_{\bk-\bq} \\
  & + G_{\bk-\bq'} D_{\bq'} \Gamma^{(4) \bk}_{-\bq, -\bq'} D_{\bq}  G_{\bk-\bq} \bigg)\\
    & \times \Lambda_{\bk-\bq,\bk}, \\
  & = -\frac{1}{\beta^2A^2}\sum_{\bq m \bq'm'} \Lambda_{\bk,\bk-\bq'} \Xi_{\bq',\bq}(\bk)\Lambda_{\bk-\bq,\bk},
\end{aligned}
  \label{eq:selfenergyintermsofXi_finalsupp}
\end{equation}
as claimed.

\subsection{Phase space approximations to the SDEs}

Let us step back and comment on the nature of the Schwinger Dyson equations we have derived; specifically Eqs \eqref{eq:two_gap_eqns}, which relates the self energy to the three-point vertex $\Gamma^{(3)}$, and \eqref{eq:vertexSDE}, which relates the three-point vertex to itself and higher-point vertices. We could of course keep going and obtain equations relating higher point vertices to successively higher point vertices. While such equations are generically difficult to solve, here we are fortunate to posess a self-consistent solution to the SDEs associated with the trivial classical model
\begin{equation}
  \begin{aligned}
 \H_{\rm{cl}} & =   \sum_{\bR, \bR'} U(\bR-\bR') (n_\bR - 4)(n_{\bR'} - 4), \\
 & U(\bR) = \frac{1}{2A} \sum_\bq V_\bq \abs{\xi_\bq}^2 e^{i \bq \cdot(\bR-\bR')},
\end{aligned}
  \label{eq:classicalmodel_supp}
\end{equation}
which corresponds to formally replacing $\Lambda_\bq(\bk) \to \xi_\bq$. Since \eqref{eq:classicalmodel_supp} is commuting, and especially simple at integer filling, in principle all correlation functions and irreducible vertices associated with it can be computed, and they will satisfy the various SDEs self-consistently. We now explain how we make use of this classical solution to solve for the 1PI vertices of the topological model, which of course differ non-perturbatively.

In each SDE, \eqref{eq:two_gap_eqns}, \eqref{eq:vertexSDE} there is a sum over a momentum transfer, $\bq$ or $\bq'$, on the right hand side. This momentum transfer shifts the momentum of one of the external legs by adding an additional bare vertex. The external vertex that the leg initially touched becomes internal. As such, this process of going higher in the SDE hierarchy is capable of making external vertices internal. This is useful because to leading order in $Us^2 \ll T$, we can neglect the contributions from internal momenta near the $\Gamma$ point; such contributions are suppressed by factors of $s^2$ due to small phase space. In \eqref{eq:two_gap_eqns} and \eqref{eq:vertexSDE}, this results in
\begin{equation}
\begin{aligned}
\tilde{\Sigma}_{\bk,n} & = -\frac{1}{\beta A}\sum_{\bq m} D_{\bq,m} \Gamma^{(3)}_{-\bq -m}(\bk,i\omega_n) G_{*,n} \Lambda_{\bk-\bq,\bk}) \\
-i&\Gamma^{(3)}_{-\bq}(\bk)  = -i\Lambda_{\bk,\bk-\bq} \\
& + \frac{i}{A}\sum_{\bq'}\Lambda_{\bk,\bk-\bq'} G_{*} D_{\bq'}  \Gamma^{(3),*}_{-\bq} G_{*} \Gamma^{(3), *}_{\bq'} \\
& -\frac{1}{A}\sum_{\bq'}\Lambda_{\bk,\bk-\bq'} G_{*} D_{\bq'} \Gamma^{(3) *}_{\bq'-\bq} D_{\bq-\bq'} \tilde{\Gamma}^{(3)}_{\bq-\bq',\bq'} \\
    & -\frac{i}{A}\sum_{\bq'}\Lambda_{\bk,\bk-\bq'} G_{*} D_{\bq'} \Gamma^{(4) *}_{-\bq, -\bq'} \\
    & = -i \Lambda_{\bk-\bq',\bk} \Xi_{\bq \bq'}(*).
    \end{aligned}
    \label{eq:approxSDE}
\end{equation}
In the second equation, we additionally assumed that $\bk - \bq = *$ is not near $\Gamma$ since we forecast plugging into the first equation where $\bk-\bq$ is an internal momentum (summed over).

We may now solve \eqref{eq:approxSDE}, and further SDEs in the hierarchy approximated in the same way, in two steps. First, we argue that quantities with Bloch momenta $\bk = *$ can be evaluated in the trivial classical model \eqref{eq:classicalmodel_supp}. To see this, we set $\bk \to *$. Then \emph{all} form factors and effective action coefficients in \eqref{eq:approxSDE} are evaluated far away from $\bk \approx \Gamma$, not just internal ones. This exactly reproduces the SDEs associated with \eqref{eq:classicalmodel_supp}, which we can obtain the solution to through solving the classical model in real space. Intuitively, from the perspective of the topological model, the area of the BZ away from the $\Gamma$ point only rarely scatters into $\Gamma$; to leading order in $s^2$ it functions as a large $\bk$-space bath that predominantly interacts with itself. These bath-bath interactions can be solved for exactly in real space using \eqref{eq:classicalmodel_supp}, since they are independent of the topological $\Gamma$-point region.

Second, quantities with Bloch momenta $\bk$, which can be near $\Gamma$, can be solved for in terms of irreducible vertices of the classical model; the region away from $\Gamma$ acts as a bath that controls the self energy and dynamics of $\Gamma$ electrons. Indeed, the right hand side of the second SDE in \eqref{eq:approxSDE} enables us to compute the three point vertex in terms of form factors and classical vertices on the right hand side, because right hand side vertices only involve starred BZ momenta. We can then obtain the self energy from the first equation in \eqref{eq:approxSDE}. This procedure can in principle be carried out, but rather than compute all of the classical vertices individually it is easier to compute $\Xi_{\bq,\bq'}(*)$, which includes all of the classical vertices in the correct combination, and then directly evaluate the self energy in terms of $\Xi$ as we did in the main text.

\end{document}